\documentclass[preprint]{elsarticle}
\usepackage{amsmath}
\usepackage{amssymb}
\usepackage{fullpage}
\usepackage{natbib}
\usepackage{subcaption}

\def\dint{\int}

\newcommand{\amsize}{0.05}

\newcommand{\asize}{0.12}
\newcommand{\bsize}{0.15}
\newcommand{\bcsize}{0.2}

\newcommand{\abheight}{3}

\newcommand{\cheight}{6}
\newcommand{\dheight}{8}
\newcommand{\eheight}{10}

\journal{Computer Methods in Applied Mechanics and Engineering}
\begin{document}

\begin{frontmatter}
\title{Bending models of lipid bilayer membranes:
\\ spontaneous curvature and area-difference elasticity}
\author[cse]{Xin Bian}
\author[cse]{Sergey Litvinov}
\author[cse]{Petros Koumoutsakos\corref{correspondingauthor}}
\ead{petros@ethz.ch}
\address[cse]{Computational Science and Engineering Laboratory,  ETH Z\"urich,
Clausiusstrasse 33, Z\"urich CH-8092, Switzerland}
\cortext[correspondingauthor]{Corresponding author}
\begin{abstract}
We preset a computational study of bending models for the curvature
elasticity of lipid bilayer membranes  that are relevant for simulations
of vesicles and red blood cells.  We compute bending energy and forces
on triangulated  meshes and evaluate and extend four well established
schemes for their approximation: Kantor and Nelson~\cite{Kantor1987a},
J\"ulicher~\cite{Juelicher1996}, Gompper and Kroll~\cite{Gompper1996}
and Meyer et. al.~\cite{Meyer2003}, termed A, B, C, D.  We present a
comparative study of these four schemes on the minimal bending model and
propose extensions  for schemes B, C and D. These extensions incorporate
the reference state and non-local energy to account for the spontaneous
curvature, bilayer coupling, and area-difference elasticity models.
Our results indicate that the proposed extensions enhance the models
to account for shape transformation including budding/vesiculation as
well as for non-axisymmetric shapes.  We find that the extended scheme
B is superior to the rest in terms of accuracy, and robustness as well
as simplicity of implementation. We demonstrate the capabilities of this
scheme on several benchmark problems including  the budding-vesiculating
process and the reproduction of the phase diagram of vesicles.
\end{abstract}


\end{frontmatter}

\section{Introduction}

Lipid bilayers are fundamental structural elements in biology.
Vesicles are compartments enclosed by lipid bilayer
membranes and they have been studied extensively over the last  four
decades~\cite{Deuling1976a, Seifert1997, Deserno2015, Tu2018}. Notably
vesicles have been the subject of investigations that were awarded
the 2013 Nobel prize in the physiology or medicine (James Rothman,
Randy Schekman and Thomas S\"udhof).  A vesicle has a relatively simple
structure when compared to other entities at cellular level and they
have been the subject of several detailed theoretical~\cite{Canham1970,
Helfrich1973a, Sheetz1974, Evans1974, Svetina1989, Seifert1992, Bozic1992,
Wiese1992} and experimental studies~\cite{Sackmann1986, Kaes1991}.
Models of vesicles are readily transferable to more complex structures,
as lipid bilayer membranes enclose most cells and their nuclei as well
as   many viruses. Importantly,  a lipid bilayer and a spectrin network
constitute a red blood cell (RBC) membrane~\cite{Evans1980, Mohandas1994}.

Lipid membranes are a few nanometers thin, but they span surfaces
that are in the order of micrometers. 
The bending elasticity of the quasi two-dimensional
structures is caused by relative extension/compression of
outer/inner surface, which can be expressed as a  function of the membrane
curvature~\cite{Seifert1997, Deserno2015, Tu2018}.  Canham proposed the
first form of membrane bending energy, that depends quadratically on
the two principal curvatures of the surface~\cite{Canham1970}. This
representation, along with constraints on total area and volume,
constitutes the {\it minimal} model for a vesicle. Helfrich formulated
a bending energy that includes a Gaussian curvature as well as a
"spontaneous" curvature $2H_0$~\cite{Helfrich1973a} that is attributed
to chemical differences in two lipid layers.
This is the {\it spontaneous curvature} (SC) model and by setting
$H_0=0$, it is equivalent to the minimal model.  In turn, Sheetz \&
Singer~\cite{Sheetz1974} and Evans~\cite{Evans1974} accounted for the
density difference of the two lipid layers.  Their work led to the
{\it bilayer couple}~(BC) model~\cite{Svetina1989} that introduces
a constraint on the area difference so that $\Delta A = \Delta A_0$,
where $\Delta A$ and $\Delta A_0$ are the instantaneous and targeted
area differences between the two lipid layers.  A few years later three
groups (\cite{Seifert1992, Bozic1992, Wiese1992}) proposed that the
non-local bending energy has a quadratic form $(\Delta A - \Delta
A_0)^2$.  Here $\Delta A_0$ reflects reference of the area difference
between the two layers.  This is {\it area-difference elasticity}
(ADE) model, which degenerates to the minimal, SC or BC model as a
special case.  The ADE model has been validated experimentally on
vesicles~\cite{Doebereiner1997,Sakashita2012}.

Despite the well established formulation of these membrane models,
simulations of vesicles and red-blood cells remain a challenging task. The
key difficulty lies in the evaluation of the resulting forces. The bending
force depends on the Laplace-Beltrami operator $\nabla^2_s$ on the mean
curvature, which further relates to $\nabla^2_s$ of the coordinates
on a surface.  Hence, the evaluation of  the bending force requires
approximations of the fourth derivatives of the surface coordinates.
A recent review showcased how computation of the bending energy
and especially the force remain elusive, even for the minimal
model~\cite{Guckenberger2017}.  These  computations have similar
difficulties as those met when  calculating Willmore flow
in differential geometry~\cite{Deserno2015, Guckenberger2017},
which are of particular interest to the computer graphics
community~\cite{crane2013robust}.

One approach has focused on axi-symmetric shapes
that reduces the calculations to solving Ordinary Differential
Equations (ODEs)~\cite{Deuling1976a, Seifert1991,Miao1994,Tu2018}.
Recent works have employed either spherical harmonic expansions/quadratic approximation
to represent the surface~\cite{Heinrich1993, Khairy2008a, Veerapaneni2011, Yazdani2012} 
or phase field~\cite{Du2004} models. We note in
particular level set methods for the description of the membrane
surface~\cite{Maitre2009,bergdorf2010lagrangian,Mietke2019} thus allowing
for volume based calculations of the membrane dynamics and 
large deformations even with topological changes.

In this work, we represent the surface as a triangulated mesh,
which has been used extensively to model vesicle and RBC membranes and they have been coupled to fluid
solvers~\cite{Noguchi2005a, Pivkin2008, Fedosov2010b, Tsubota2014, Li2013, Guckenberger2016}. 
However, the majority of vesicle and RBC simulations have employed only the minimal model. 
Exceptions are Monte Carlo simulations from groups
of Seifert~\cite{Jaric1995} and Wortis~\cite{Lim2002} and the finite
element method from Barrett et al~\cite{Barrett2008a}.
We examine and extend four established schemes for the discretization of the general bending
model on a triangulated mesh~\cite{Kantor1987a,  Juelicher1996, Gompper1996, Meyer2003}.
We perform a number of benchmark problems to compare their accuracy, robustness
and stability.

The paper is structured as follows: we describe energy functionals of the
minimal, SC, BC and ADE models in Section~\ref{sec_continuum} and present
an expression for the force density.  In Section~\ref{sec_scheme},
we present the four discretization schemes (A, B, C, D) which have
been  previously applied to the minimal model~\cite{Kantor1987a,
Juelicher1996, Gompper1996, Meyer2003}.  We extend schemes B, C, D to
incorporate the spontaneous curvature and non-local bending energy thus
accessing the  SC, and BC/ADE models.  The details of the derivations
on Sections~\ref{sec_continuum} and \ref{sec_scheme} can be found
in~\ref{sec_appendix_force_variational_formulation}--\ref{sec_appendix_volume}.
In Section~\ref{sec_numerics1}, we evaluate the proposed schemes
on shapes with known analytical expressions for the energy and force.
In Section~\ref{sec_numerics2} we compute the phase diagrams for
oblate-stomatocyte, oblate-discocyte, prolate-dumbbell, prolate-cigar
shapes described by the minimal model with the four discretization
schemes, from different initial shapes such as prolate, oblate and sphere. 
Thereafter,  we generate slices of phase diagrams of SC, BC
and ADE models.  
We compare our results with solutions obtained by  axi-symmetric ODEs, spherical harmonic as
well as results from the program "Surface Evolver" and experiments. 
In Section~\ref{sec_numerics3} we present results from dynamic simulations.
We summarize our findings in Section~\ref{sec_summary}.
\section{Continuous energy and force}
\label{sec_continuum}

\subsection{Energy functionals}
\label{sec_continuum_energy}
The primary model for the two-dimensional membrane elasticity is based on
an extension of one dimensional beam theory by Canham~\cite{Canham1970}
with an energy functional: \begin{equation} E^C = \frac{\kappa_b}{2}\dint
(C^2_1 +C^2_2) dA, \label{eq_energy_canham} \end{equation} where
$\kappa_b$ is the bending elastic constant, $C_1$ and $C_2$ are the two
principal curvatures\footnote{We take a convention that $C_1$ and $C_2$
are positive for a sphere.  With this convention, the surface unit normal
vector points inwards.} and the  integral  is taken over a two-dimensional
parametric surface embedded in three-dimensional space.

The {\it Canham energy functional} is scale-invariant, with  smaller
vesicles corresponding to larger curvatures. This energy functional
is known as the {\it minimal model} of a vesicle membrane.  Its phase
diagram is determined by the reduced volume between a vesicle and a sphere
$v=3V/(4\pi R^3)$, with the same area and $R$ is the sphere's radius.

A few years later, inspired by research in liquid crystal,
Helfrich~\cite{Helfrich1973a} developed an alternative free energy
functional for lipid bilayers
\begin{eqnarray}
E^H &=& 2{\kappa_b}  \dint
(H-H_0)^2   dA  +  \kappa_g \dint  G  dA,
\label{eq_energy_helfrich}
\end{eqnarray}
where $H=(C_1+C_2)/2$ is the mean curvature and $2H_0$ is the {\it
spontaneous curvature}, reflecting the asymmetry in the chemical
potential on the two sides of the lipid bilayer.  $G=C_1C_2$ is the
Gaussian curvature, and $\kappa_g$ is a bending elastic constant.
According to Gauss-Bonnet theorem $\dint  G  dA=4\pi(1-g)$, where $g$
is the genus of the surface.  For the spherical topologies considered
here, $g=0$ and the energy term $\kappa_g\dint  G dA$ is omitted.
The remaining term in Eq. \ref{eq_energy_helfrich} constitutes the {\it
spontaneous curvature}~(SC) model~\cite{Helfrich1973a}. We note that
for $H_0=0$, Eq.~(\ref{eq_energy_canham}) has the same dynamics as that
of Eq.~(\ref{eq_energy_helfrich}).  The Helfrich energy with $H_0=0$
is also scale-invariant.  However, a non-zero $H_0$ introduces a length
scale and we define $h_0 = H_0/R$.

It is important to remark that most simulations of vesicles or
RBCs~\cite{Li2013, Tsubota2014, Guckenberger2016} use $H_0=0$, thus
assuming no chemical differences between the two sides of the lipid
bilayer. However it has been noted~\cite{Seifert1997} that
such an assumption lacks a physical realization.  We denote the energy
due to spontaneous curvature as
\begin{eqnarray}
 E^S =E^H-E^H_{H_0=0}=2\kappa_b \dint {H_0}^2 \,dA - 2\kappa_b \dint
 H\,H_0  \, dA,
\label{eq_energy_spontaneous}
\end{eqnarray}
The first term resembles surface tension energy encountered in models
of  multiphase flow~\cite{Landau1987} with $2\kappa_bH^2_0$ is
a "surface tension" constant \footnote{Ideally, a lipid membrane is
incompressible and the total area $A=\dint  dA$ of the neutral surface is
constant.  However, numerically we treat the constraint on area
as penalization and therefore, we also consider the energies
and forces due to total area arising from SC and ADE models.}.

In the early 90s, three groups~\cite{Seifert1992, Bozic1992, Wiese1992}
proposed an additional term to the  energy  functional that reflects
the area difference between the outer and inner leaflets of the lipid
membrane.  This non-local term is {\it area-difference
elasticity (ADE)} and it is expressed as:
\begin{eqnarray}
E^{AD}=\frac{\alpha \kappa_{b}  \pi}{2AD^2}\left(\Delta A - \Delta
A_0\right)^2.  \label{eq_energy_area_difference}
\end{eqnarray}
Here $\alpha \kappa_b$ is a bending elastic constant due to area
difference and $D$ is the thickness of the bilayer.  The ratio $\alpha$ is
in the order of unity~\cite{Bozic1992, Seifert1992, Miao1994, Waugh1995,
Lim2002} and depends on the properties of the lipid.

In summary, the Helfrich model extended by the ADE terms, is the so
called {\it ADE model}~\cite{Seifert1997} that can be expressed as:
\begin{eqnarray}
E &=& E^H + E^{AD} =  2\kappa_b  \dint  (H-H_0)^2   dA  
+  \frac{ \alpha \kappa_{b} \pi}{2AD^2}\left(\Delta A - \Delta A_0\right)^2,
\label{eq_energy_total}
\end{eqnarray}

Setting $H_0=0$ in the ADE model and constrain $\Delta A = \Delta
A_0$ we obtain the {\it bilayer-couple (BC)} model~\cite{Svetina1989}
that includes the area difference as a constraint.  We simplify the
formulation by introducing the zero, first and second moments of the
mean curvature $H$
\begin{align}
\mathcal{M}_0= A=\dint dA, \quad
 \mathcal{M}_1=\dint H dA, \quad
 \mathcal{M}_2=\dint H^2 dA.
 \label{eq_moments_continuum}
\end{align}
We rewrite the total energy as
\begin{align}
E = \underbrace{2\kappa_b  \mathcal{M}_2}_{E^H}
+  \underbrace{\frac{2  \alpha \kappa_b \pi}{A} \mathcal{M}^2_1}_{E^{AD}} 
- \underbrace{ 4\kappa_b H_0   \mathcal{M}_1}_{E^{H}} 
-  \underbrace{ \frac{2\alpha \kappa_b \pi}{A} \frac{\Delta A_0}{D}  \mathcal{M}_1  }_{E^{AD}}
+ \underbrace{2 \kappa_b H^2_0 A}_{E^H}  
+ \underbrace{\frac{\alpha \kappa_b \pi}{2A} \left(\frac{ \Delta A_0}{D}\right)^2}_{E^{AD}},
  \label{eq_energy_continuum_group1}
\end{align}
Here  the "$\underbrace{}$" indicates that the terms originate from
either $E^H$ or $E^{AD}$, and we have used the expression $\Delta A  =2D\mathcal{M}_1$.

\subsection{Force from calculus of variation}
\label{sec_continuum_force}
We use the energy functional and the calculus of variations to derive
(details are shown in~\ref{sec_appendix_force_variational_formulation})
the force magnitude corresponding to each moment as
\begin{align}
\mathcal{F}_0 = 2H, \quad 
\mathcal{F}_1 = G, \quad 
\mathcal{ F}_2 = -2H \left(H^2-G\right) - \nabla^2_s H,
\label{eq_moments_force}
\end{align}
Here $\nabla_s$ is the surface gradient operator and $\nabla^2_s=\nabla_s
\cdot \nabla_s$ is the Laplace-Beltrami operator.

We use the above expressions to write the magnitude of force density
acting along the normal ${\bf n}$  as
\begin{align}
f &= \underbrace{ 2\kappa_b  \mathcal{F}_2 }_{f^H}   
+ \underbrace{ \frac{4 \alpha \kappa_b \pi}{A} \mathcal{M}_1 \mathcal{F}_1 
- \frac{2 \alpha \kappa_b \pi}{A^2} \mathcal{M}^2_1 \mathcal{F}_0 }_{f^{AD}}  
 - \underbrace{ 4\kappa_{b} H_0  \mathcal{F}_1 }_{f^H}   \nonumber \\
& - \underbrace{ \frac{2 \alpha \kappa_b \pi}{A} \frac{\Delta A_0}{D} \mathcal{F}_1   
+  \frac{2 \alpha \kappa_b \pi}{A^2} \frac{\Delta A_0}{D} \mathcal{M}_1 \mathcal{F}_0 }_{f^{AD}}
+ \underbrace{  2 \kappa_b H^2_0   \mathcal{F}_0 }_{f^H}   
- \underbrace{ \frac{\alpha \kappa_b \pi}{2A^2}   \left(\frac{ \Delta A_0}{D}\right)^2 \mathcal{F}_0 }_{f^{AD}},
 \label{eq_force_density_group1}
\end{align}.

These terms can be rearranged to show the components of the force
density corresponding to the energy functionals as
\begin{align}
{f}^H & = -2\kappa_b \left[ 2(H-H_0)(H^2+H_0H-G)+\nabla^2_sH \right], \nonumber \\
{f}^{AD} &=\alpha \kappa_{b} \pi 
\left[
\left( 2\mathcal{M}_1 - \frac{\Delta A_0}{D} \right) \frac{2G}{A}  
- \left( 2\mathcal{M}_1 - \frac{\Delta A_0}{D} \right)^2 \frac{H}{A^2}
\right].
\label{eq_force_density_individual}
\end{align}
\section{Discretization schemes}
\label{sec_scheme}
\begin{figure}
\centering
\begin{subfigure}{0.48\textwidth}
\includegraphics[width=\columnwidth]{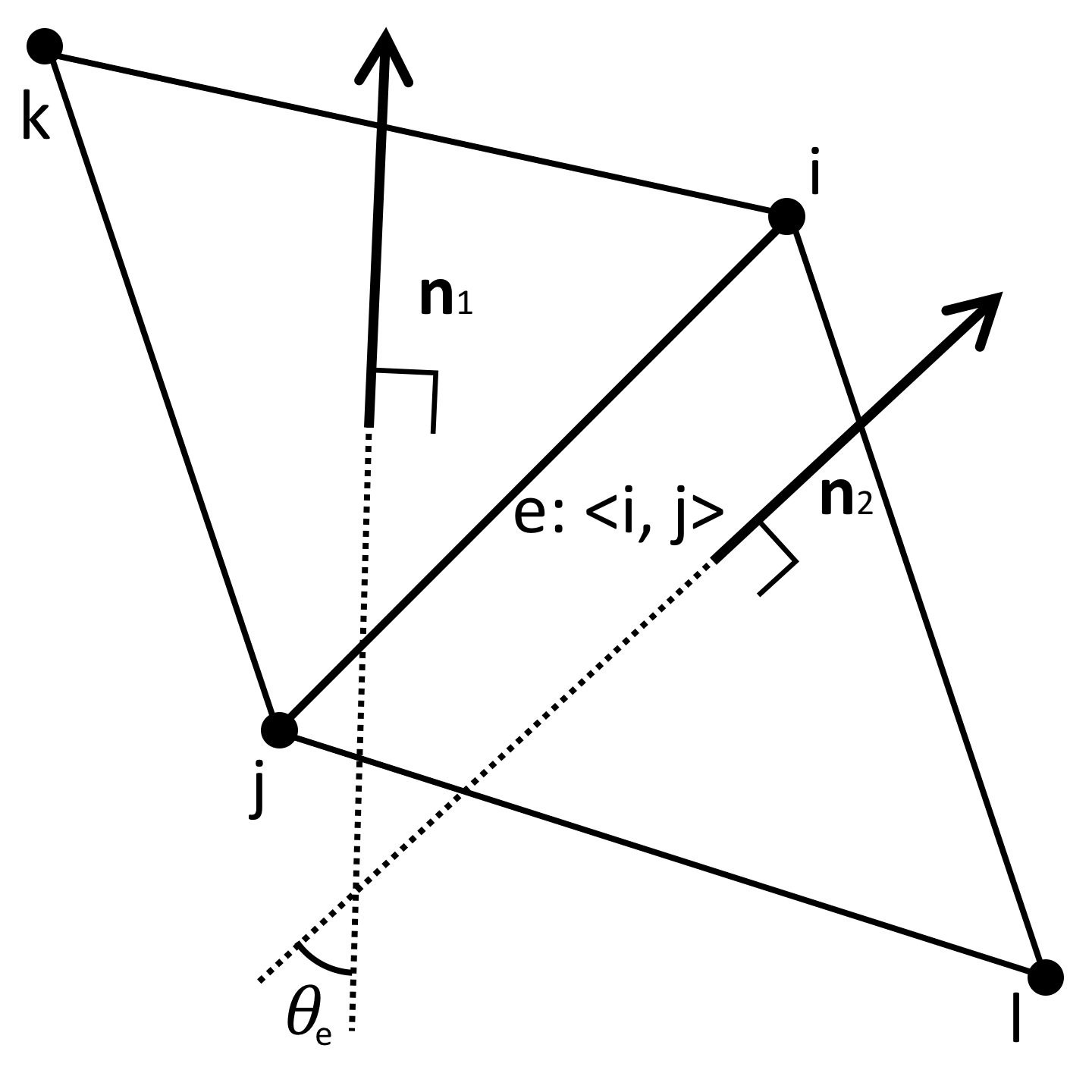}
\caption{}
\end{subfigure}
\begin{subfigure}{0.48\textwidth}
\includegraphics[width=\columnwidth]{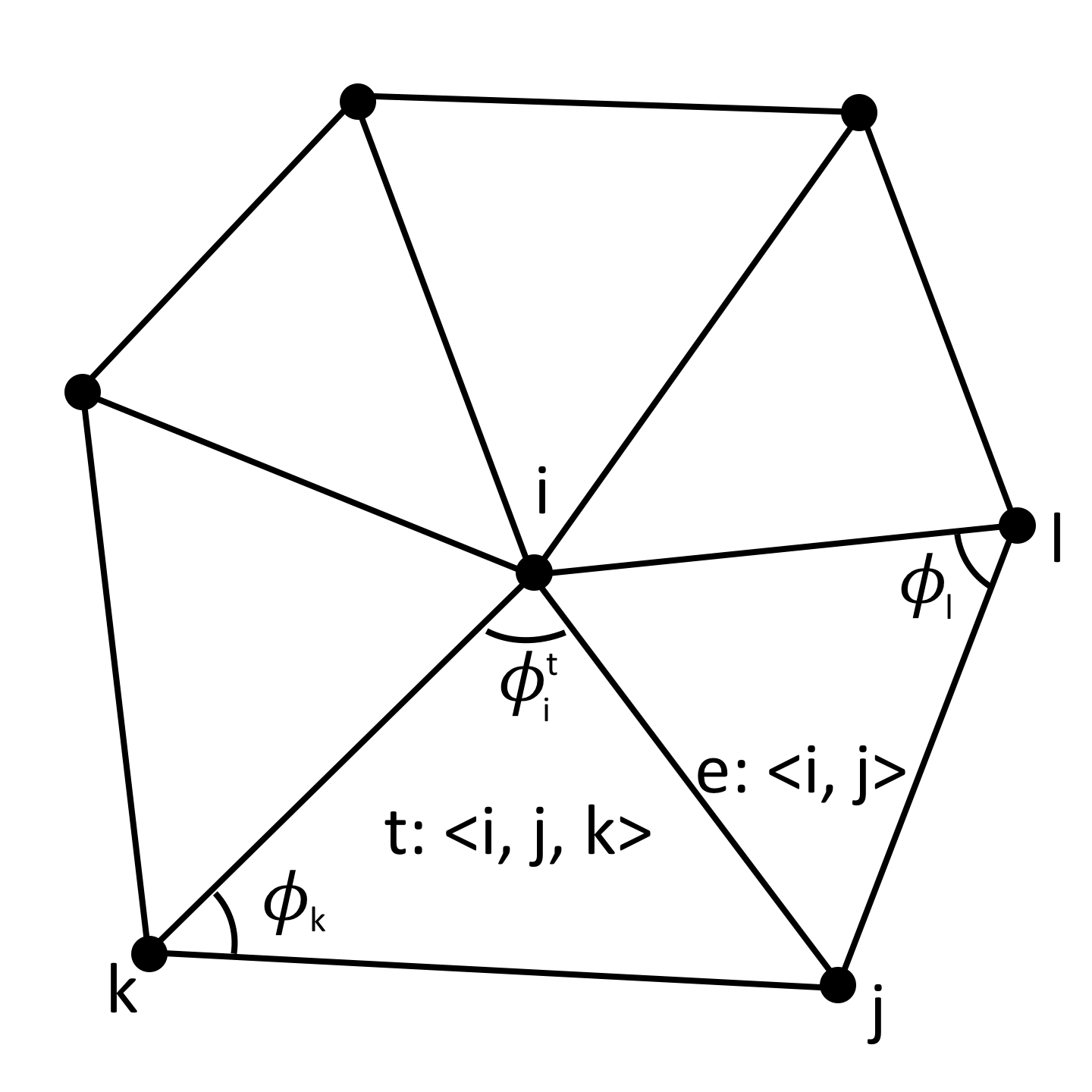}
\caption{}
\end{subfigure}
\caption{(a) An angle $\theta_e$ is formed by ${\bf n}_1$ and ${\bf n}_2$,
which are unit normal vectors of two triangles
sharing the same edge $e:<i,j>$, angle $\theta_e$ is relevant for energy and force calculation of scheme A and B.
(b) notations for the indices of vertices, edges, angles and triangles 
are relevant for all four schemes.}
\label{fig_mesh_surface}
\end{figure}
We  represent the membrane as a triangulated mesh and use the following
notation (see Fig.~\ref{fig_mesh_surface}):
\begin{itemize}
\item $i,j,k,l$ are indices for vertices
\item $e:\left<i,j\right>$ is index for the edge connecting vertices $i$ and $j$.
\item $t:\left<i,j,k\right>$ is index for the triangle connecting vertices $i,j$ and $k$.
\item $\theta_e$ is the angle between the normal vectors of two triangles sharing edge $e$.
\item $\phi_k$ and $\phi_l$ are angles associated with vertices $k$ and $l$.
 $\phi^t_k$ is the angle associated with vertex $k$ within triangle $t:\left<i,j,k\right>$.
\item $N_v, N_e, N_t$ are the numbers of vertices, edges and triangles.
\end{itemize}

The total energy is
\begin{eqnarray}
E= E^H + E^{AD} = 2\kappa_b \sum^{N_v}_i \left(H_i-H_0\right)^2A_i
+\frac{ \alpha \kappa_b \pi}{2AD^2} \left(  2D\sum^{N_v}_i H_i A_i - \Delta A_0 \right)^2.
\label{eq_energy_discrete}
\end{eqnarray}
where $\Delta A=2D \sum^{N_v}_i H_i A_i$ is the area difference.

We  introduce the discrete moments as
\footnote{We do not differentiate continuous and discrete symbols
if they are clear from the context.}
\begin{align}
\mathcal{M}_0 =  A = \sum^{N_v}_i A_i, \quad
\mathcal{M}_1 = \sum^{N_v}_i H_i A_i, \quad
\mathcal{M}_2 = \sum^{N_v}_i  H^2_i A_i.
 \label{eq_moments_discrete}
\end{align}
so that the total energy becomes
\begin{align}
E = 2\kappa_b  \mathcal{M}_2   
   +  \frac{2  \alpha \kappa_b \pi}{A} \mathcal{M}^2_1 
  - \frac{2\pi\kappa_{ad} \Delta a'_0}{D}   \mathcal{M}_1   + 2  \kappa_b H^2_0 A  
  + \frac{\alpha \kappa_b \pi}{2A}   \left(\frac{ \Delta A_0}{D}\right)^2,
\label{eq_energy_discrete_group2}
\end{align}
where  $\Delta a'_0= \frac{2DH_0}{\alpha \pi} +\frac{\Delta A_0}{A}$
is non-dimensional.

It is apparent that $H_0$ from Helfrich energy and $\Delta A_0$ from
ADE energy may compensate each other's mechanic effects, although the
origins of the two are different.  Therefore, we take their combined
effect represented by $\Delta a'_0$ \cite{Lim2002, Khairy2008a}.
We assume $H_0=0$ and $\Delta a'_0= \frac{\Delta A_0}{A}$ in the ADE
model so that the resultant $\Delta A$ from a simulation of the ADE
model can be non-dimensionalized as $\Delta a'=\Delta A/A$.  Therefore,
for a given area $A$, we have three parameters $v$, $\alpha$, and
$\Delta a'_0$.  Other work have adopted $\Delta a_0=\Delta A_0/(4\pi R
D)$ and  $\Delta a=\Delta A/(4\pi R D)$ to non-dimensionalize the area
differences, where the denominator corresponds to the area difference of
a sphere~\cite{Seifert1991, Ziherl2005}.  This approach has the advantage
of eliminating the membrane thickness $D$.  To compare with different
references, we keep both means of non-dimensionalization and they are
related as $\Delta a = \Delta a'R/ D$ and $\Delta a_0 = \Delta a'_0R/ D$.

We consider four discretization schemes which have been applied to
the minimal model in the past.  Note that scheme A defines the energy
without defining $H_i$ and $A_i$.  In turn, for schemes B, C, and D,
there are explicit definitions of $H_i$ and $A_i$. In the present
work we  extend these three schemes to account for all energy terms in
Eq.~(\ref{eq_energy_discrete_group2}) and their corresponding forces.

\subsection{Discrete force}
\begin{table}
\centering
\caption{Summary of discretization schemes: $H$ is the mean curvature; $G$ is
the Gaussian curvature; $H_0$ is half the spontaneous curvature; ADE
stands for area-difference elasticity; Momenta denote both linear and
angular momenta.  Symbol ``$\surd$" denote that the original scheme
has a definition of $H$, or has a definition of $G$, or allows for
incorporating $H_0$, or allows for incorporating energy and force due
to ADE, or conserve both linear and angular momenta.}
\begin{tabular}{cccccc}
& $H$ & $G$ & $H_0$ & ADE & momenta \\
A \cite{Kantor1987a} &  &   &  &  &  $\surd$  \\
B \cite{Juelicher1996}  & $\surd$ &  & $\surd$  & $\surd$  &   $\surd$  \\
C \cite{Gompper1996} & $\surd$  &  & $\surd$ & $\surd$ &   $\surd$ \\
D \cite{Meyer2003} & $\surd$ & $\surd$  & $\surd$  & $\surd$ & \\
\end{tabular}
\label{table_schemes_summary}
\end{table}

For schemes A, B, and C,
the force at a vertex ${\bf x}_m$ is
\begin{align}
{\bf F}_m=- \frac{\partial E}{\partial {\bf x}_m},
\end{align}
We may convert between force and force density associated with each vertex as
\begin{align}
{\bf f}_m={\bf F}_m/A_m, \quad {\bf F}_m = {\bf f}_m A_m,
\end{align}
where $A_m$ is the area associated with vertex $m$.
The former is used by scheme A, B, and C, to convert force
to force density for comparison with calculus of variation in
Section~\ref{sec_continuum_force}.  For scheme D, the force density
is computed directly and converted to force to use in the energy
minimization process.  A summary of the discretization schemes is shown
in Table~\ref{table_schemes_summary} and we elaborate each scheme in
the following.

\subsection{Scheme A}
\label{section_scheme_a}
This scheme was invented by Kantor \& Nelson ~\cite{Kantor1987a} to
simulate planar polymeric network in Monge form. This model was later
broadly adopted to model generally curved membranes, in particular for
red blood cells~\cite{Discher1998, Li2005, Pivkin2008, Fedosov2010b,
Vliegenthart2011}. The energy depends on each angle $\theta_e$
formed by the normal vectors of two triangles,
which share the same edge $e:\left<i,j\right>$ (Fig.~\ref{fig_mesh_surface}(a)).
The total energy is
\begin{align}
E = 2 \tilde{\kappa}_b \sum^{N_e}_{e: \left<i,j\right>}  \left[1-\cos \left( \theta_{e} - \theta_0 \right) \right],
\label{eq_energy_a}
\end{align}
where $\tilde{\kappa}_b$ is the bending modulus.
We split the local energy evenly onto two vertices $i$ and $j$.
Alternatively, splitting  the energy evenly onto the four vertices $i, j, k, l$,
has marginal differences.
There is no difference on the force between the two ways of splitting.
$\theta_0$ is the spontaneous angle,
that is, for $\theta_e \equiv \theta_0$, $E \equiv 0$.
If $\theta_e-\theta_0 \approx 0$,
by Taylor expansion $\cos(\theta_e-\theta_0) \approx  1-(\theta_e-\theta_0 )^2/2$.
Therefore, the approximation of the energy reads~\cite{Krueger2012}
\begin{eqnarray}
E^{1st}= \tilde{\kappa}_b \sum^{N_e}_{e: \left<i,j\right>}  \left( \theta_{e} - \theta_0 \right)^2.
\label{eq_energy_a1st}
\end{eqnarray}

A few notes are in order; for a triangulated cylinder surface, $\tilde
{\kappa}_b = \sqrt{3} \kappa_b$ can be derived~\cite{Seung1988},
but it is not true for other shapes~\cite{Gompper1996}, as
demonstrated later.  For a mesh of uniform equilateral triangles on
a sphere, the $\theta_0$ is related to the $H_0$ via a resolution
parameter (e.g, the edge length of the triangle).  However, $\theta_0$
and $H_0$ have no correspondence  in general.  Furthermore,  there
is no definition of $H_i$ or $A_i$ in this scheme and therefore,
it is ambiguous to extend it to include ADE.

We will consider scheme A in its original form with
$\theta_0$ (and its approximation) and compare it with the
minimal model.  The expression for the discrete force is given in
\ref{sec_appendix_force_scheme_a} and the only primitive derivative to
compute is $\frac{\partial \theta_e}{\partial {\bf x}_m}$.

\subsection{Scheme B}
\label{section_scheme_b}
\begin{figure}
\centering
\includegraphics[width=0.5\textwidth]{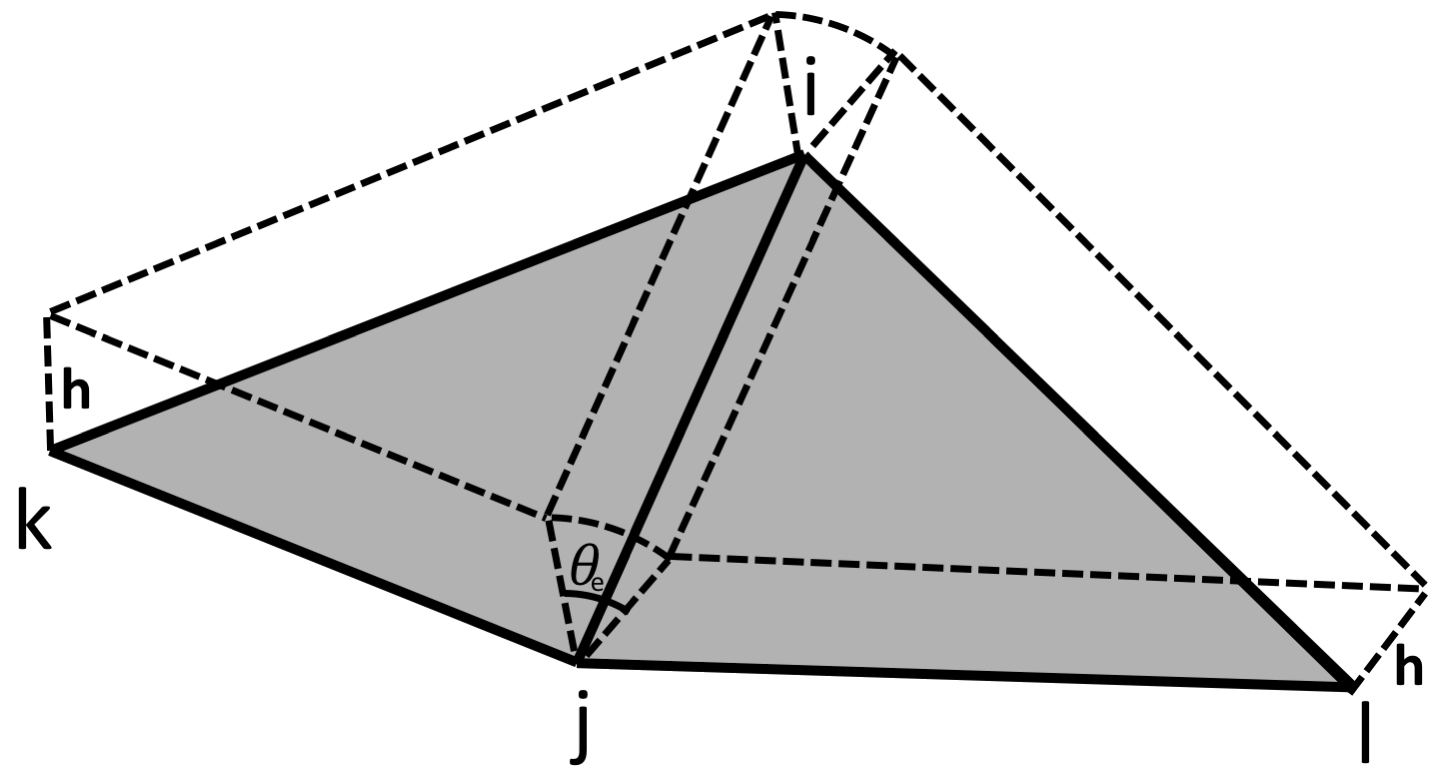}
\caption{A sketch of two imaginary triangles parallel to the two correspondingly
neighboring triangles and a cylinder so formed.  The local first moment
of mean curvature is concentrated on the cylinder.}
\label{fig_two_triangles_parallel}
\end{figure}
\begin{figure}
\centering
\begin{subfigure}{0.3\textwidth}
\includegraphics[width=\columnwidth]{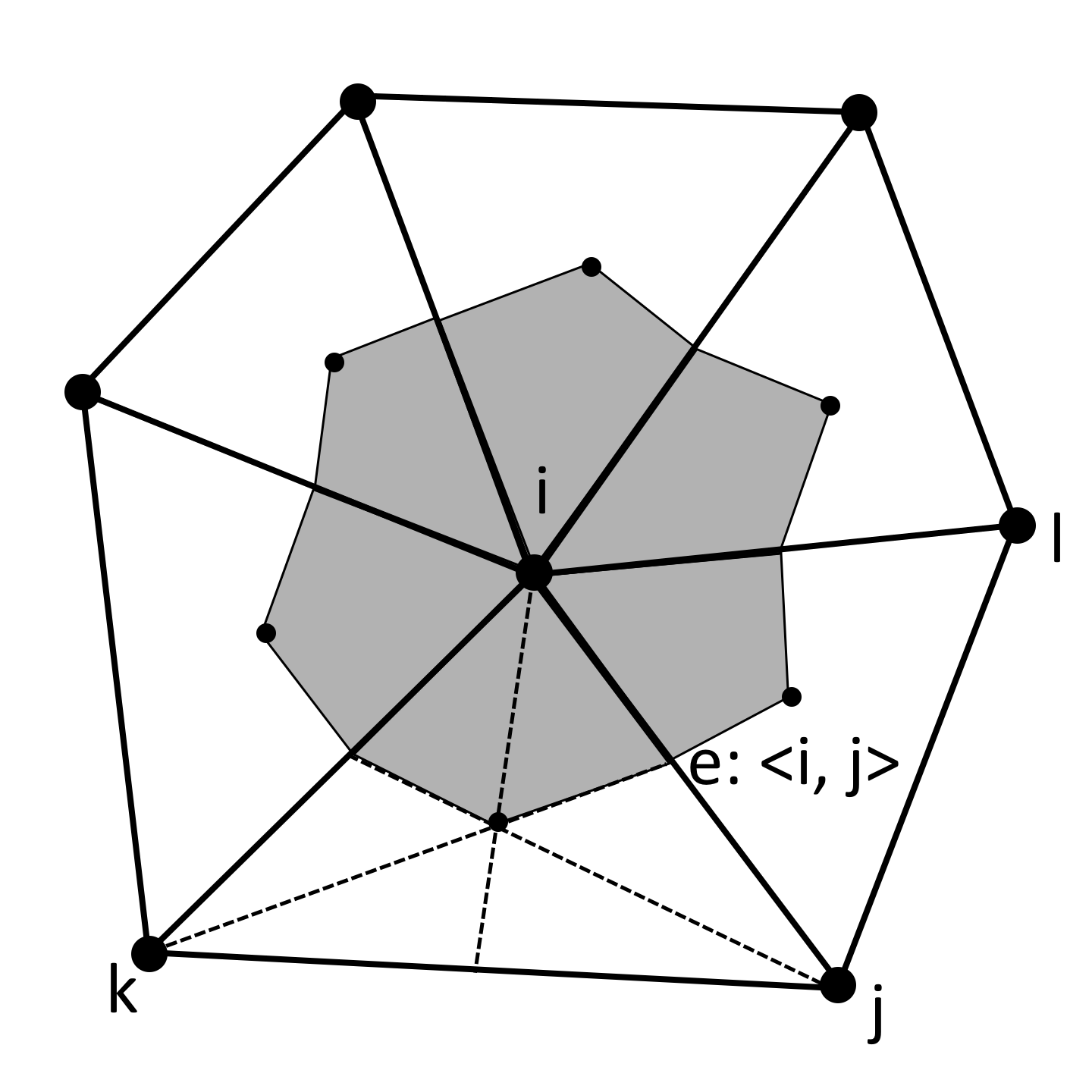}
\caption{}
\end{subfigure}
\begin{subfigure}{0.3\textwidth}
\includegraphics[width=\columnwidth]{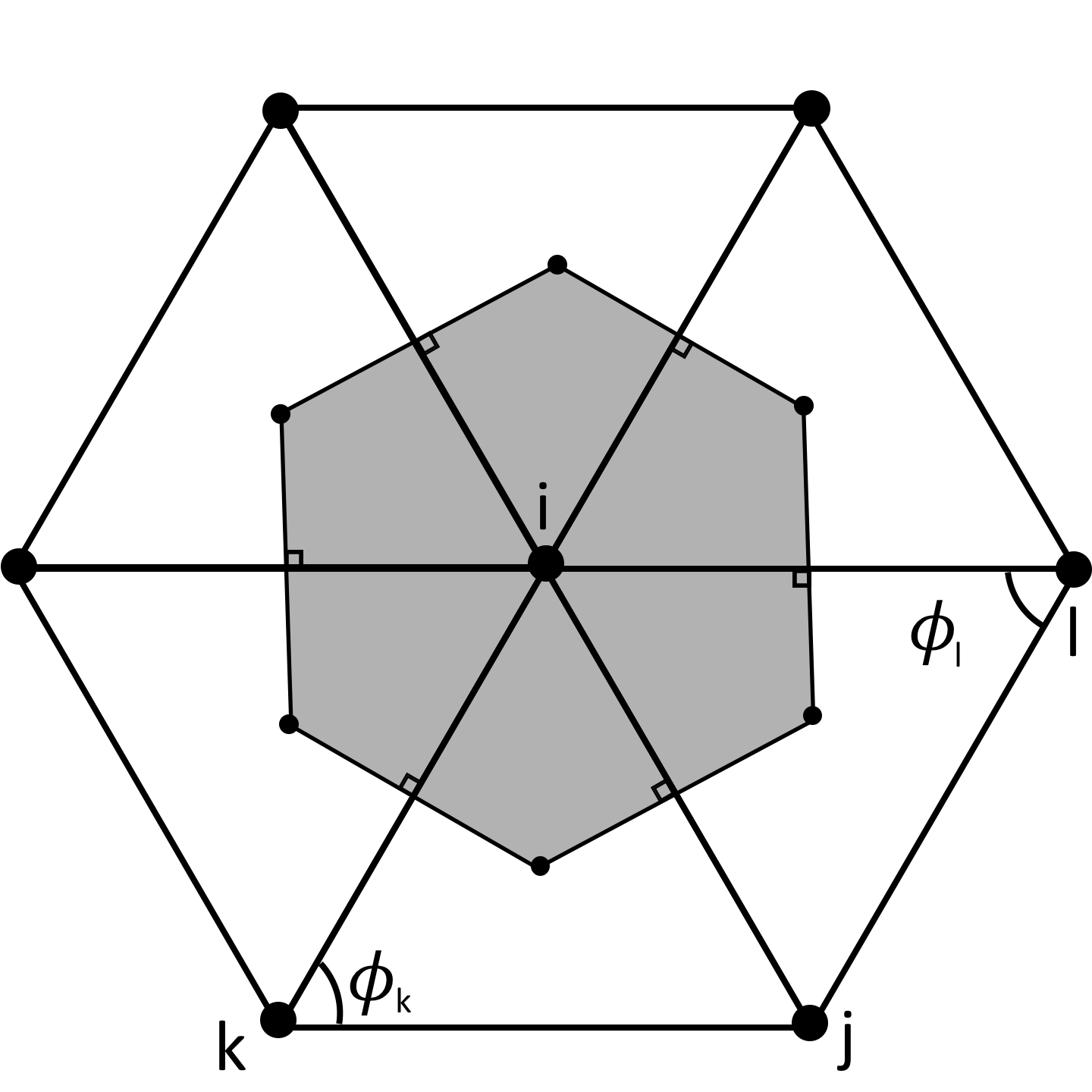}
\caption{}
\end{subfigure}
\begin{subfigure}{0.33\textwidth}
\includegraphics[width=\columnwidth]{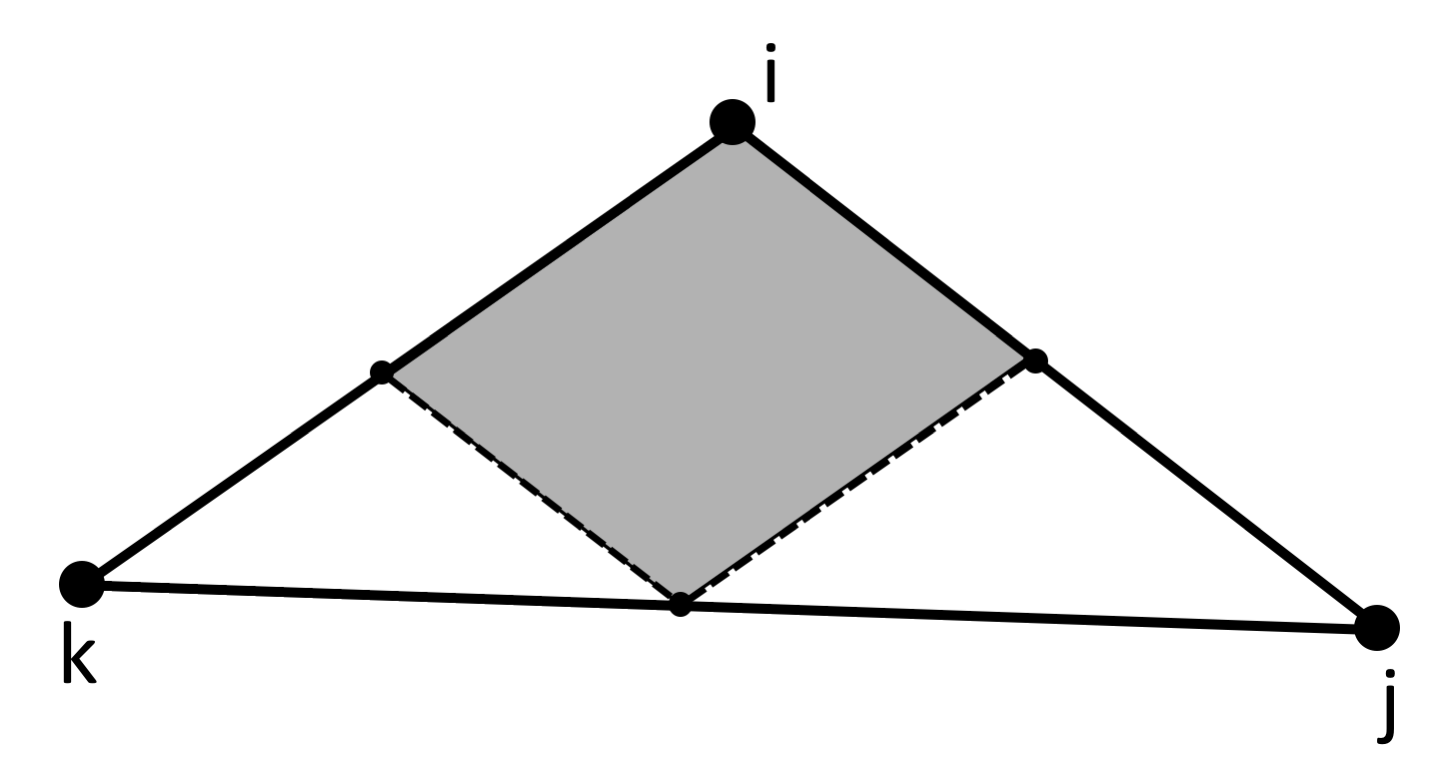}
\caption{}
\end{subfigure}
\caption{A sketch of area associated with vertex $i$.
(a) Barycentric tessellation of area using barycentric centers of each triangle around vertex $i$
by scheme B.
(b) Voronoi tessellation of area using circumcenters of each triangle around vertex $i$
by scheme C and D. 
(c) Tessellation of an obtuse triangle using the middle point of each edge by scheme D.}
\label{fig_area}
\end{figure}

This scheme was proposed by J\"ulicher~\cite{Juelicher1994a,
Juelicher1996} to calculate bending energy of a vesicle with topological
genus more than zero. 
To motivate the scheme we consider two triangles
parallel to the original two neighboring triangles as sketched on
Fig. \ref{fig_two_triangles_parallel}.  Each parallel triangle is
separated from its counterpart with $h$ distance in the normal direction.
The shared edge $e_{ij}$ of the two triangles and the two neighboring
edges of the imaginary triangles form a fraction of a cylinder, as
indicated on Fig. \ref{fig_two_triangles_parallel}.  We define the first
moment of mean curvature on the fraction of the cylinder as
\begin{align}
\mathcal{M}^e_1 = \frac{1}{2}\left(0+\frac{1}{h}\right) h\theta_e l_e=\frac{1}{2}l_e\theta_e.
\end{align}
The two principal curvatures of the cylinder are $0$ and $1/h$.
The surface area for the fraction of the cylinder is $h\theta_e l_e$,
where $l_e$ is the length of the edge $e_{ij}$
and $\cos \theta_e = {\bf n}_1 \cdot {\bf n}_2$ as indicated on Fig \ref{fig_mesh_surface}(a).
We note that $\mathcal{M}^e_1$ does not depend on $h$ and has a proper limit as $h \rightarrow 0$. On the contrary, if we had defined local mean curvature or second moment of mean curvature directly instead, they would both behave singular as 1/h.
The total first moment of mean curvature reads
\begin{align}
\mathcal{M}_1= \frac{1}{2} \sum^{N_e}_{e:<i,j>} l_e \theta_e. 
\end{align}
For the local first moment associated with a vertex instead of edge we have
\begin{align}
\mathcal{M}_1=\sum^{N_i}_{i} \mathcal{M}^i_1 = \frac{1}{4} \sum^{N_i}_{i} l_e \theta_e,
\end{align}
where the prefactor is $1/4$ to account for the total summation. 
For each vertex $i$, we have relation $\mathcal{M}^i_1=H_iA_i$.
This leads to a definition of $H_i$ as
\begin{eqnarray}
H_i=\frac{1}{4A_i} \sum^{N^{i}_e}_{e:\left<i, j\right>} l_{e} \theta_{e},
\label{eq_cm_b}
\end{eqnarray}
where summation runs over all $N^i_e$ neighboring edges around vertex $i$.
The local area associated with vertex $i$ is based on the barycentric centers
of the surrounding triangles as sketched on Fig. \ref{fig_area}(a) and it reads
\begin{eqnarray}
A_i=\frac{1}{3}\sum^{N^{i}_t}_{t:\left<i,j,k\right>} A^{t},
\label{eq_area_b}
\end{eqnarray}
where $A^{t}$ is the area of a neighboring triangle $t:\left<i,j,k\right>$.
This means that each triangle $t$ split its area evenly onto its three vertices.
The summation runs over all $N^i_t$ neighboring triangles around vertex $i$.

With the definitions of local mean curvature and area in Eqs.~(\ref{eq_cm_b}) and (\ref{eq_area_b}),
we may calculate the three moments defined in Eq.~(\ref{eq_moments_discrete}).
The expression for the discrete force is given in \ref{sec_appendix_force_scheme_b} 
and it consists of three primitive derivatives 
\begin{eqnarray}
\frac{\partial l_e}{\partial {\bf x}_m}, \quad  
\frac{\partial \theta_e}{\partial {\bf x}_m}, \quad
\frac{\partial A^t}{\partial {\bf x}_m}.
\end{eqnarray}

\subsection{Scheme C}
\label{section_scheme_c}
This scheme was proposed by Gompper \& Kroll~\cite{Gompper1996} to
simulate a triangular network of fluid vesicles and later applied to
model general membranes~\cite{Noguchi2005a}.  Its discrete definition
of Laplace-Beltrami operator reads~\cite{Itzykson1986, Pinkall1993,
Meyer2003}:
\begin{align}
\left(-\nabla^2_s w\right)_i 
= \frac{\sum^{N^{i}_e}_{e:\left<i,j\right>} \left(\cot \phi_k + \cot \phi_l\right)\left(w_i-w_j\right)}{2A^{voro}_i},
\label{eq_lb_operator_discrete}
\end{align}
where $w$ is an arbitrary function. The Voronoi area associated to vertex $i$
indicated on Fig. \ref{fig_area}(b) and is defined as
\begin{align}
A^{voro}_i = \frac{1}{8} \sum^{N^{i}_e}_{e:\left<i,j\right)} \left(\cot \phi_k + \cot \phi_l\right)\left|{\bf x}_i-{\bf x}_j\right|^2.
\label{eq_area_voronoi}
\end{align}
If we consider the discrete counterpart of an expression for mean
curvature known from differential geometry, that is,
\begin{eqnarray}
 H_i = \frac{1}{2}\left(\nabla^2_s {\bf x}\right)_i \cdot {\bf n}_i,
\label{eq_mc_equal_lb_norm_discrete}
\end{eqnarray}
and take function $w$ as the surface coordinate ${\bf x}$ in
Eq. (\ref{eq_lb_operator_discrete}), we have an explicit discrete
definition of $H_i$.  When only the Helfrich energy/force with
$H_0=0$ is considered, $H^2_i=\frac{1}{4}\left[\left(\nabla^2_s {\bf
x}\right)_i\right]^2$ \cite{Gompper1996, Guckenberger2016, Hoore2018},
as $\nabla^2_s {\bf x}$ is along ${\bf n}$ direction.
However, for the general energy/force in the SC and BC/ADE models
considered in this work, we need an explicit discrete definition of
${\bf n}_i$.  We consider a discrete ${\bf n}_i$ ~\cite{Thuerrner1998,
Jin2005a, Guckenberger2016}.  where the unit normal vector for vertex
$i$ is the sum of neighboring normal vectors weighted  by incident
angles~\cite{Thuerrner1998, Jin2005a}.

\begin{eqnarray}
{\bf n}_i = \frac{\sum^{N^{i}_t}_{t: \left<i, j, k\right>} \phi^{t}_i {\bf u}^t}
{|\sum^{N^{i}_t}_{t: \left<i, j, k\right>} \phi^{t}_i {\bf u}^t|},
\label{eq_normal_vector_discrete}
\end{eqnarray}
where $\phi^t_i$ is the angle at vertex $i$ of triangle $t:
\left<i,j,k\right>$, ${\bf u}^t$ is the unit normal vector of triangle
$t$.

With the definitions of local mean curvature and area in
Eqs.~(\ref{eq_mc_equal_lb_norm_discrete}) and (\ref{eq_area_voronoi}),
we calculate the three moments in Eq.~(\ref{eq_moments_discrete}), and
the total energy in Eq.~(\ref{eq_energy_discrete_group2}).  The expression
of the force is given in \ref{sec_appendix_force_scheme_c}.

\subsection{Scheme D}
\label{section_scheme_d}
This scheme was derived from an effort of developing a "unified and
consistent set of flexible tools to approximate important geometric
attributes, including normal vectors and curvatures on arbitrary
triangle meshes" \cite{Meyer2003}.  It has been adopted by several groups
to study the bending mechanics of the minimal model for vesicles and
RBCs \cite{Boedec2011, Tsubota2014, Guckenberger2016}.  The definition
of the discrete Laplace-Beltrami operator reads as~\cite{Meyer2003},
\begin{eqnarray}
\left(-\nabla^2_s w\right)_i = \frac{\sum^{N^{i}_e}_{e:\left<i,j\right>} \left(\cot \phi_k + \cot \phi_l\right)\left(w_i-w_j\right)}{2A^{mix}_i},
\label{eq_lb_operator_discrete2}
\end{eqnarray}
which has almost identical expression as the one from scheme C, except
that the area for vertex $i$ is calculated as a mixture of two approaches
\cite{Meyer2003}(see Fig. \ref{fig_area}(c)).If the triangle $t:<i,j,k>$
is non-obtuse then local areas $A_i$, $A_j$, and $A_k$ on three vertices
take the contribution from Voronoi tessellation of $A^t$, If the triangle
$t:<i,j,k>$ is obtuse then the tessellation  relies on the middle point
of each edge. Therefore, the triangle $t:<i,j,k>$ contributes $A^t/2$
to $A_i$, and $A^t/4$ to $A_j$ and $A_k$ each~\cite{Meyer2003}.
 
The key difference of scheme D from the other three schemes is in terms
of the force calculation.  The force density of scheme D relies on
the variational expression in Eq. (\ref{eq_force_density_group1})
or (\ref{eq_force_density_individual}).  Once the discrete
mean curvature $H_i$ is calculated at each vertex, we apply
Eq. (\ref{eq_lb_operator_discrete2}) {\it second time} on discrete
values of $H$ to get $(\nabla^2_s H)_i$.  Furthermore, only scheme
D needs a discrete definition of Gaussian curvature at vertex
$i$~\cite{Polthier1998}
\begin{eqnarray}
G_i = \frac{1}{A^{mix}_i}\left(2\pi - \sum^{N^{i}_t}_{t:\left<i,j,k\right>} \phi^t_i \right),
\end{eqnarray}
which employs all incident angles around vertex $i$ and its
area definition.  Therefore, each discrete force density in
Eq. (\ref{eq_moments_force}) is readily available and thereafter the
total force density can also be obtained.

\section{Results}

We present the results of the comparative study for all four schemes
and the proposed extensions. We distinguish applications on prescribed
shapes and on dynamically equilibrated shapes.

\subsection{Prescribed configurations: sphere and bi-concave oblate}
\begin{figure}
\centering
\begin{subfigure}{0.45\textwidth}
\includegraphics[width=\columnwidth]{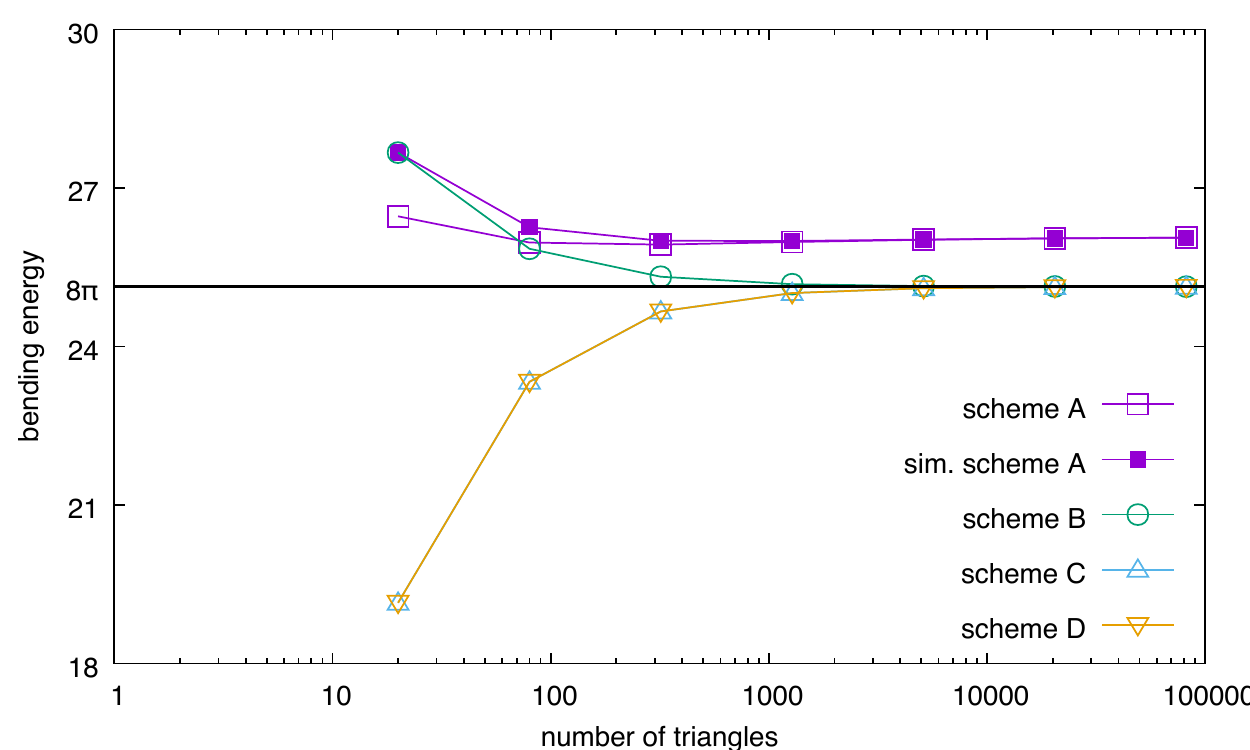}
\caption{}
\label{fig_eng_vs_tot_tri_a}
\end{subfigure}
\begin{subfigure}{0.45\textwidth}
\includegraphics[width=\columnwidth]{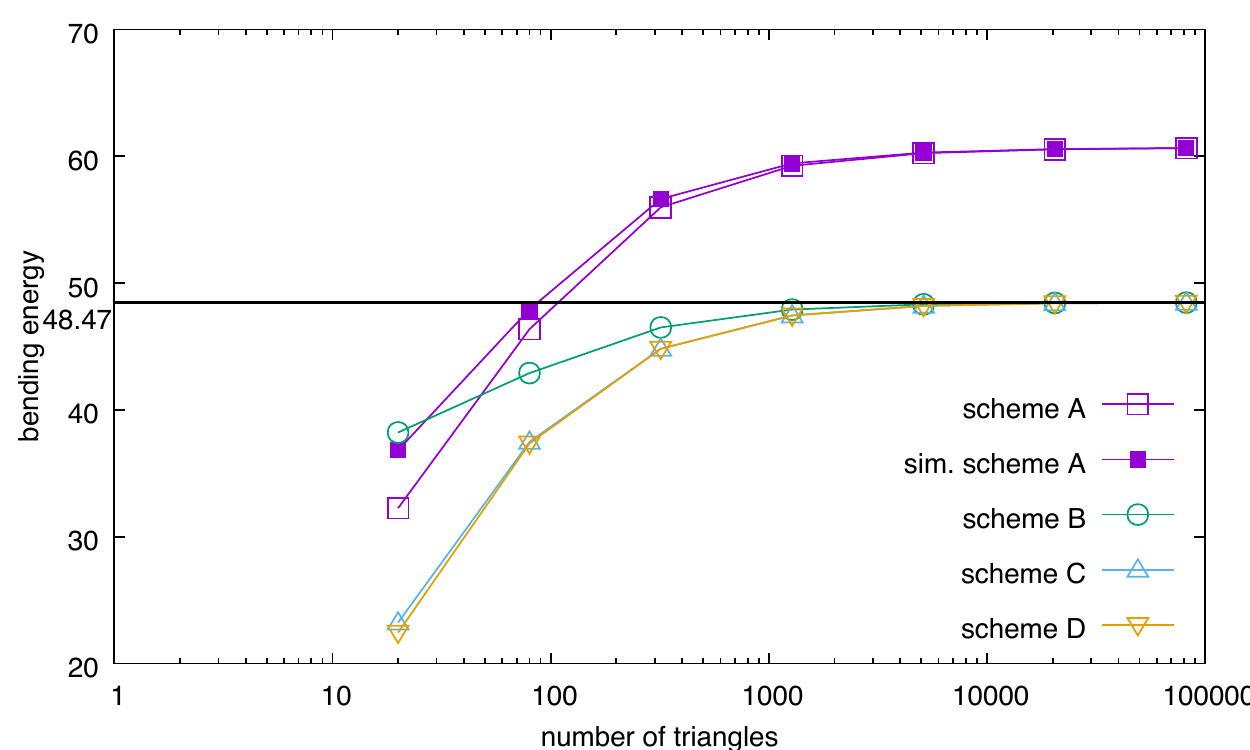}
\caption{}
\label{fig_eng_vs_tot_tri_b}
\end{subfigure}
\caption{Helfrich energy from four discretization schemes with $\kappa_b=1$ and $h_0=0$. (a) sphere; (b) biconcave oblate. 
For scheme A, $\tilde{\kappa}_b=\sqrt{3}\kappa_b$ and $\theta_0=0$,
and solid square is from the simplified scheme A with
first-order approximation.}
\label{fig_eng_vs_tot_tri}
\end{figure}

We first examine the performance of the four schemes on the calculation
of the Helfrich energy  for a sphere and a biconcave oblate  described by
an empirical function \cite{Evans1972} using $\kappa_b=1$ and $h_0=0$ .
The  sphere is approximated with  an icosahedron  and higher resolutions
are obtained by applying  Loop's subdivision scheme~\cite{Loop1987}
for triangulated meshes with $N_t=80$, $320$, $1280$, $5120$.  The same
resolution is obtained  for a biconcave oblate by transforming the surface
coordinates of the sphere to the empirical function of the biconcave
shape.  We present the results in Fig.~\ref{fig_eng_vs_tot_tri} verifying
the total energy converges (albeit to different values !)  with increasing
the number of triangles.  For the sphere, all four schemes converge with
increasing the number of triangles (Fig. ~\ref{fig_eng_vs_tot_tri_a}).
The first-order approximation of scheme A deviates significantly from
scheme A when $N_t<320$ when the angle $\theta$ between two neighboring
triangles is not small.  Schemes B, C and D converge to a correct
value $8\pi$.

For the biconcave-oblate configuration we compute the reference solution
analytically (see \ref{sec_appendix_maxima}).  We find that all four
schemes converge to the reference solution with increasing number of
triangles ( Fig. ~\ref{fig_eng_vs_tot_tri_b}).  We note that the results
from the first-order approximation for scheme A deviate significantly
from those of the complete scheme A for $N_t<1280$.  Moreover, with
$\tilde{\kappa}_b=\sqrt{3}\kappa_b$, scheme A and its simplified version
again converge to a higher value than the one given by the reference
solution.  The value of the total energy for the prescribed shapes does
not depend on the details of the triangulated mesh,  as long as the mesh
is regular.

\label{sec_numerics1}

\begin{figure}
\centering
\begin{subfigure}{0.45\textwidth}
\includegraphics[width=\textwidth]{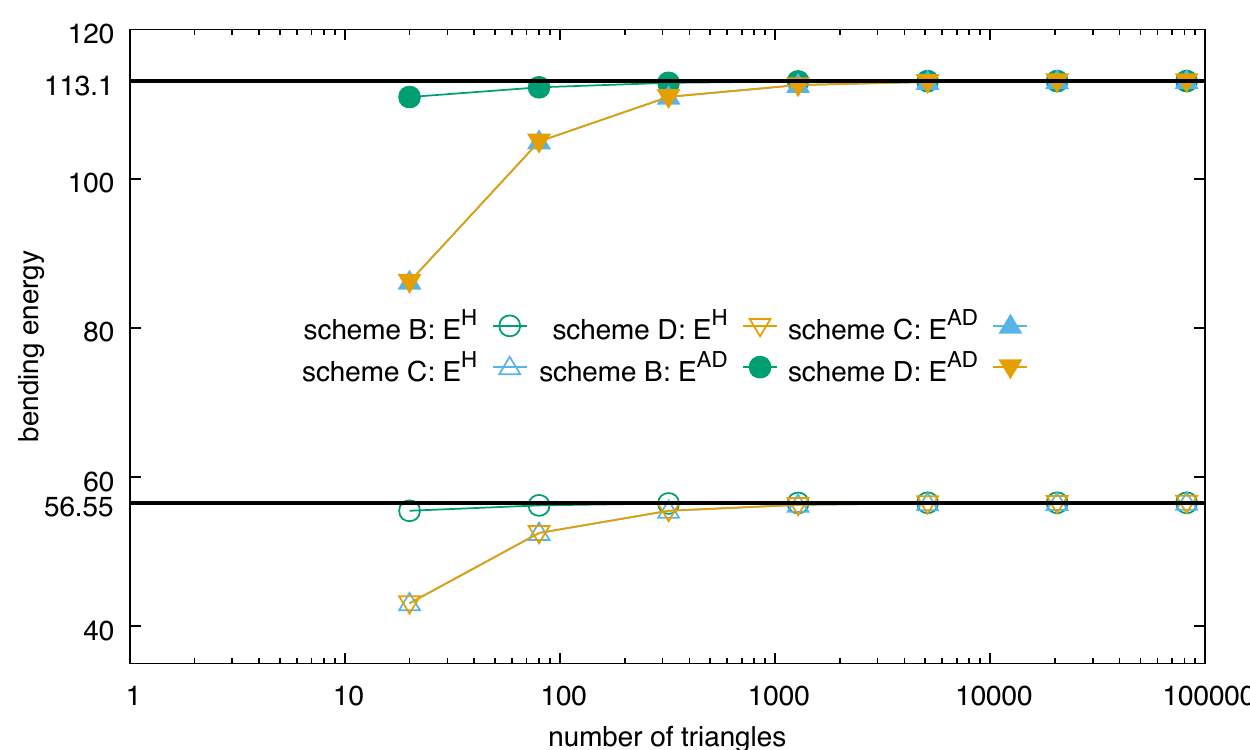}
\caption{}
\label{fig_eng_vs_tot_tri_ext_a}
\end{subfigure}
\begin{subfigure}{0.45\textwidth}
\includegraphics[width=\textwidth]{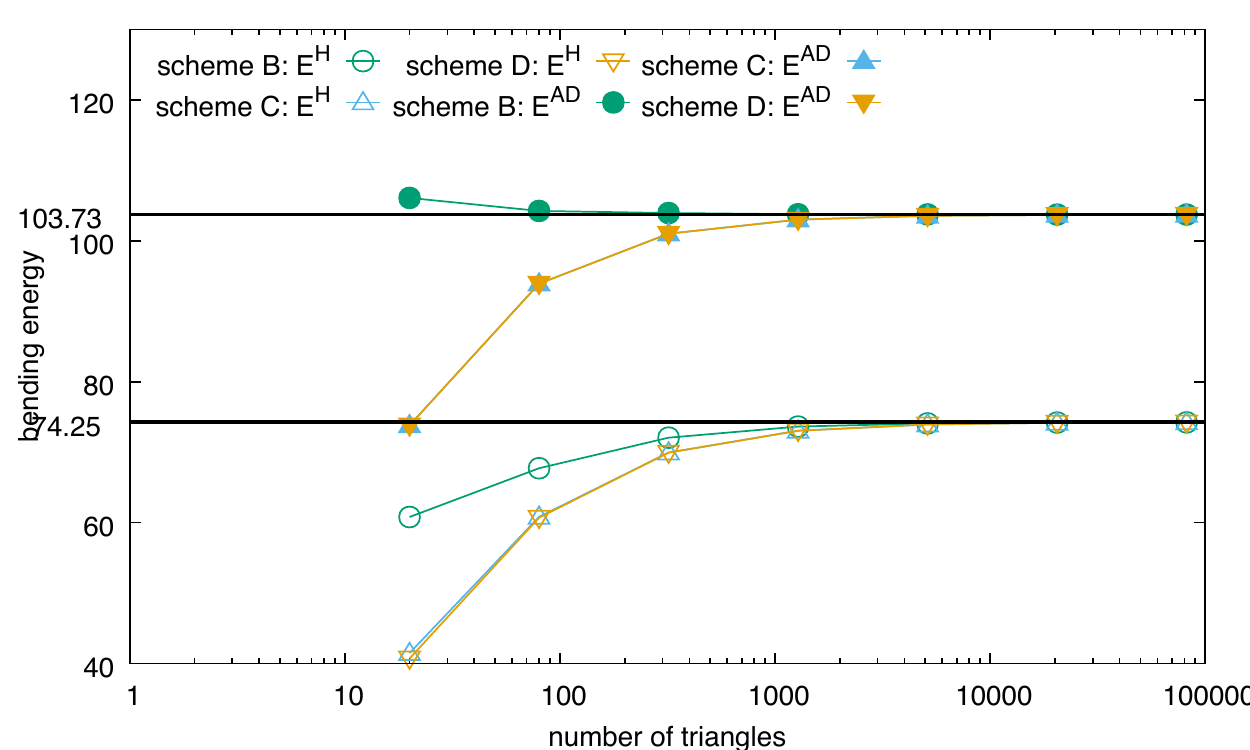}
\caption{}
\label{fig_eng_vs_tot_tri_ext_b}
\end{subfigure}
\caption{Total Helfrich and area-difference elasticity (ADE) energy from three discretization schemes with 
$\kappa_b=1$,  $h_0=-0.5$, $\alpha=2/\pi$
and $d=D/R=0.001$.
$E^H$; Helfrich energy; $E^{AD}$: ADE energy.
 (a) sphere with $\Delta a_0=\Delta A_0/(4\pi RD)=-1$; 
 (b) biconcave oblate with $\Delta a_0=\Delta A_0/(4\pi RD)=-0.7$.}
\label{fig_eng_vs_tot_tri_ext}
\end{figure}

Subsequently, we compute total energy with $\kappa_b=1$,
$h_0=-0.5$ and $\alpha=2/\pi$ using scheme B, C and D (
Fig.~\ref{fig_eng_vs_tot_tri_ext}).  All three schemes converge with
increasing the number of triangles to the reference values.  As shown in
Fig. ~\ref{fig_eng_vs_tot_tri_ext}, for low resolution $N_t<320$, scheme
B deviates much less than schemes C and D from the reference values.

\begin{figure}
\centering
\begin{subfigure}{0.45\textwidth}
\includegraphics[width=\columnwidth]{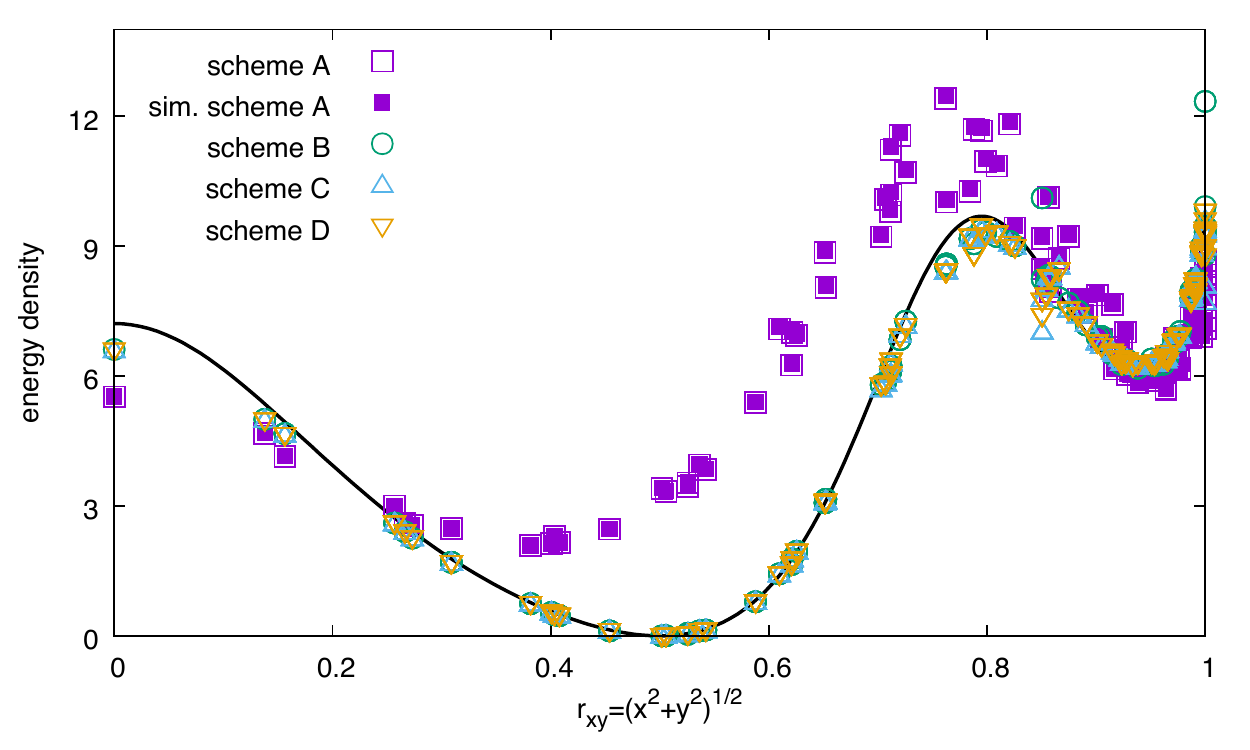}
\caption{}
\end{subfigure}
\begin{subfigure}{0.45\textwidth}
\includegraphics[width=\columnwidth]{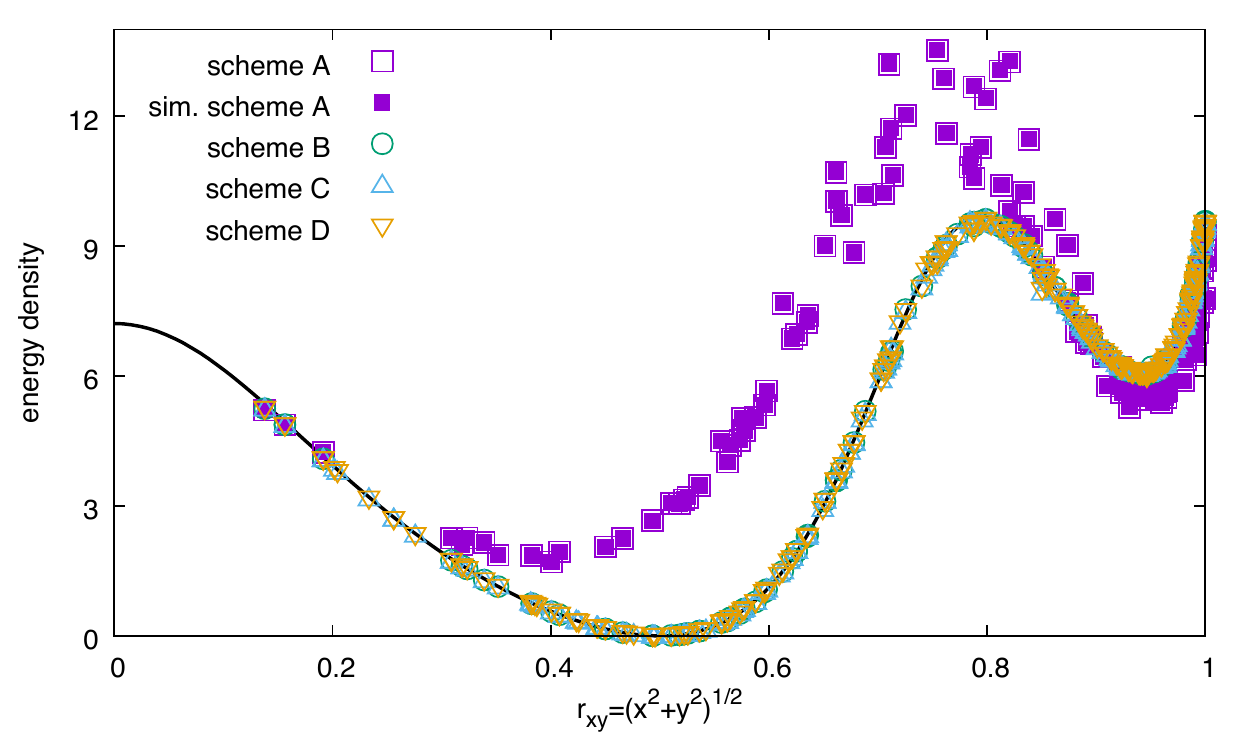}
\caption{}
\end{subfigure}
\caption{Energy density versus radial distance of vertices from axis of symmetry for a biconcave oblate with $\kappa_b=1$ and $h_0=0$. 
For scheme A and its simplified version $\tilde{\kappa}_b = \sqrt{3}\kappa_b$ and $\theta_0=0$. 
 (a) $N_t=1280$; (b) $N_t=5120$ and plotted for clarity with every $16$ vertices in arbitrary order.}
 \label{fig_eng_density_vs_rxy}
\end{figure}
Furthermore, for the biconcave-oblate configuration we plot energy
density versus the radial distance from the axis of symmetry (
Fig.~\ref{fig_eng_density_vs_rxy}).  Results from scheme B, C and D with
$N_t=1280$ coincide with the reference line except a few discrepancies,
which disappear with high resolution $N_t=5120$.  Scheme A and its
simplified version show no difference, but deviate from the reference
values.  Increasing the resolution from $N_t=1280$ to $5120$ does not
improve the results of scheme A or those of its simplified version.
The results indicate  that scheme A is not an effective discretization
of the Helfrich energy,consistent with the findings  of two recent works
\cite{Tsubota2014, Guckenberger2016}.

\begin{figure}
\centering
\begin{subfigure}{0.45\textwidth}
\includegraphics[width=\columnwidth]{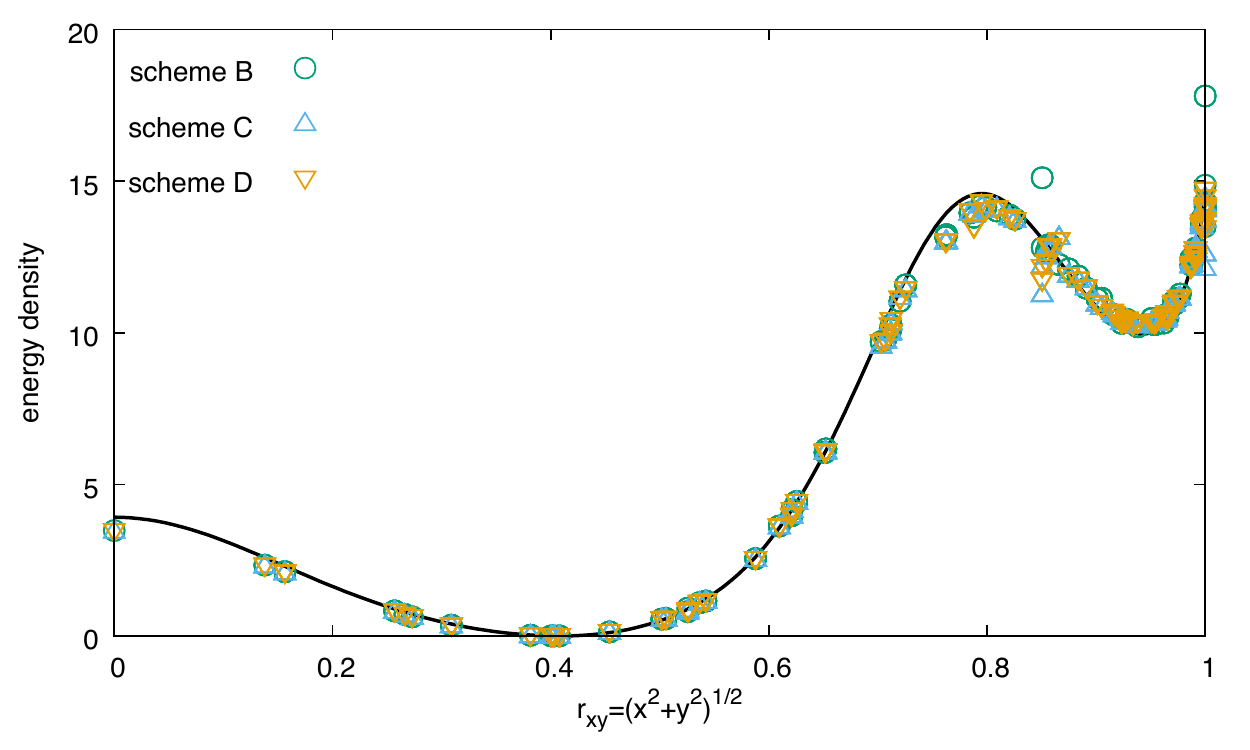}
\caption{}
\end{subfigure}
\begin{subfigure}{0.45\textwidth}
\includegraphics[width=\columnwidth]{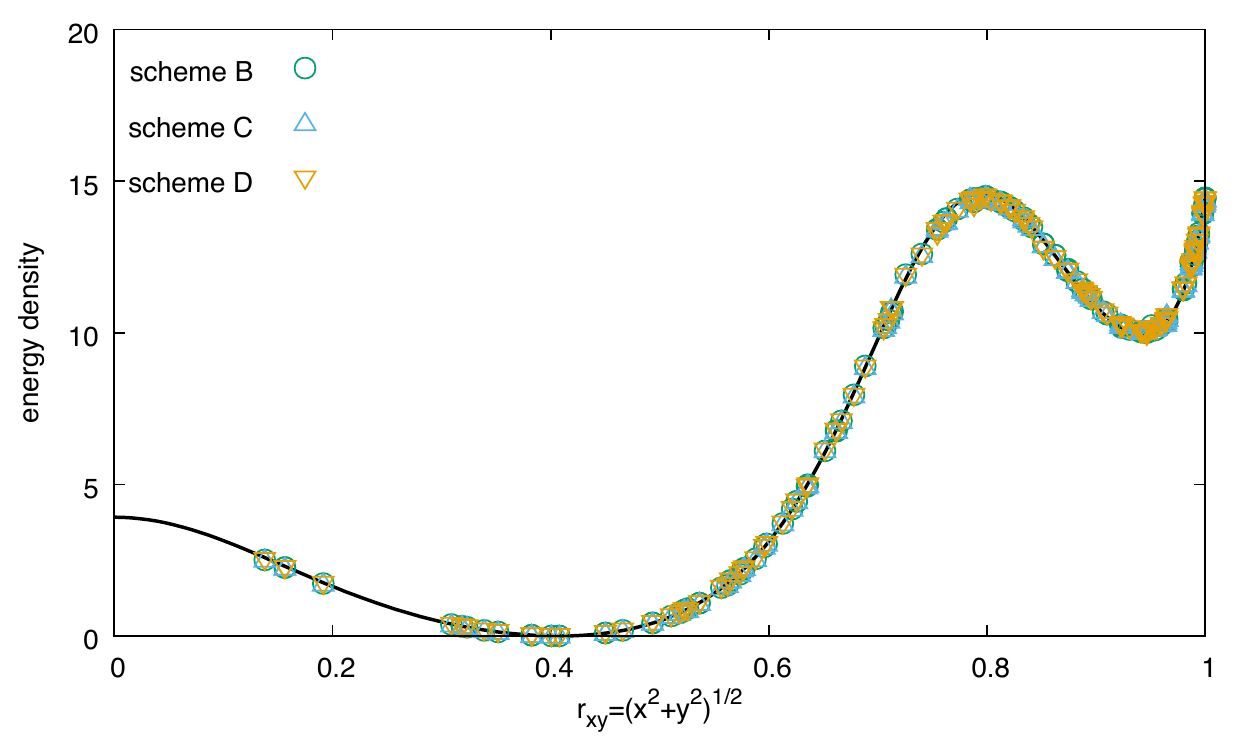}
\caption{}
\end{subfigure}
\caption{Energy density versus radial distance of vertices from axis for a
biconcave oblate with $\kappa_b=1$ and $h_0=-0.5$.  (a) $N_t=1280$;
(b) $N_t=5120$ and plotted for clarity with every $16$ vertices in
arbitrary order.}
\label{fig_eng_density_vs_rxy_ext}
\end{figure}
Finally, we examine the energy density for the Helfrich energy with
$h_0\neq 0$, as computed by schemes B, C and D.  We present the results
in Fig. \ref{fig_eng_density_vs_rxy_ext}.  All three schemes capture
accurately the profile of the reference line.  Some outliers showing for
$N_t=1280$ also disappear for higher resolution $N_t=5120$.  Note that
the ADE energy is non-local and has only one global value, which was
already presented in Fig. \ref{fig_eng_vs_tot_tri_ext}.

\subsection{Numerical results on equilibrium shapes}
\label{sec_numerics2}

Here we consider the  equilibrium shapes of a closed membrane. The
energy minimization process is initialized from a prescribed shapes
such as a sphere, prolate and oblate ellipsoids, or from the equilibrium
shape (e.g. stomatocyte) of an optimization using different parameters.

The minimization process relies on the calculation of the forces
corresponding to the associated energy.  The  energy and the force
are both calculated on the triangulated mesh. We set $\kappa_b=0.01$
and $\alpha=2/\pi$ \cite{Lim2002, Khairy2008a}.
We penalize the constraints on global area with $\kappa_{ag}=2$
(\ref{sec_appendix_area}) and volume with $\kappa_v=1$
(\ref{sec_appendix_volume})~\cite{Du2004, Li2005, Pivkin2008,
Fedosov2010b}. We also add an in-plane viscous damping force between
any two neighboring vertices connected by an edge in order to dampen
the kinetic energy.  To explore the energy landscape efficiently, we
also add a stochastic force of white noise between any two neighboring
vertices connected by an edge.  The pair of viscous and stochastic forces
between vertices resemble the pairwise thermostat~\cite{Espanol1995},
which conserves momenta and has a proper thermal equilibrium.  The time
integration is performed by the explicit velocity Verlet method.  As the
minimization evolves, the configuration of the triangulated mesh may
deteriorate, e.g., elongate in on direction or generate obtuse angles. We
remedy this mesh distortion by introducing two regularization schemes:
by  introducing  a constraint on the local area with $\kappa_{al}=1$
as penalization (\ref{sec_appendix_area}) \cite{Li2005, Pivkin2008,
Fedosov2010b} or by triangle equiangulation \cite{Brakke1992}.  We note
that as we perform multiple runs from different initial shapes we
may reach different local minima of the energy landscape.  Thereafter,
we select the smallest among all available local minima as the global
minimum, that is, the ground state.

\subsubsection{The minimal model}
\label{sec_numerics2_canham}

We first consider the minimal model described by Helfrich energy with
$h_0=0$ ~\cite{Deuling1976a, Seifert1991}.  As the equilibrium shapes are
axi-symmetric, the computations are reduced to solving two-dimensional
Euler-Lagrangian ODEs. We aim to reproduce the known phase-diagram with
configurations and energies, in particular for small reduced volumes
\cite{Peng2010}.  Furthermore, as the minimal model is widely used to
simulate vesicles and RBCs we examine the performance of all four schemes
in this situation.

\begin{figure}
\centering
\includegraphics[width=0.8\columnwidth]{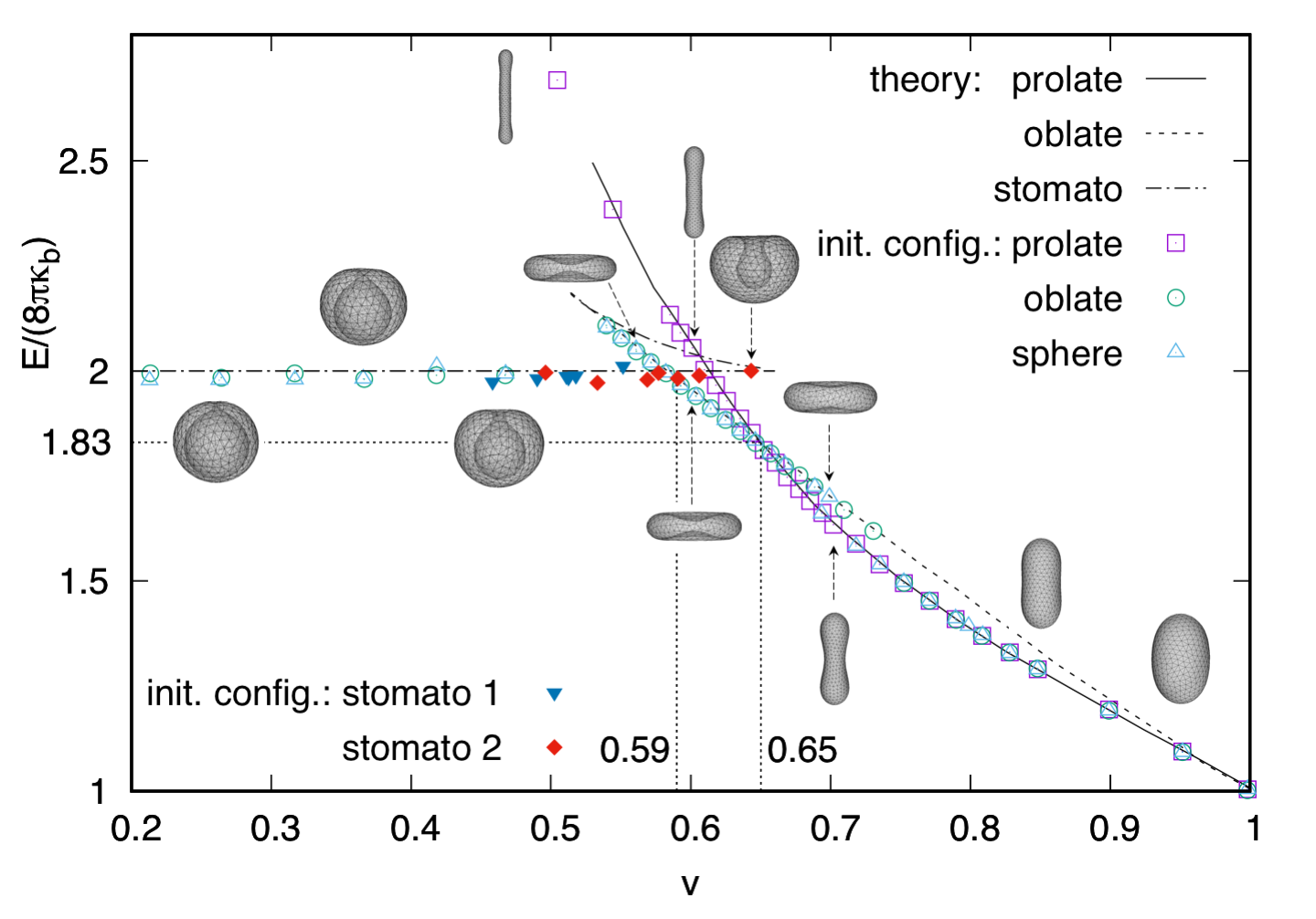} 
\caption{Local energy minima and corresponding shapes at different branches in the
minimal model: symbols for energy and some selective configurations are by
scheme B with $N_t=1280$.  The axis of symmetry for each configuration
always points upwards.  Reference lines of theory are adapted from
Seifert et al.~\cite{Seifert1991}.}
\label{fig_energy_canham_B}
\end{figure}

We present the phase diagram of the reduced volume $v$ and the
normalized energy of the vesicles as generated by scheme B in
Fig.~\ref{fig_energy_canham_B}.  Results from schemes A, C and D are
quite similar to those obtained from scheme B,  so in the following we
only emphasize their discrepancies.  We adopt the reference lines from
the work of Seifert et al~\cite{Seifert1991} and denote our results
with symbols.  Each symbol corresponds to the energy minima
as obtained from our minimization procedure.

The reference lines correspond to three types of configurations.
With decreasing $v$ in the prolate branch, the shape changes from
sphere to prolate, dumbbell, and long caped cylinder. With decreasing
$v$ in the oblate branch the shape changes from a sphere to
a famous biconcave-oblate shape at around $v=0.6$. For $v=0.51$
the oblate branch bifurcates to form two stomatocyte sub-branches.
One sub-branch has constant energy for $0 < v < 0.66$ while the other
has energy which depends on volume.

The simulation results denoted by symbols in
Fig. \ref{fig_energy_canham_B}, are obtained by  three initial shapes:
sphere, prolate and oblate ellipsoids. The volume for the initial prolate
and oblate ellipsoids are the same as that of the target $v$.  However,
the prolate and oblate branches generated in this work are different
from the reference work~\cite{Seifert1991}.  The reference restricts
the prolate branch to only prolate-like shapes and the oblate branch
to oblate-like shapes.  Here, we do not impose such constraint on the
shape during the minimization procedure.  More specific, all the squared
symbols are from initial shape of prolates and they all stay on the
prolate branch when reaching the local energy minimum.  Similarly, all
the circle symbols are from initial shapes of oblates, with the
corresponding $v$, that stay on the oblate branch for $v \lesssim 0.75$
and jump onto the prolate branch for $v \gtrsim 0.75$.  This results
corroborates an earlier finding \cite{Jaric1995} where the oblate branch
is only locally stable for $v  \lesssim  0.75$.  For the minimization
from a spherical initial shapes (up triangle symbols), the
energy minima are found on the prolate branch for $v \gtrsim 0.75$,
on the oblate branch for $v \lesssim 0.65$ and scatter between prolate
and oblate branches for $0.65 \lesssim v \lesssim 0.75$.

For the constant energy  sub-branch of the stomatocyte, we can not explore
the partial ($v \gtrsim 0.51$) energy landscape directly.  Therefore, we
take two final shapes of stomatocyte, which are obtained by minimization
with initial shapes of oblates.  One final shape is from $v=0.4$ and
the other is from $v=0.45$.  We use these two shapes as initial shapes
to run minimization procedure with targeted reduced volume $v$ ranging
from $0.45$ to $0.66$.  The results from $v=0.4$ are denoted with solid
down triangles while from $v=0.45$ are solid diamonds.

\begin{figure}
\centering
\includegraphics[width=0.65\columnwidth]{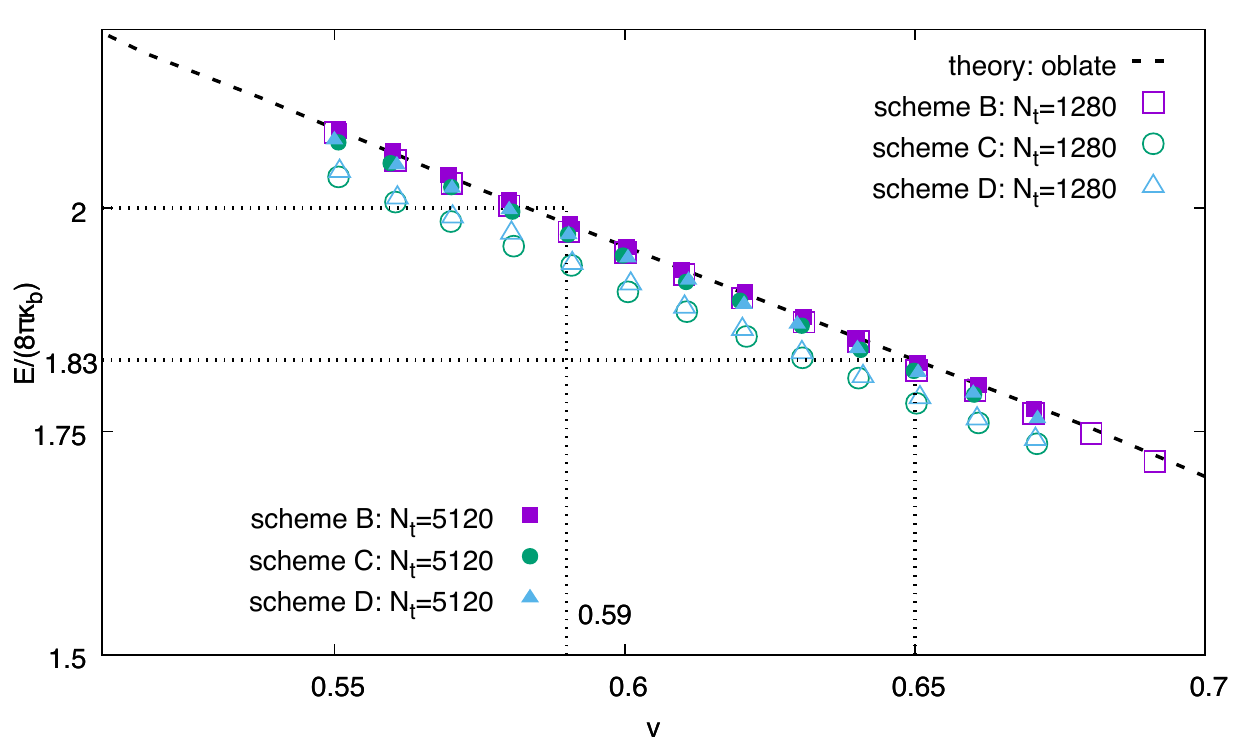} 
\caption{Resolution study for schemes B, C and D with $N_t=1280$ and $5120$:
initial shapes are sphere.  The reference lines of theory are
adapted from Fig. 8 in Seifert et al. \cite{Seifert1991}.}
\label{fig_energy_canham_sphere}
\end{figure}

We demonstrate the convergence of the minimization process for the
calculated energies, by examining the results in $0.55 < v < 0.7$, as this
range contains the regime of oblate-like biconcave shapes ($0.59<v<0.65$)
that is relevant for the  simulations of RBCs.  We present results
from schemes B, C and D with two resolutions $N_t=1280$ and $5120$ in
Fig.~\ref{fig_energy_canham_sphere}.  It is apparent that scheme B is
superior, as it already captures the reference values with $N_t=1280$,
whereas scheme C and D reproduce the same results with $N_t=5120$.
Moreover, the energies at delimiting points, that is, $2$ for $v=0.59$
and $1.83$ for $v=0.65$ are reproduced quite well by all three schemes.
The minimal energies computed by scheme A are removed, as they are
completely out of range.  This should not be surprising, given its
performances on two static configurations in section~\ref{sec_numerics1}.
Nevertheless, we present the equilibrium shapes by scheme A along with
the other three schemes.

\begin{table}
\centering
\caption{Equilibrium shapes of the minimal model: comparison among four schemes for $0.65 < v < 1$ with $N_t=1280$. }
\begin{tabular}{ c  c  c  c  c}
$v$ & $0.67$ & $0.75$ & $0.85$ & $0.95$  \\
\raisebox{\cheight\height}{A} &
\includegraphics[width=\bsize\columnwidth]{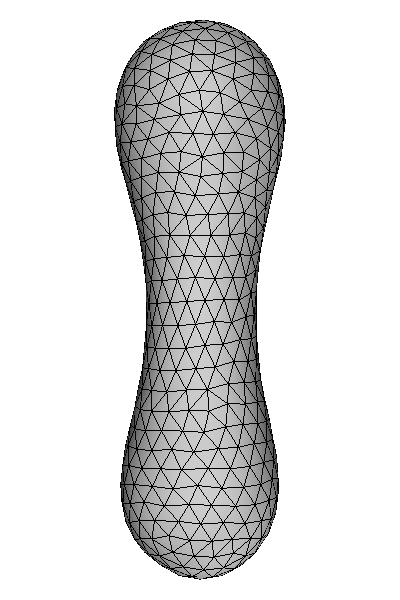}  &
\includegraphics[width=\bsize\columnwidth]{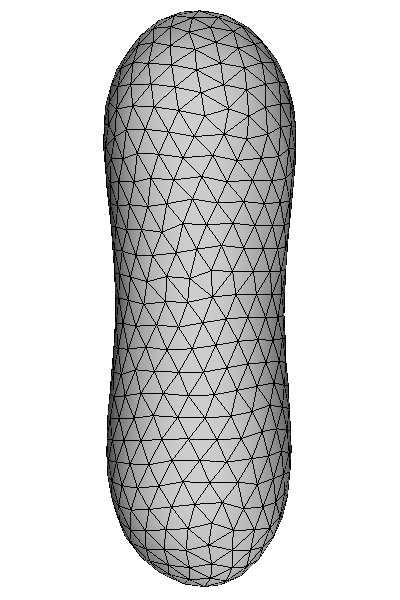}  &
\includegraphics[width=\bsize\columnwidth]{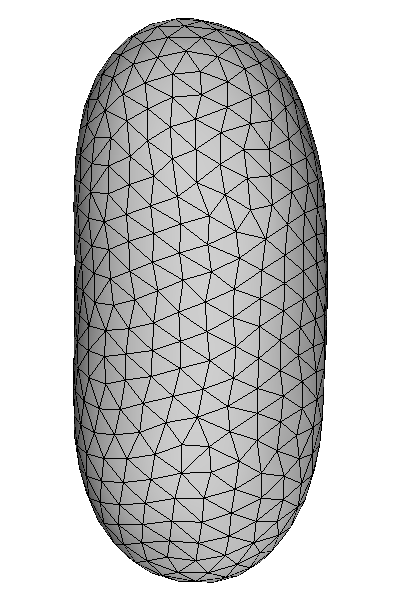}  &
\includegraphics[width=\bsize\columnwidth]{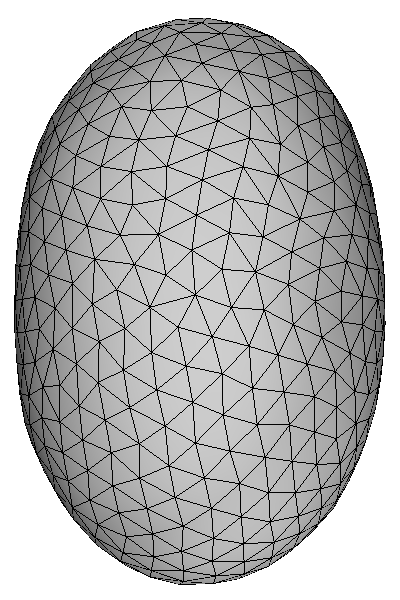}  \\
\raisebox{\cheight\height}{B} &
\includegraphics[width=\bsize\columnwidth]{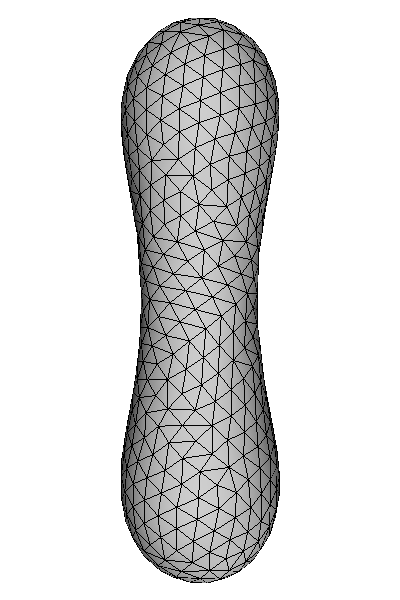} &
\includegraphics[width=\bsize\columnwidth]{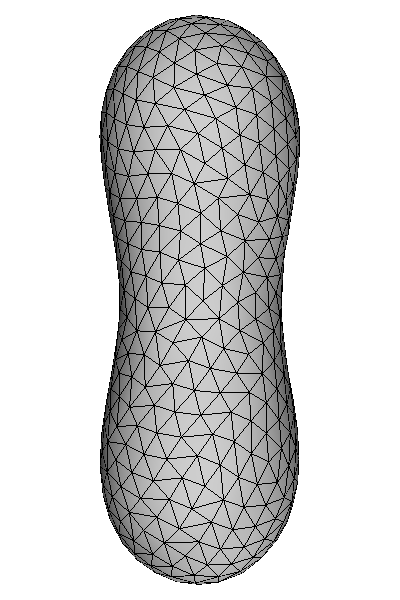}  &
\includegraphics[width=\bsize\columnwidth]{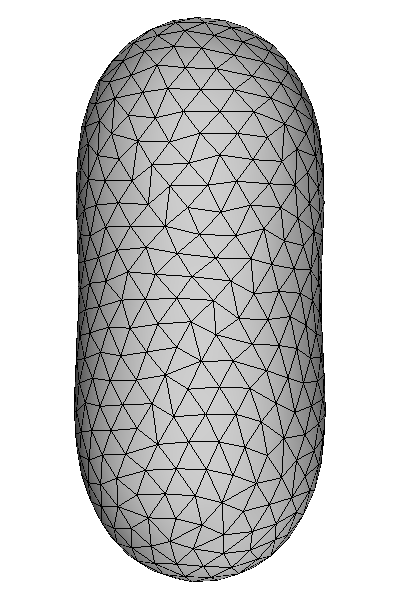}  &
\includegraphics[width=\bsize\columnwidth]{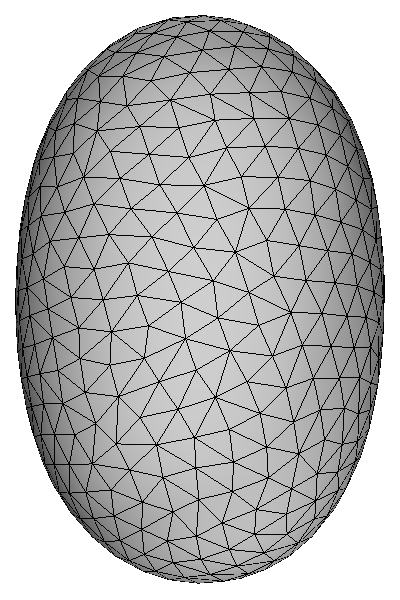}  \\
\raisebox{\cheight\height}{C} &
\includegraphics[width=\bsize\columnwidth]{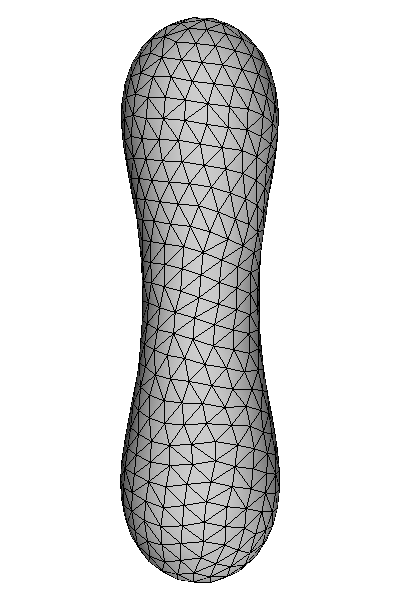}  &
\includegraphics[width=\bsize\columnwidth]{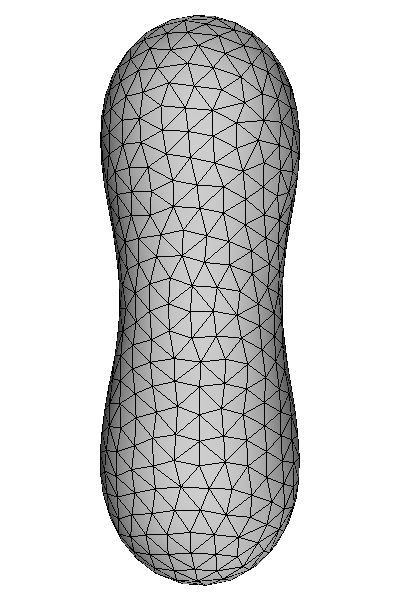}  &
\includegraphics[width=\bsize\columnwidth]{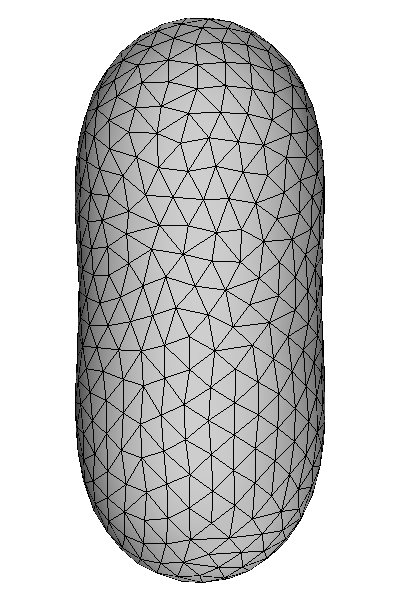}  &
\includegraphics[width=\bsize\columnwidth]{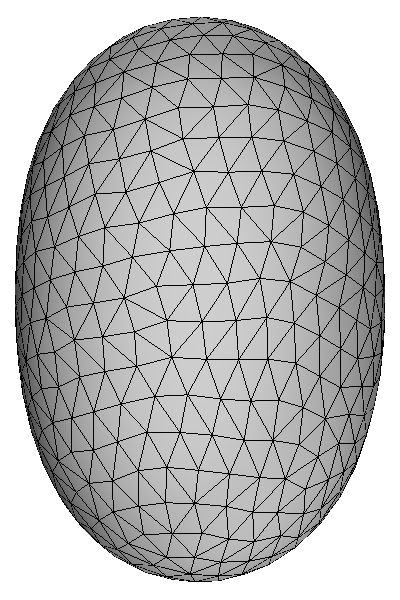}  \\
\raisebox{\cheight\height}{D} &
\includegraphics[width=\bsize\columnwidth]{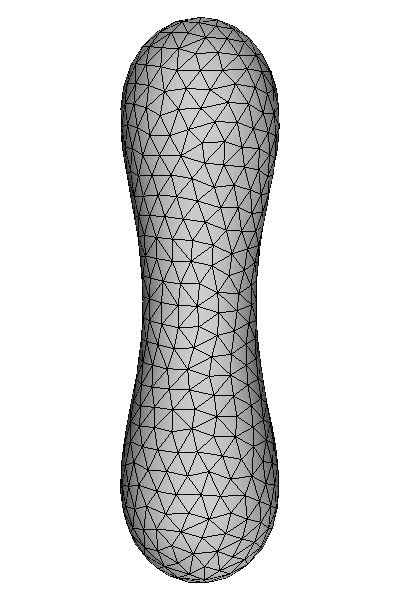}  &
\includegraphics[width=\bsize\columnwidth]{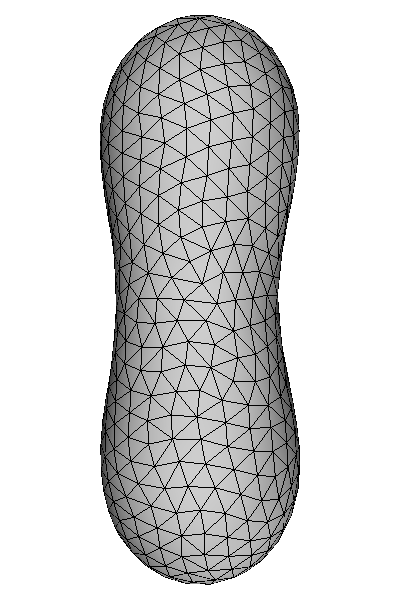}  &
\includegraphics[width=\bsize\columnwidth]{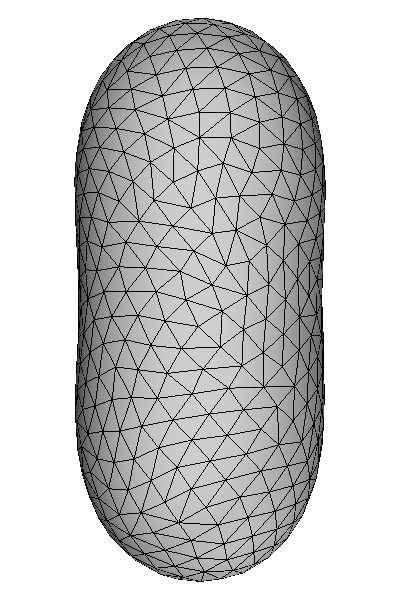}  &
\includegraphics[width=\bsize\columnwidth]{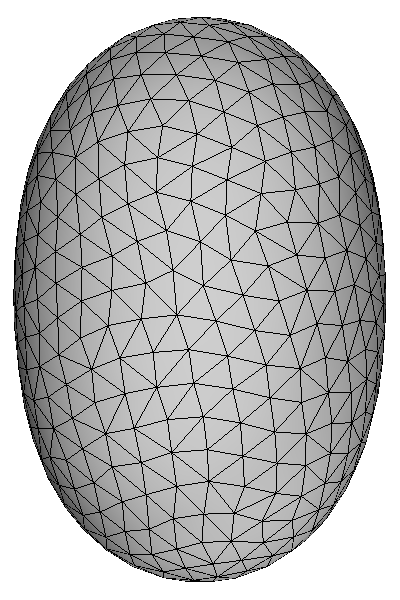}  \\
\end{tabular}
\label{table_canham1}
\end{table}
We select four equilibrium configurations in the
prolate-like/dumbbell/cigar regime with $0.65 < v < 1$ generated by the
four schemes with $N_t=1280$ in Table \ref{table_canham1}. We find very small  differences in the shapes obtained by the four schemes.
\begin{table}
\centering
\caption{Equilibrium shapes of the minimal model:  comparison among four schemes for $0.59 < v < 0.65$ with $N_t=1280$.}
\begin{tabular}{ c  c  c  c}
$v$ & $0.6$ & $0.62$ & $0.64$  \\
\raisebox{\eheight\height}{A} &
\includegraphics[width=\bcsize\columnwidth]{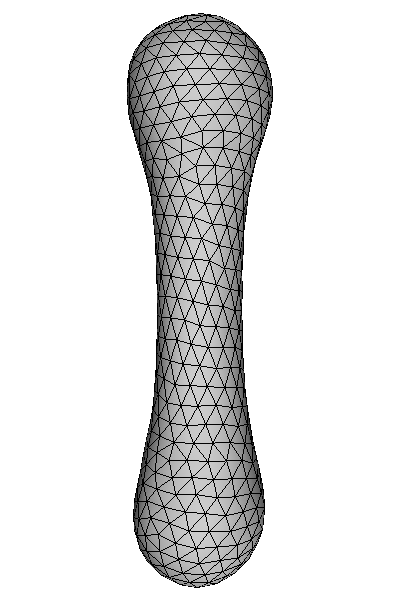}  &
\includegraphics[width=\bcsize\columnwidth]{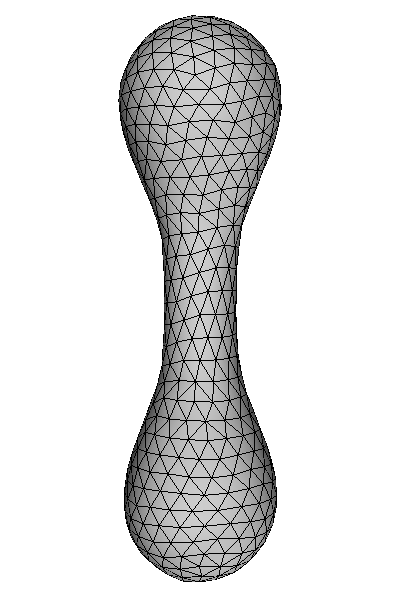}  &
\includegraphics[width=\bcsize\columnwidth]{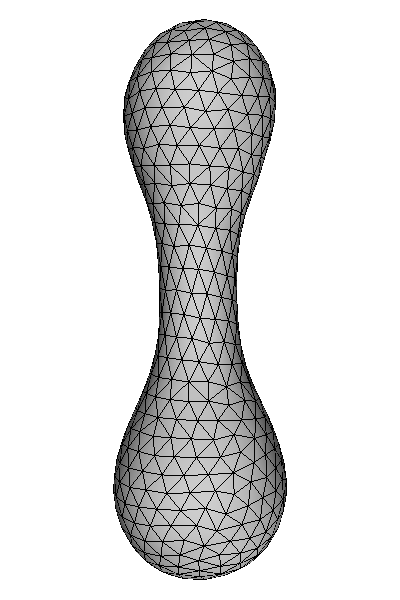}  \\
\raisebox{\abheight\height}{B} &
\includegraphics[width=\bcsize\columnwidth]{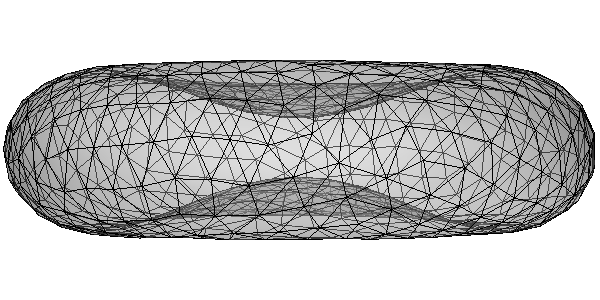}  &
\includegraphics[width=\bcsize\columnwidth]{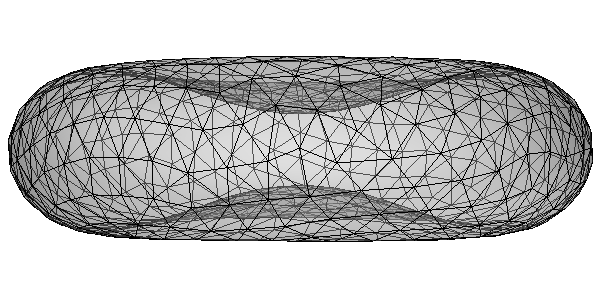} &
\includegraphics[width=\bcsize\columnwidth]{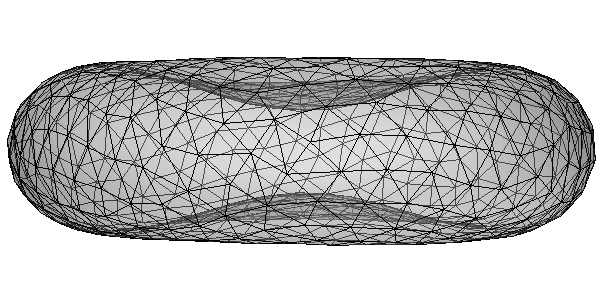} \\
\raisebox{\abheight\height}{C} &
\includegraphics[width=\bcsize\columnwidth]{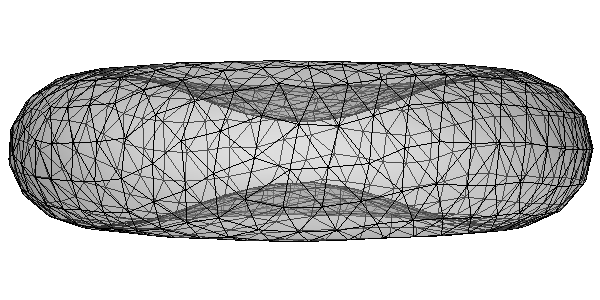}  &
\includegraphics[width=\bcsize\columnwidth]{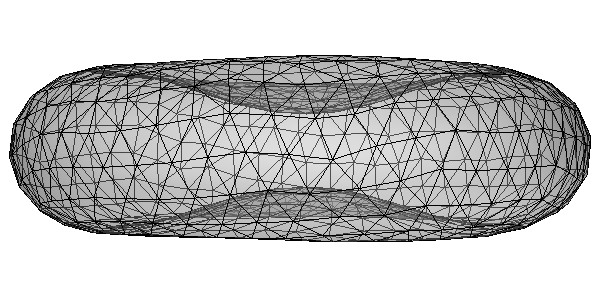}  &
\includegraphics[width=\bcsize\columnwidth]{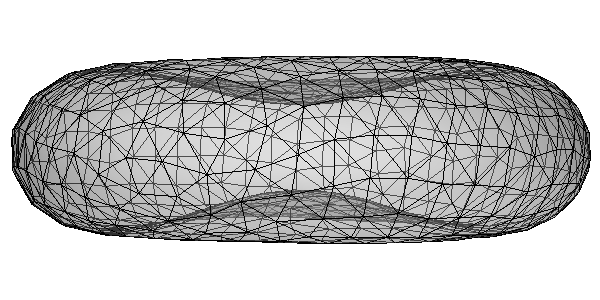}  \\
\raisebox{\abheight\height}{D} &
\includegraphics[width=\bcsize\columnwidth]{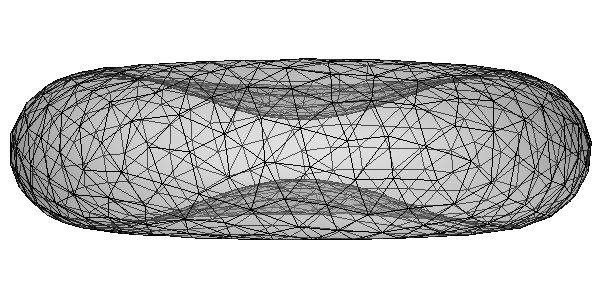}  &
\includegraphics[width=\bcsize\columnwidth]{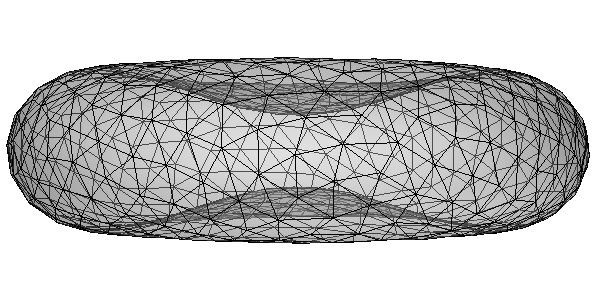}  &
\includegraphics[width=\bcsize\columnwidth]{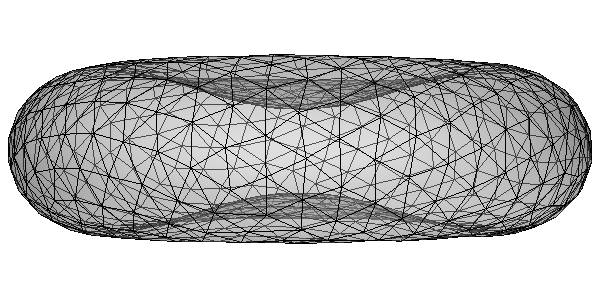} \\
\end{tabular}
\label{table_canham2}
\end{table}
However,  in the biconcave-oblate regime for $0.59 < v < 0.65$
( Table~\ref{table_canham2}), shapes obtained by scheme A are
different from those of the other three schemes,  consistent with
recent work~\cite{Hoore2018}.  With an oblate-like biconcave shape
as initial shapes, we reproduce the same sequence of shapes by
scheme A as those shown in the supplementary movie of~\cite{Hoore2018}.
However, the run time of~\cite{Hoore2018} is too short to observe the
equilibrium configurations.  Moreover, the final (local) equilibrium
shape by scheme A is also dependent on the initial shapes. Each
shape by scheme A in Table \ref{table_canham2} is selected as the smallest
energy among steady states of three minimizations, each of which has one
initial shapes among sphere, prolate and oblate.  The equilibrium
shapes by scheme A are always prolate-like/dumbbells/cigars, which are
completely different from the analytical results in the reference study
\cite{Seifert1991}. Schemes B, C, and D  all  produce the biconcave
oblates with no distinct differences among them.

\begin{table}
\centering
\caption{Equilibrium shapes of the minimal model: comparison among four schemes
for $0.25 \le  v \le 0.45$.}
\begin{tabular}{c  c  c  c  c  c}
$v$ & $0.25$ &  $0.3$ & $0.35$ & $0.4$ & $0.45$   \\
\raisebox{\abheight\height}{B: $N_t=320$} &
\includegraphics[width=\amsize\columnwidth]{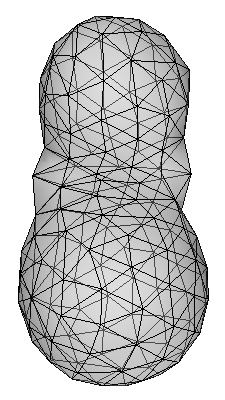}
\includegraphics[width=\amsize\columnwidth]{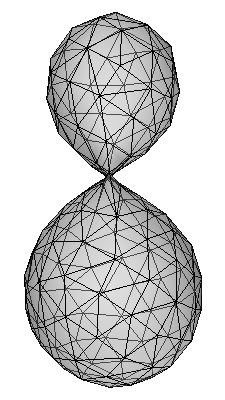} &
\includegraphics[width=\amsize\columnwidth]{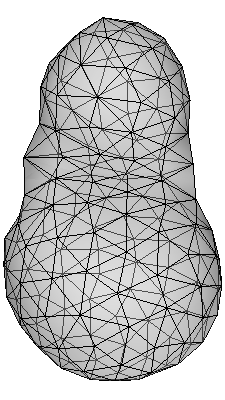}
\includegraphics[width=\amsize\columnwidth]{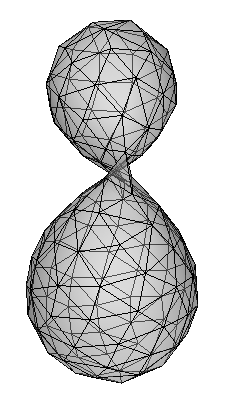} &
\includegraphics[width=\asize\columnwidth]{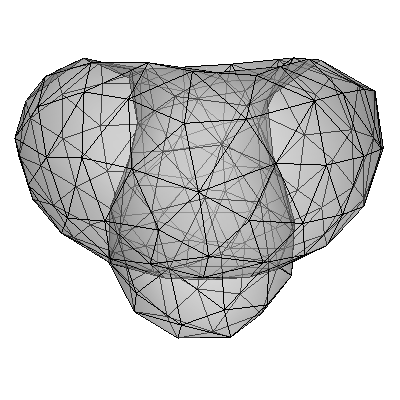}  &
\includegraphics[width=\asize\columnwidth]{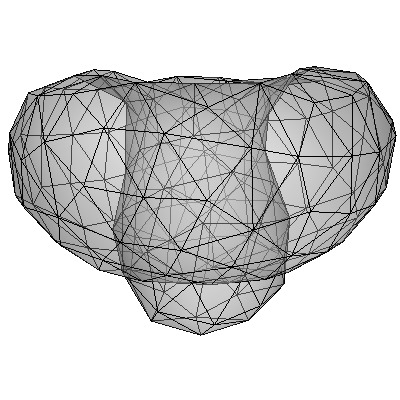}  &
\includegraphics[width=\asize\columnwidth]{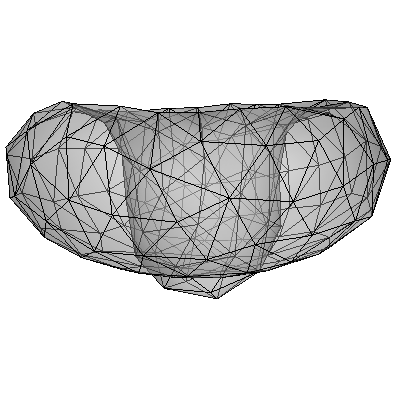}  \\
\raisebox{\abheight\height}{B: $N_t=1280$} &
\includegraphics[width=\asize\columnwidth]{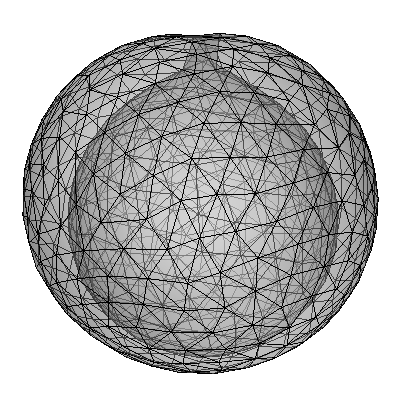}  &
\includegraphics[width=\asize\columnwidth]{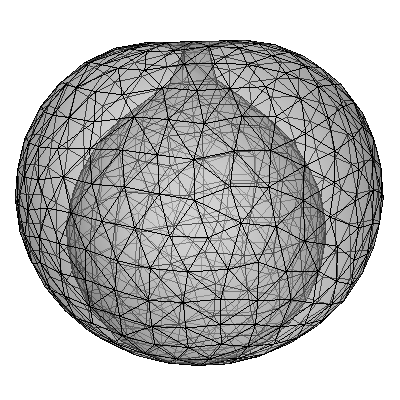}  &
\includegraphics[width=\asize\columnwidth]{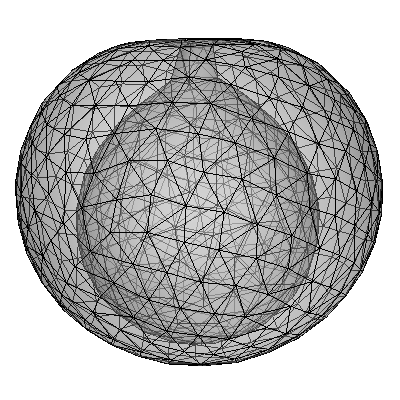}  &
\includegraphics[width=\asize\columnwidth]{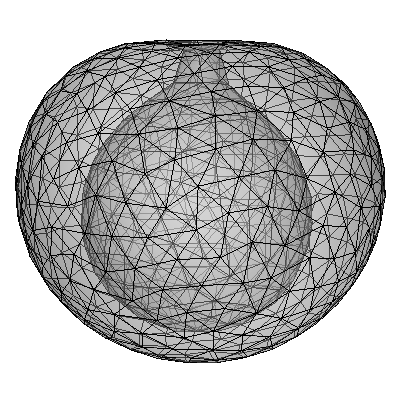}  &
\includegraphics[width=\asize\columnwidth]{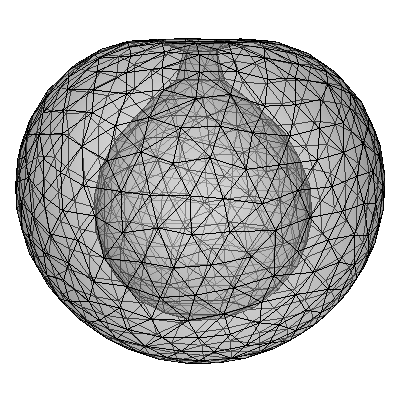}  \\
\raisebox{\abheight\height}{ B: $N_t=5120$} &
\includegraphics[width=\asize\columnwidth]{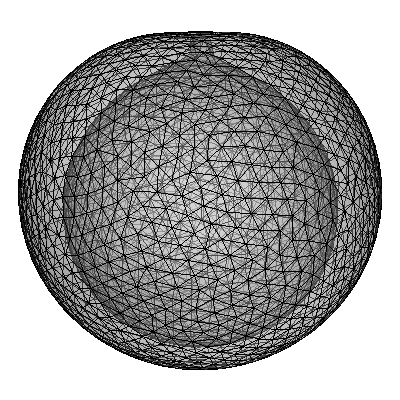}  &
\includegraphics[width=\asize\columnwidth]{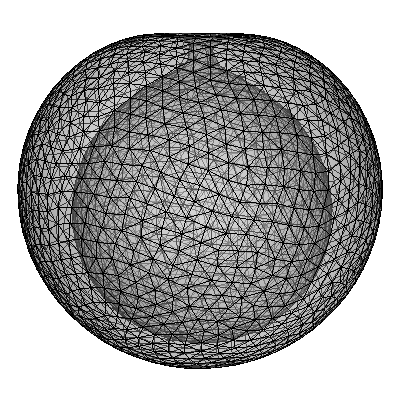}  &
\includegraphics[width=\asize\columnwidth]{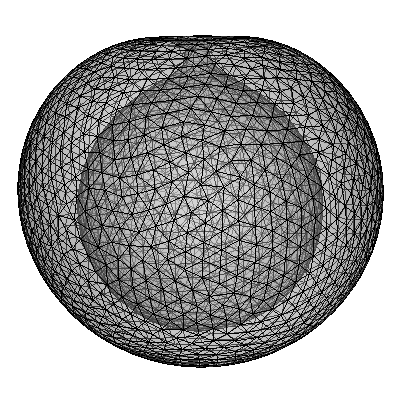}  &
\includegraphics[width=\asize\columnwidth]{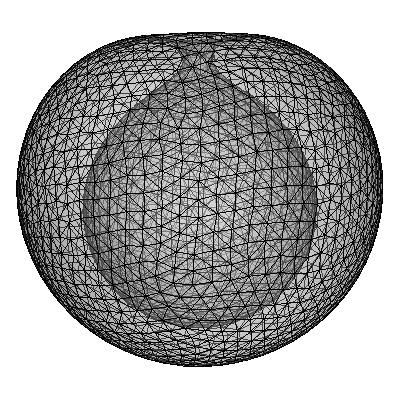}  &
\includegraphics[width=\asize\columnwidth]{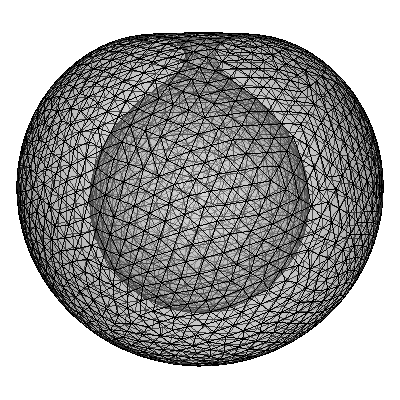}  \\
\raisebox{\abheight\height}{C: $N_t=1280$} &
\includegraphics[width=\asize\columnwidth]{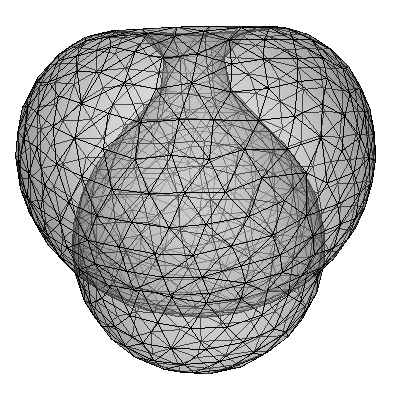}  &
\includegraphics[width=\asize\columnwidth]{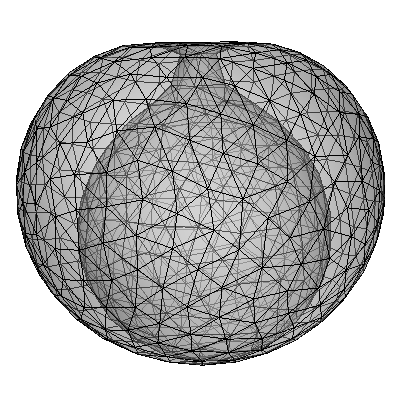}  &
\includegraphics[width=\asize\columnwidth]{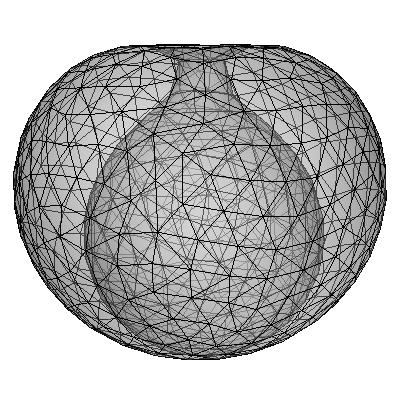}  &
\includegraphics[width=\asize\columnwidth]{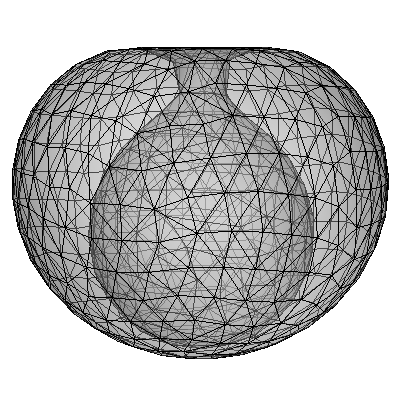}  &
\includegraphics[width=\asize\columnwidth]{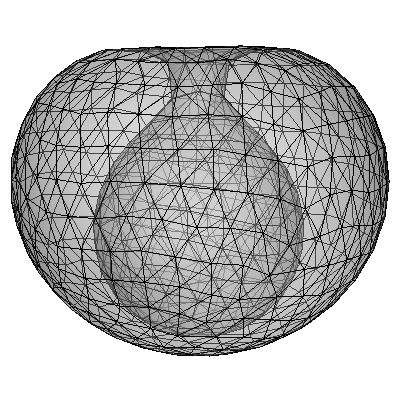}  \\
\raisebox{\abheight\height}{C: $N_t=5120$} &
\includegraphics[width=\amsize\columnwidth]{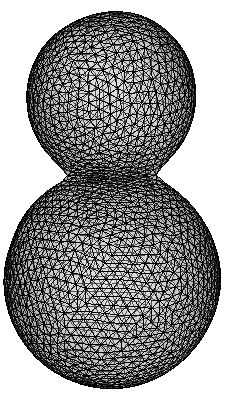}
\includegraphics[width=\amsize\columnwidth]{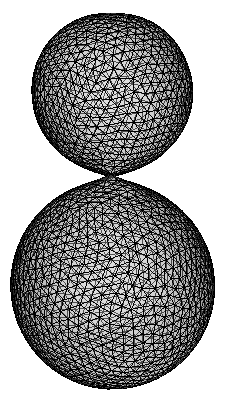} &
\includegraphics[width=\asize\columnwidth]{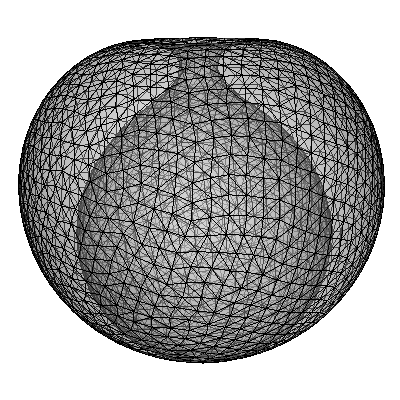}  &
\includegraphics[width=\asize\columnwidth]{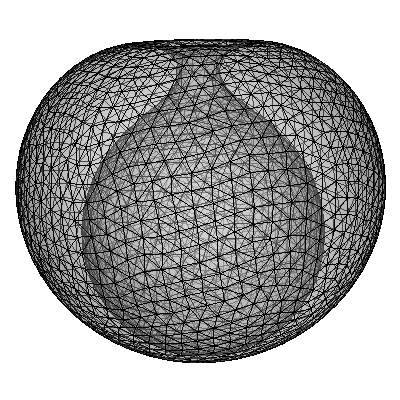}  &
\includegraphics[width=\asize\columnwidth]{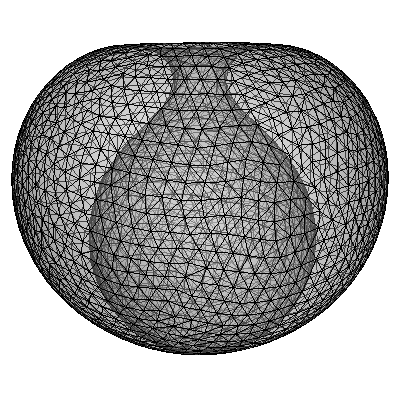}  &
\includegraphics[width=\asize\columnwidth]{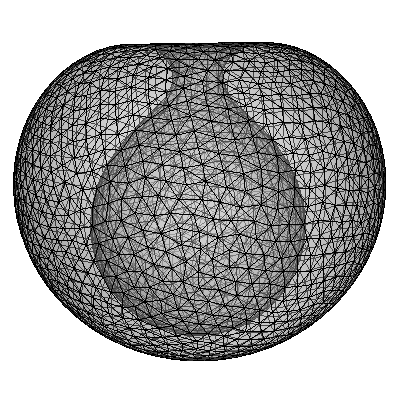}  \\
\end{tabular}
\label{table_canham3}
\end{table}
At smaller reduced volume $v < 0.59$, scheme A does not generate
stomatocyte, but always ends up with vesiculations, that is, two sphere
like shapes  connected via thin tubes (not shown here).   Schemes B
and C generate equilibrium shapes consistent with what is expected by
the reference study as shown in Table~\ref{table_canham3} for two and
three different resolutions.  Scheme B is stable even at low resolution
of $N_t=320$, although the equilibrium shapes are not as accurate.
Especially for $v=0.25$ and $0.3$, scheme B with $N_t=320$ does not
lead to the proper stomatocyte.  Instead the two spheres are connected
externally via a line segment, which can be seen from two orthogonal views
(the first two entries in the first row of Table~\ref{table_canham3}).
These two shapes are caused by the under-resolved neck connecting
the two spheres at this low resolution.  All equilibrium shapes
generated by scheme B with$N_t=1280$ are accurately axi-symmetric
stomatocyte and show virtually almost no difference from the results
of $N_t=5120$.  These results are comparable to the two-dimensional
contours reported in the reference~\cite{Seifert1991}.  The results of
scheme C are  shown in the last two rows of Table \ref{table_canham3}
for $N_t=1280$ and $5120$.  Especially for the most challenging case
of $v=0.25$ considered here, scheme C leads to  unphysical shapes.
Even for $v \ge 0.3$, scheme C cannot separate the inner sphere from
the outer sphere with a reasonable distance, due to poor resolution at
the neck.  The results of scheme D are very similar to those of scheme
C as the minimization procedure evolves for each target $v$.  However,
once scheme D reaches quasi-steady state with a proper stomatocyte shape,
the overall configuration oscillates and the shape becomes unstable.
We tried to employ regularization and smoothing to stabilize scheme D,
but we were  unsuccessful.  We speculate  that  since scheme D does
not enforce the conservation of momentum or energy this may cause
this instability.  To the best of our knowledge,  there are no reports
from previous work using triangulated mesh to simulate such low volumes.
The exact reason why scheme D is unstable remain obscure.

\subsubsection{Spontaneous curvature model}
We compute the minimal energies determined by the SC model and the
corresponding equilibrium configurations.  In particular, we consider two
spontaneous curvature values $h_0=1.2$ and $1.5$ so that we can compare
with results of Seifert et al \cite{Seifert1991}.  The possible shapes
generated by the SC model are still axi-symmetric.  However, the SC model
enables quite a few more configurations that are numerically challenging. These include pear-like shapes which break the up-down symmetry, budding
transition with a narrow neck connecting the neighboring compartments of
the same size, budding transitions with the neighboring compartments of
different sizes, vesiculation where connection between the neighboring
compartments of the same size is lost and vesiculation where connection
between the neighboring compartments of different sizes is lost.  There
are also discontinuous transitions for $h_0 \neq 0$, but the minimization
procedure of exploring the energy landscape is quite similar to the case
of $h_0=0$ and we do not elaborate on this finding here.

\subsubsection{$h_0 = 1.2$}
\begin{figure}
\centering
\begin{subfigure}{0.49\textwidth}
\includegraphics[width=\columnwidth]{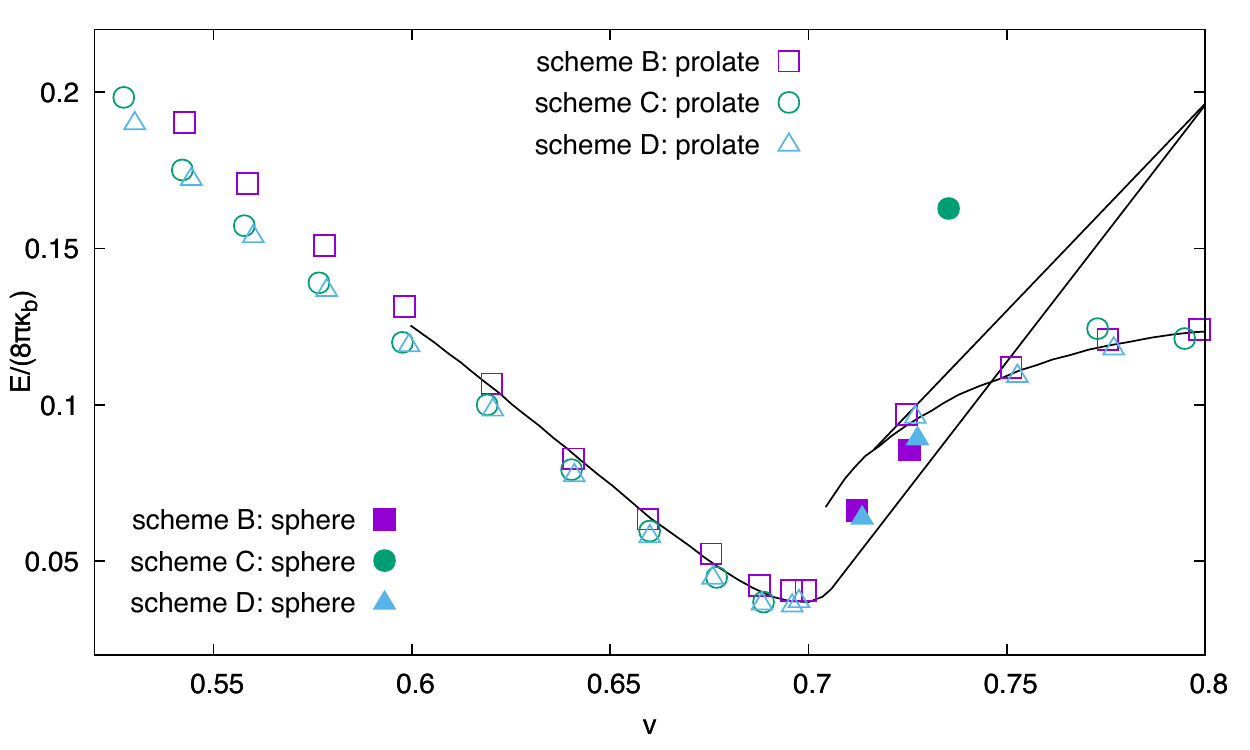} 
\caption{}
\label{fig_energy_sc_h12}
\end{subfigure}
\begin{subfigure}{0.49\textwidth}
\includegraphics[width=\columnwidth]{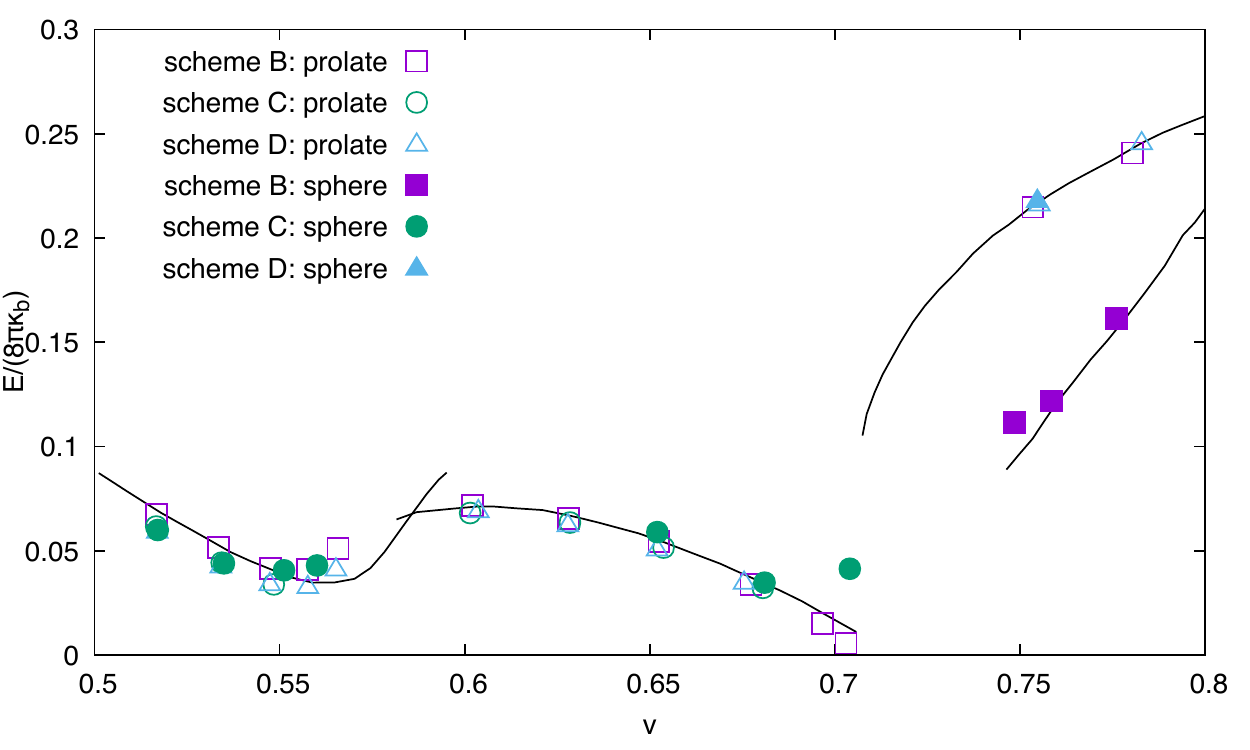} 
\caption{}
\label{fig_energy_sc_h15}
\end{subfigure}
\caption{Minimized Helfrich energies for each $v$ by schemes B, C, and D with
$N_t=1280$.  The initial shapes are prolates or spheres .
(a) $h_0=1.2$ and (b) $h_0=1.5$.  The reference lines are adapted from
Seifert et al. \cite{Seifert1991}.}
\label{fig_energy_sc}
\end{figure}

\begin{table}
\centering
\caption{Equilibrium shapes of the SC model $h_0=1.2$:  comparison among three
schemes for $0.52 < v < 0.8$ with $N_t=1280$.}
\begin{tabular}{ccccccc}
$v$ & $0.53$ &  $0.6$  &  $0.69$ &  $0.725-0.735$ & $0.75$ & $0.8$  \\
\raisebox{\dheight\height}{B} &
\includegraphics[width=\asize\columnwidth]{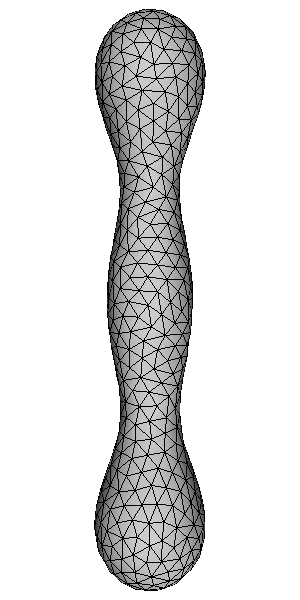}  &
\includegraphics[width=\asize\columnwidth]{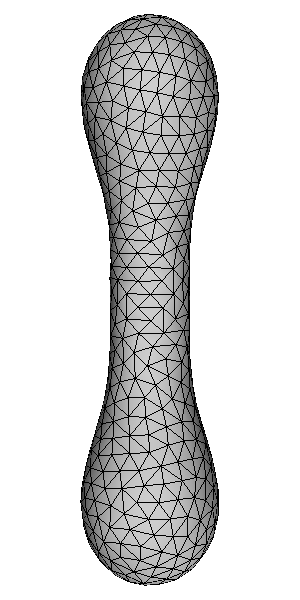}  &
\includegraphics[width=\asize\columnwidth]{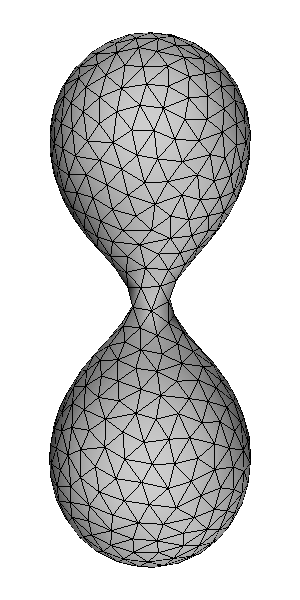}  &
\includegraphics[width=\asize\columnwidth]{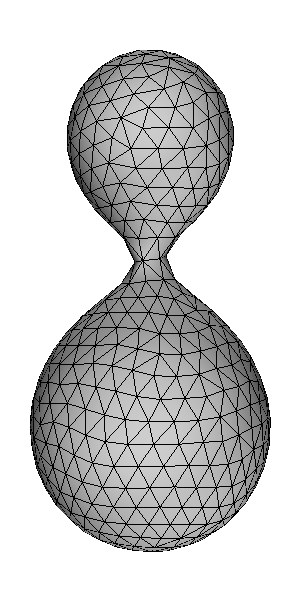}  &
\includegraphics[width=\asize\columnwidth]{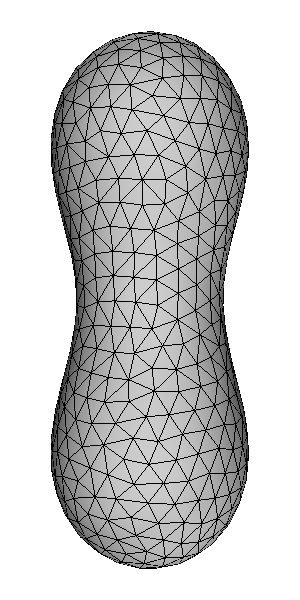}  &
\includegraphics[width=\asize\columnwidth]{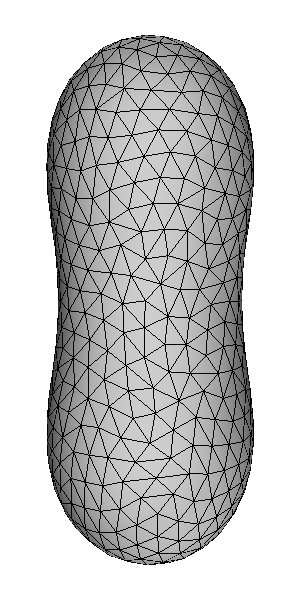}  \\
\raisebox{\dheight\height}{C} &
\includegraphics[width=\asize\columnwidth]{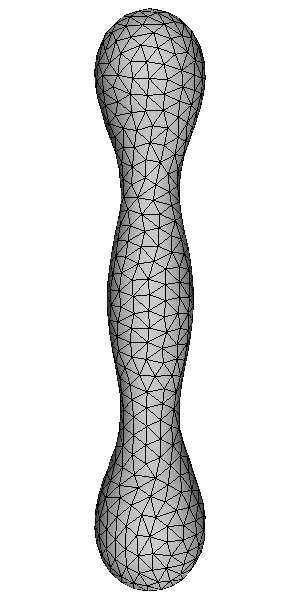}  &
\includegraphics[width=\asize\columnwidth]{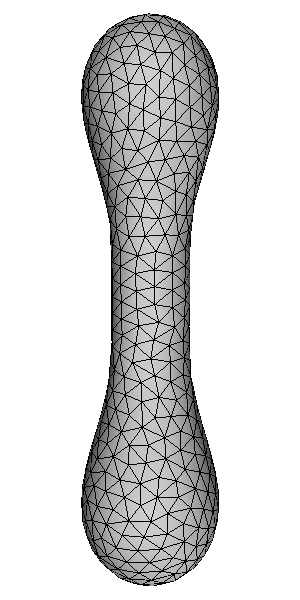}  &
\includegraphics[width=\asize\columnwidth]{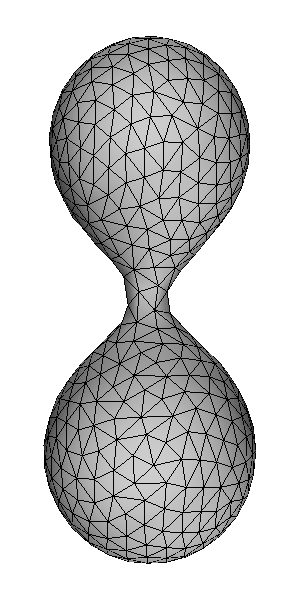}    &
\includegraphics[width=\asize\columnwidth]{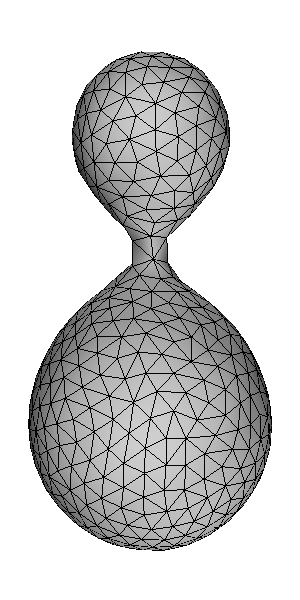}  &
\includegraphics[width=\asize\columnwidth]{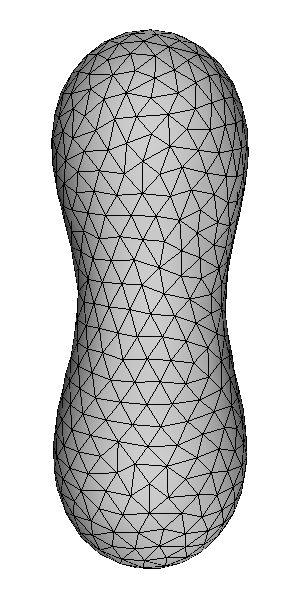}  &
\includegraphics[width=\asize\columnwidth]{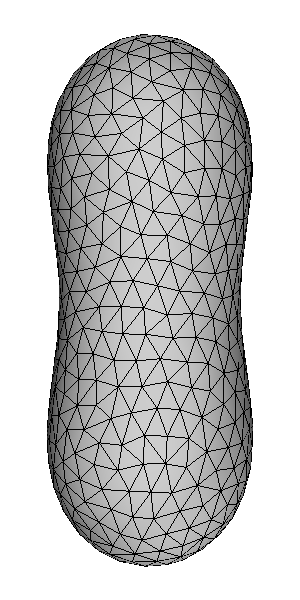}  \\
\raisebox{\dheight\height}{D} &
\includegraphics[width=\asize\columnwidth]{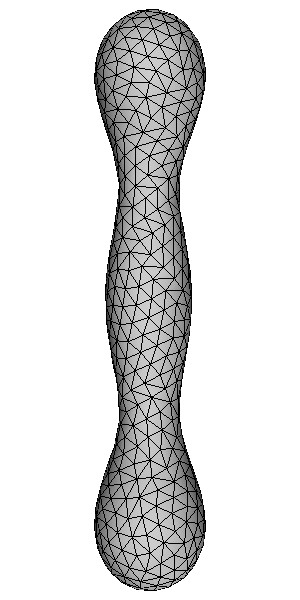}  &
\includegraphics[width=\asize\columnwidth]{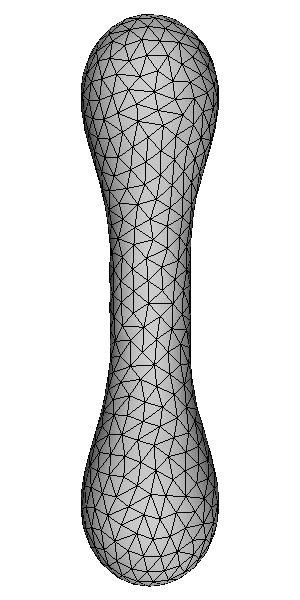}  &
\includegraphics[width=\asize\columnwidth]{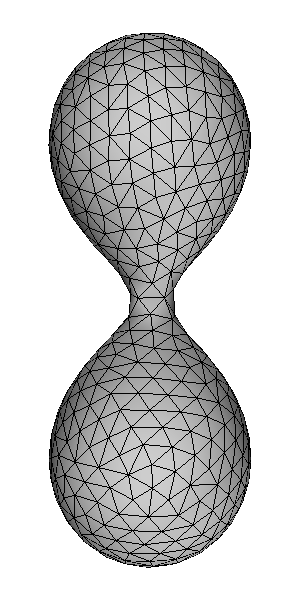}  &
\includegraphics[width=\asize\columnwidth]{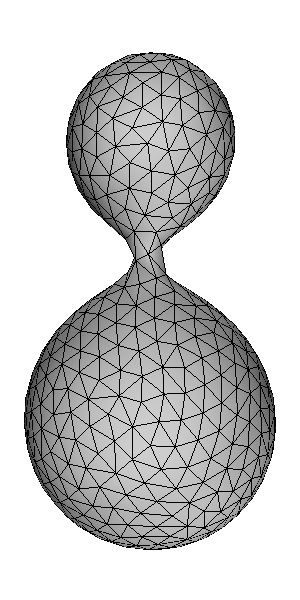}  &
\includegraphics[width=\asize\columnwidth]{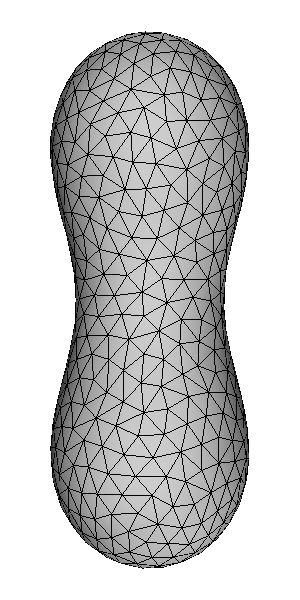}  &
\includegraphics[width=\asize\columnwidth]{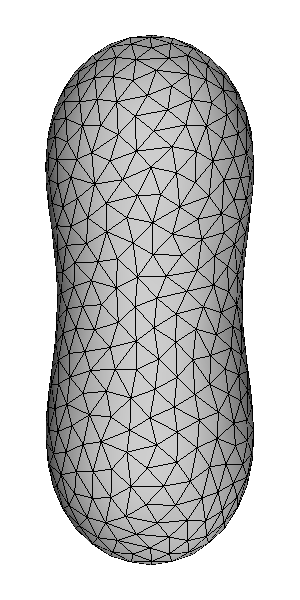}
\end{tabular}
\label{table_shape_sc_12}
\end{table}

We first consider $h_0=1.2$ and present the energy diagram in
Fig. \ref{fig_energy_sc_h12} and the representative equilibrium shapes in
Table \ref{table_shape_sc_12}.  The references are adapted from Seifert
et al.~\cite{Seifert1991}.  For the regime $v<\sqrt{2}/2 \approx 0.71$,
all three schemes accurately capture the shapes of necklace (with
two necks) at $v=0.53$, dumbbells at $v=0.6$ and budding at $v=0.69$,
as shown in the first three columns of Table \ref{table_shape_sc_12}.
The minimum energies in this regime computed by the three schemes are
also quite close to the reference line.  Especially scheme B accurately
reproduces the energy line with $N_t=1280$ as shown in the left part
of Fig. \ref{fig_energy_sc_h12}.  For the regime $v \gtrsim 0.75$,
we expect prolate/dumbbell shapes as suggested by the reference.
All three schemes B, C and D are able to get the minimum energies
accurately as shown in the right part of Fig. \ref{fig_energy_sc_h12},
and to generate the expected shapes comparable to the reference (see
figure 12 of Ref. \cite{Seifert1991}) as shown in the last two columns of
Table \ref{table_shape_sc_12}.  The most challenging regime resides in
$0.71 \lesssim v \lesssim 0.75$, since there are multiple local minima
and the energy diagram bifurcates.  However, all three schemes are able
to produce the budding, where there is only one very thin neck connecting
the two compartments of different sizes.  For this very extreme regime,
we do not expect accurate computation on the minimum energy.  However, the
energies by scheme B and D are still surprisingly close to the reference,
whereas the scheme C is completely off due to the poor representation
of the narrow neck with only a couple of triangles.

\subsubsection{$h_0 = 1.5$}
\begin{table}
\centering
\caption{Equilibrium shapes of SC model $h_0=1.5$:  comparison among three schemes for $0.52 \le v \le 0.78$ with $N_t=1280$.}
\begin{tabular}{ccccccc}
$v$ & $0.52$ &  $0.56$ &  $0.65$ &  $0.7$&  $0.76$  & $0.78$ \\
\raisebox{\dheight\height}{B} &
\includegraphics[width=\asize\columnwidth]{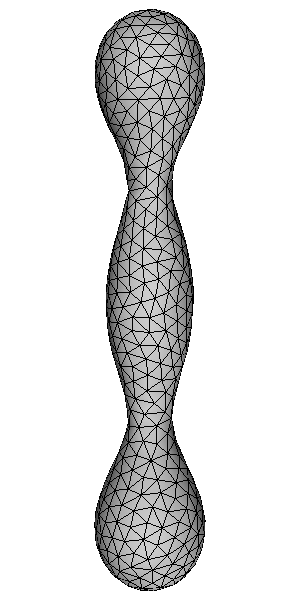}  &
\includegraphics[width=\asize\columnwidth]{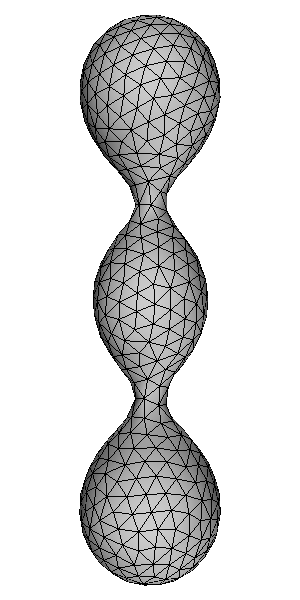}  &
\includegraphics[width=\asize\columnwidth]{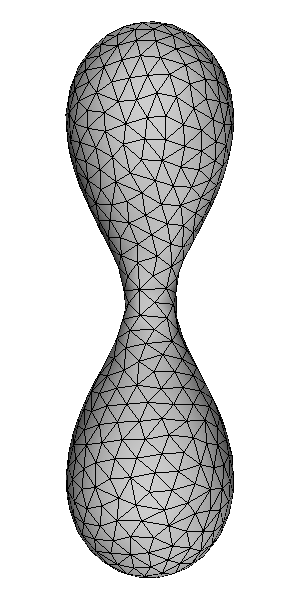}  &
\includegraphics[width=\asize\columnwidth]{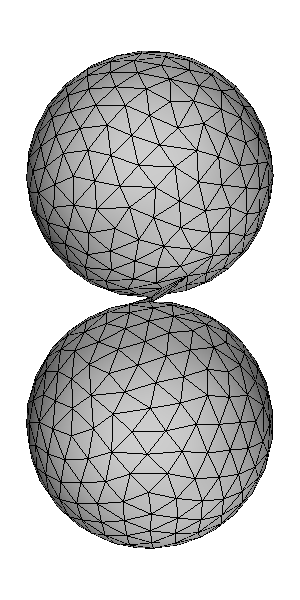}  &
\includegraphics[width=\asize\columnwidth]{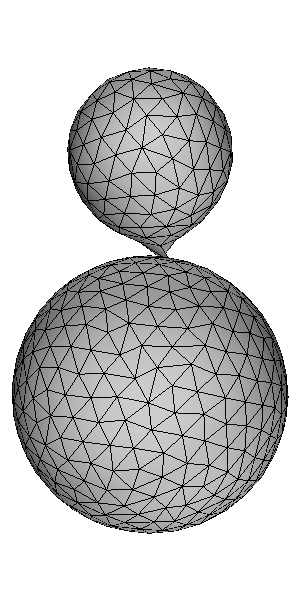}  &
\includegraphics[width=\asize\columnwidth]{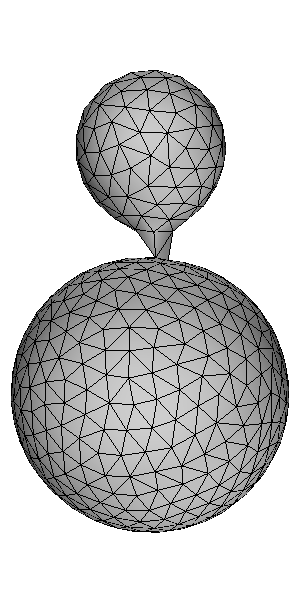}  \\
\raisebox{\dheight\height}{C} &
\includegraphics[width=\asize\columnwidth]{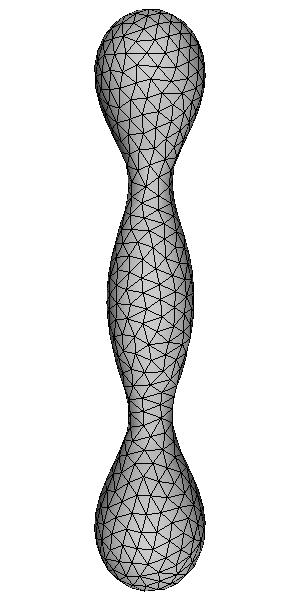}  &
\includegraphics[width=\asize\columnwidth]{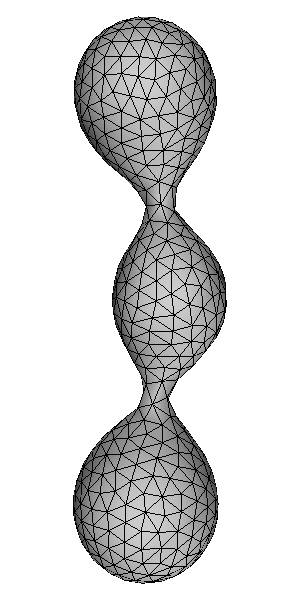}  &
\includegraphics[width=\asize\columnwidth]{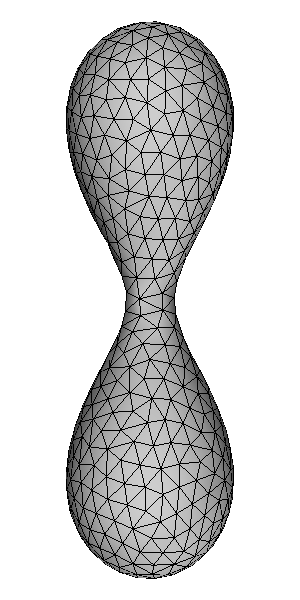}  &
\includegraphics[width=\asize\columnwidth]{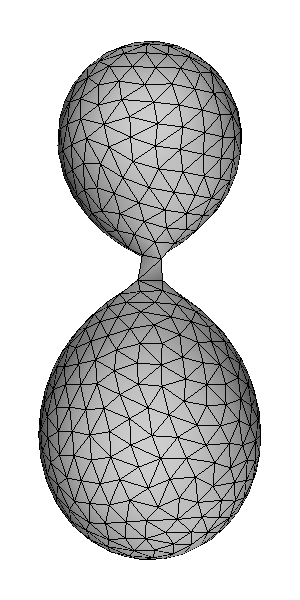}  &
\raisebox{\dheight\height}{NA} &
\raisebox{\dheight\height}{NA} \\
\raisebox{\dheight\height}{D} &
\includegraphics[width=\asize\columnwidth]{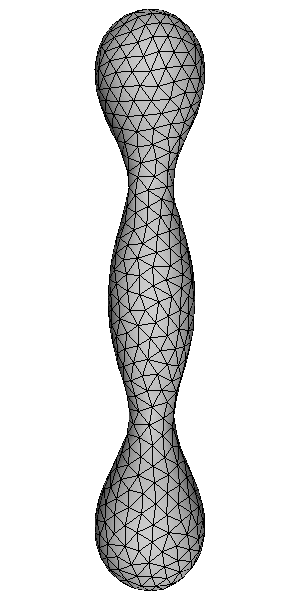}  &
\includegraphics[width=\asize\columnwidth]{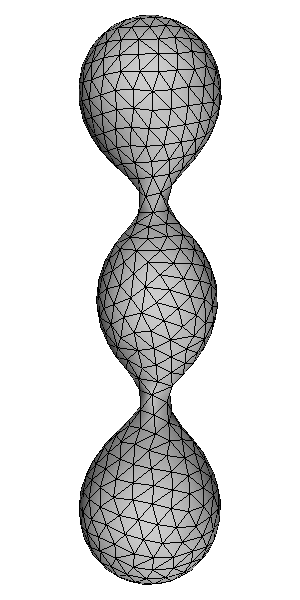}  &
\includegraphics[width=\asize\columnwidth]{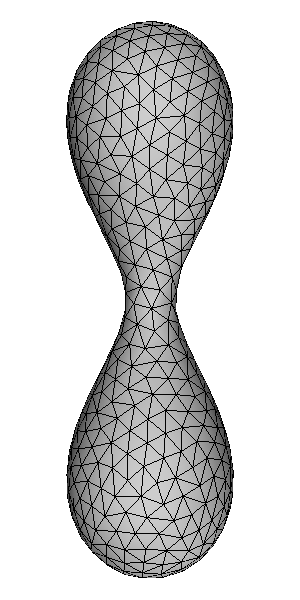}  &
\raisebox{\dheight\height}{NA} &
\raisebox{\dheight\height}{NA} &
\raisebox{\dheight\height}{NA}
\end{tabular}
\label{table_shape_sc_15}
\end{table}

Computations for the  case of  $h_0=1.5$ are  even more challenging
as, besides budding transitions, vesiculations are also expected.
For the regime $ v \lesssim 0.58$, all three schemes are able
to generate necklaces as shown in the first two columns in Table
\ref{table_shape_sc_15} and compute minimum energy accurately as shown
on the left part of Fig. \ref{fig_energy_sc_h15}.  There is notable
discrepancy for the equilibrium shape generated by scheme C at $v=0.56$,
as it cannot sustain the axis-symmetry due to its inability to resolve
the narrow necks of the budding shapes.  For the regime $ 0.6 \lesssim
v \lesssim 0.71$, all three schemes can generate dumbbell shapes at
equilibrium, for example, at $v=0.65$.  However, only scheme B can
generate vesiculation of two spheres of equal size expected from the
reference \cite{Seifert1991}, as shown by one example of $v=0.7$ on Table
\ref{table_shape_sc_15}.  Scheme C at $v=0.7$ may generate vesiculation,
but the two compartments do not have equal size and the energy differs
from the reference.  Perhaps the regime of $ v \gtrsim 0.75$ is the most
challenging one, since the two compartments are expected to have up-down
symmetry breaking for the vesiculation.  We find that both schemes C
and D to run stably in this regime of vesiculations. On the contrary,
scheme B delivers the correct results with a satisfactory accuracy on
both the energy values and configurations.

To summarize this section, scheme B with $N_t=1280$ is able to generate
the whole spectrum of equilibrium shapes including dumbbells, necklaces,
even buddings and vesiculations where only a couple of triangles are
connecting different compartments.  Furthermore, the corresponding energy
value for each equilibrium  shape is very accurate in comparison
with the reference value.  Both schemes C and D with $N_t=1280$ run
into troubles when budding or/and vesiculations are expected.  With higher
resolutions $N_t=5120$, neither scheme C nor D show significant
improvement (results not presented herein).
\subsubsection{Bilayer-couple model}
\label{sec_numerics2_bc}
We  consider the phase diagram of the BC model which can be realized
by taking $\alpha \rightarrow \infty$ in Eq. ~(\ref{eq_energy_total})
so that the area-difference elasticity becomes a quadratic penalization.
In this model, there are two non-dimensional parameters $v$ and $\Delta
a$.  We consider the BC model before presenting results from the complete
ADE model for a few reasons: 1. The BC model has only continuous phase
transitions~\cite{Svetina1989, Seifert1991}.  2. The equilibrium shapes
and possible resultant $\Delta a$ in BC model contains those of the ADE
model~\cite{Seifert1997}.  3. The phase diagram of BC has two controlling
parameters, one less than that of ADE model.  4. With $h_0=0$, we can
focus on examinations the area-difference elasticity.

\begin{figure}
\centering
\begin{subfigure}{0.464\textwidth}
\includegraphics[width=\columnwidth]{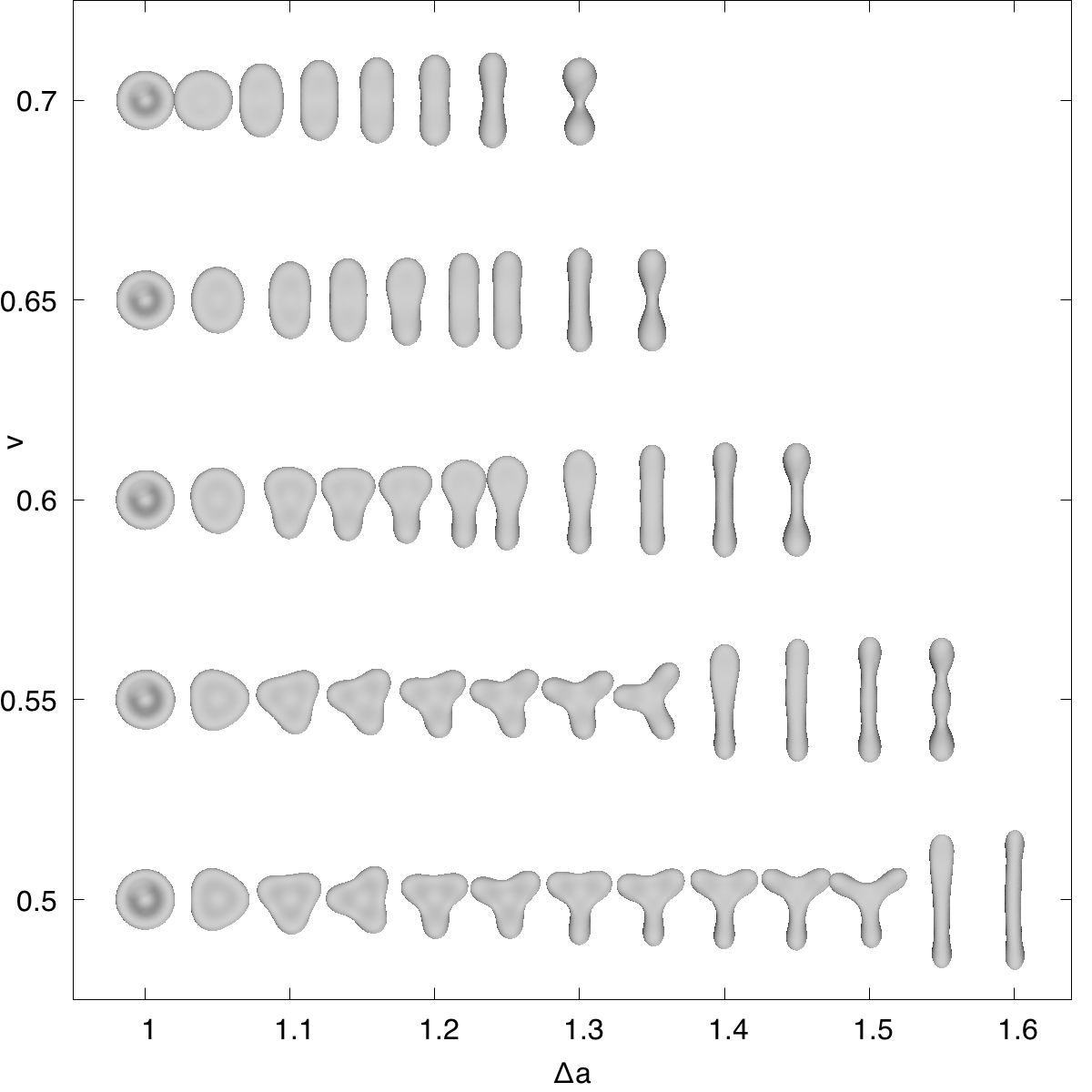} 
\caption{}
\end{subfigure}
\begin{subfigure}{0.523\textwidth}
\includegraphics[width=\columnwidth]{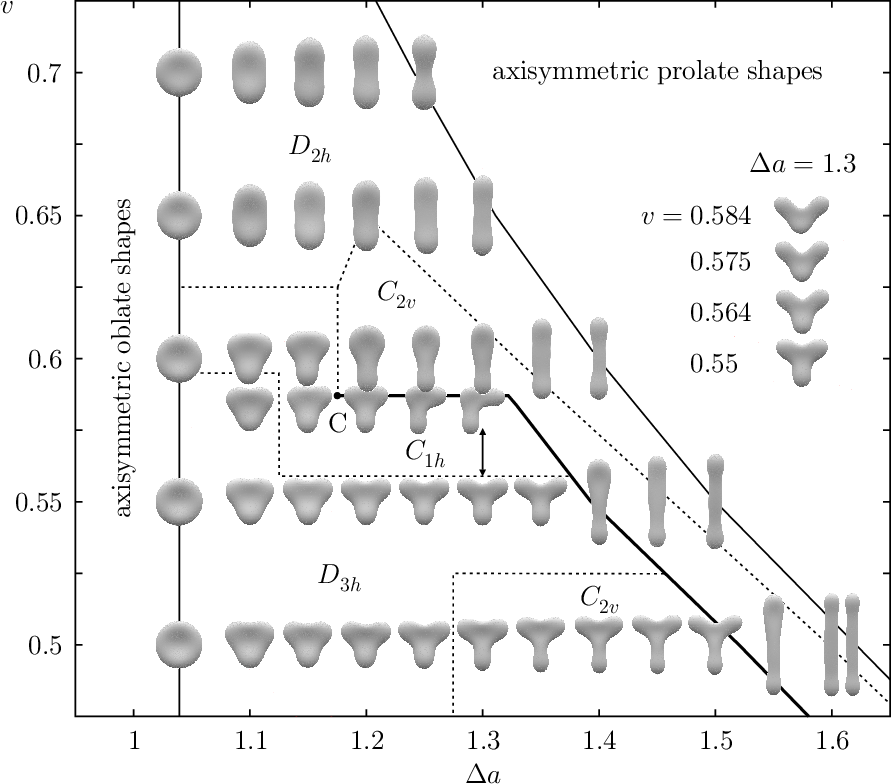} 
\caption{}
\end{subfigure}
\caption{Equilibrium shapes of the BC model.  The results are by scheme B with
$N_t=1280$ (a) and comparable to the numerical work of Ziherl and
Svetina~\cite{Ziherl2005} (b) (permission pending) and the experiments
of Sakashita et al.~\cite{Sakashita2012} (not shown here).}
\label{fig_bc_pd}
\end{figure}

We consider the same parameter ranges of $v$ and $\Delta a$ as
in Ref.~\cite{Ziherl2005}, where the program Surface Evolver was
used~\cite{Brakke1992}.  We present our phase diagram by scheme
B in Fig. \ref{fig_bc_pd}, where almost each individual shape has
one-to-one correspondence to the shape on Fig. 1 of Ziherl and Svetina
\cite{Ziherl2005}.  As in the SC model, we obtain similar axi-symmetric
shapes such as stomatocytes, prolate-like/dumbbells, budding necklaces.
Different from the SC model, we obtain in addition many non-axisymmetric
shapes from the constraint of the area difference between outer and
inner leaflets.  These shapes include elliptocytes (e.g., the fourth
one at $v=0.7$ or $0.65$), rackets (e.g., the fifth one at $v=0.65$
and sixth one at $v=0.6$), triangle oblates with equal sides (e.g.,
the sixth one at $v=0.55$), and triangle oblates with unequal sides
(e.g., the seventh one at $v=0.5$).  All these shapes have been recently
validated by experiments on self-assembled vesicles combined with image
analysis~\cite{Sakashita2012}.

\begin{figure}
\centering
\includegraphics[width=0.7\columnwidth]{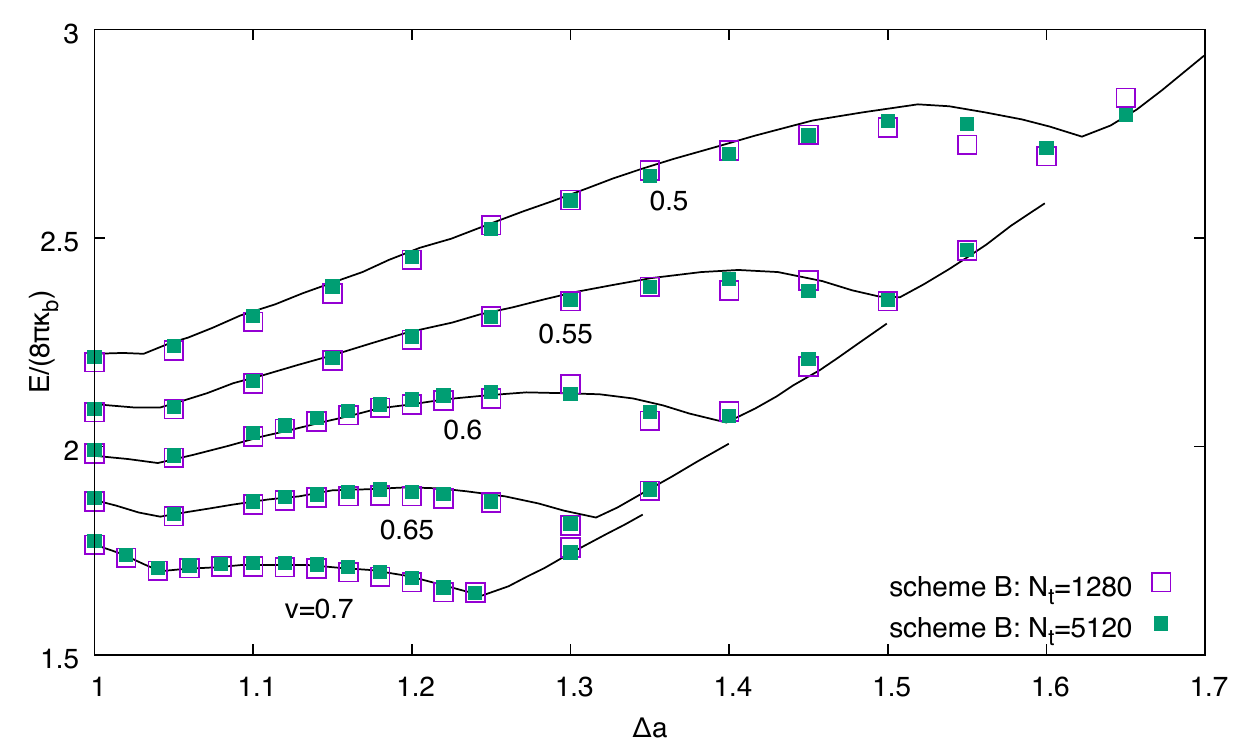} 
\caption{Minimal energies for the BC model.
The references lines are adapted from Ziherl and Svetina~\cite{Ziherl2005}.}
\label{fig_bc_energy}
\end{figure}

Furthermore, we verify the corresponding energy values as shown in
Fig.~\ref{fig_bc_energy}, where the reference values are adapted from
Ziherl and Svetina~\cite{Ziherl2005}.  We observe that energies computed
by scheme B with $N_t=1280$ and $5120$ exhibit negligible discrepancies
and they both follow closely  the reference values.  Therefore, we omit
the equilibrium configurations generated with $N_t=5120$.

We have also computed the same phase diagram by scheme C and D with
$N_t=1280$ and $5120$.  However, scheme C  and scheme D become unstable
for certain values of $\Delta a$.  We speculate that the vulnerability
of scheme C is due to the definition of the discrete normal vector,
which is not compatible with the discrete definition of the Laplace
operator.  We maintain that the instability of scheme D is due to its
lack of conservation properties.  We consider that a more rigorous
investigations on scheme C and D are beyond the scope of this work.

\subsection{Area-difference-elasticity model}
\label{sec_numerics2_ade}
Here we compute the total energy and force in the ADE model.
For comparison, we consider as reference the work of Khairy et
al. ~\cite{Khairy2008a}, which employs spherical harmonics to represent
the surface and calculate the total energy.  Therefore, we take the same
reduced volume $v\approx 0.642$ and $\alpha=2/\pi$, which represent the
parameters of a normal RBC~\cite{Lim2002, Khairy2008a}.

\begin{figure}
\centering
\centering
\begin{subfigure}{0.45\textwidth}
\includegraphics[width=\columnwidth]{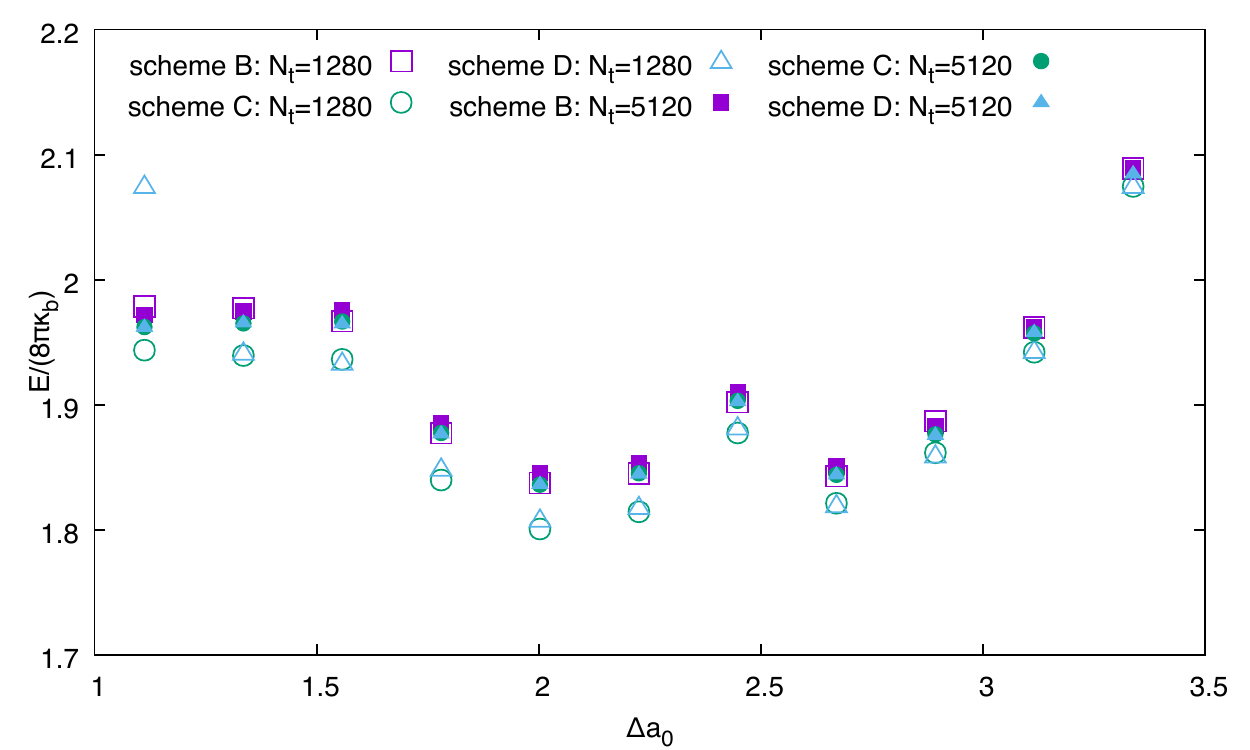} 
\caption{}
\label{fig_ade_alpha064_a}
\end{subfigure}
\begin{subfigure}{0.45\textwidth}
\includegraphics[width=\columnwidth]{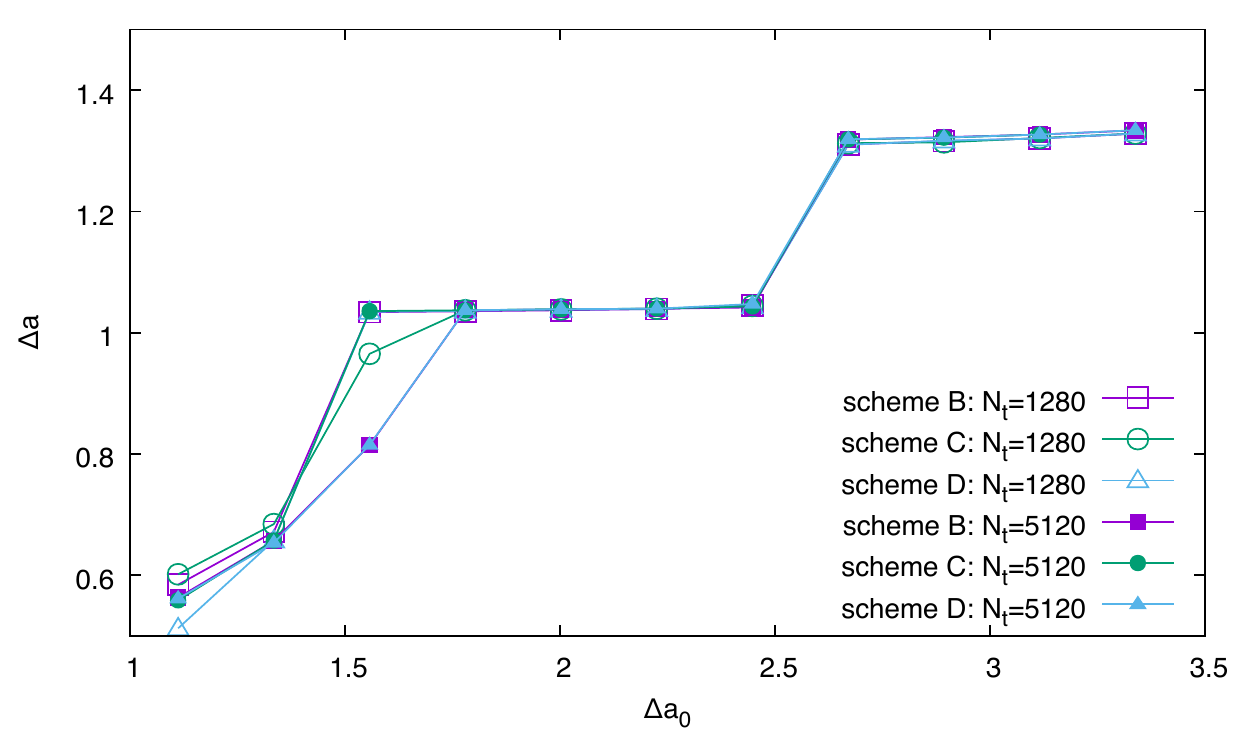} 
\caption{}
\label{fig_ade_alpha064_b}
\end{subfigure}
\begin{subfigure}{0.45\textwidth}
\includegraphics[width=\columnwidth]{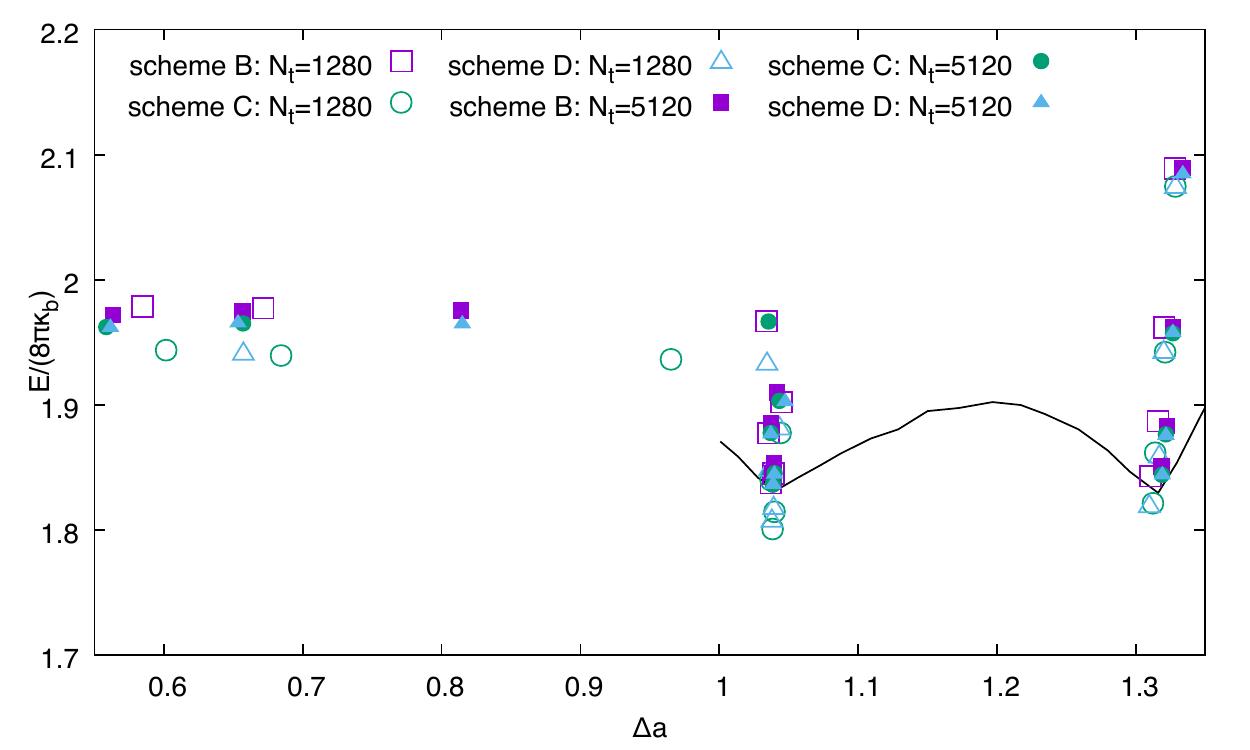} 
\caption{}
\label{fig_ade_alpha064_c}
\end{subfigure}
\caption{Minimal energies and area differences from ADE model: $\alpha=2/\pi$.}
\label{fig_ade_alpha064}
\end{figure}

We present the energy versus reference area difference $\Delta a_0$
on Fig. \ref{fig_ade_alpha064_a}, where we include results of all three
schemes and each with two resolutions of $N_t=1280$ and $5120$.  For each
$\Delta a_0$ considered from $1$ to $3.5$, there is only small discrepancy
between the schemes with a proper resolution, that is, scheme B with
$N_t=1280$ and $5120$ and scheme C and D with $N_t=5120$.  We present  the
resultant $\Delta a$ versus $\Delta a_0$ on Fig. \ref{fig_ade_alpha064_b}
and observe that $\Delta a$ increases as $\Delta a_0$ for smaller
$\Delta a_0 \lesssim 1.75$.  It stays almost at one plateau for $1.75
\lesssim \Delta a_0 \lesssim 2.5$ and at another plateau for $\Delta
a_0 \gtrsim 2.6$.  Correspondingly, we plot again the energy versus the
resultant $\Delta a$ on Fig. \ref{fig_ade_alpha064_c}.  We observe that
the associated energies span among possible area differences for $\Delta
a \lesssim 1$ and cluster at two discrete states of area differences
for $\Delta a \gtrsim 1$.  These characteristics of area differences and
minimal energies reflect the possible equilibrium configurations of the
vesicles, which will be presented next.

\begin{table}
\centering
\caption{Equilibrium shapes of the ADE model:  comparison among four schemes for
$N_t=1280$ and $5120$, $\alpha=2/\pi$.  Both values of $\Delta a_0$ and
$\Delta a'_0$ are listed to compare with Khairy et al \cite{Khairy2008a}.}
\begin{tabular}{c c  c  c  c  c}
$\Delta a_0$ & $1.11$ & $1.34$ & $1.78$ & $2.34-2.56$ & $3.34$ \\
$100\Delta a'_0$ & $0.1$ & $0.12$ & $0.16$ & $0.21-0.23$ &  $0.3$ \\
\raisebox{\cheight\height}{B: $N_t=1280$} &
\raisebox{0.2\height}{\includegraphics[width=\asize\columnwidth]{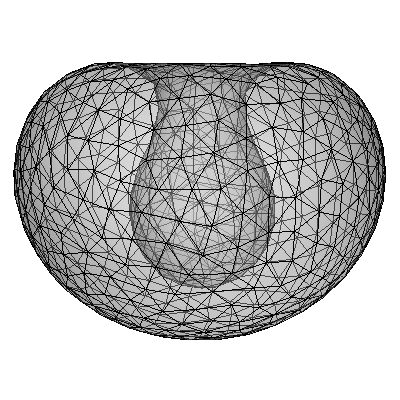} } &
\raisebox{0.2\height}{\includegraphics[width=\asize\columnwidth]{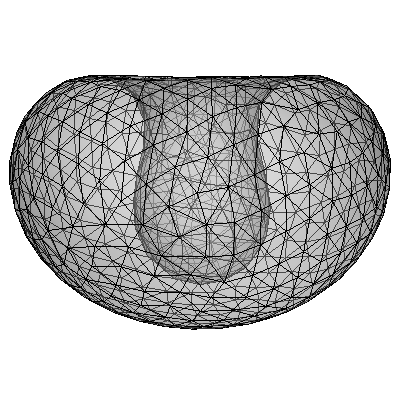} } &
\raisebox{\height}{\includegraphics[width=\asize\columnwidth]{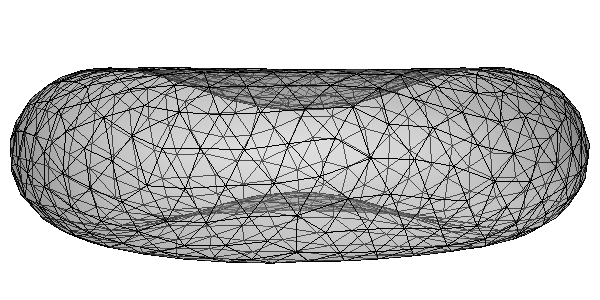} } &
\raisebox{0.2\height}{\includegraphics[width=\asize\columnwidth]{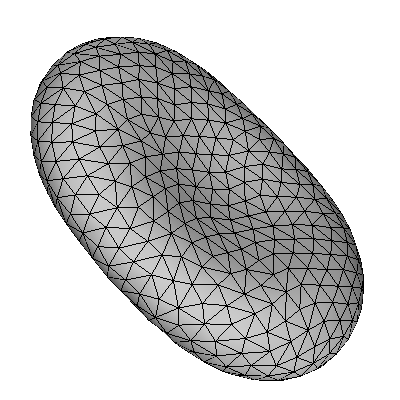} } &
\includegraphics[width=\asize\columnwidth]{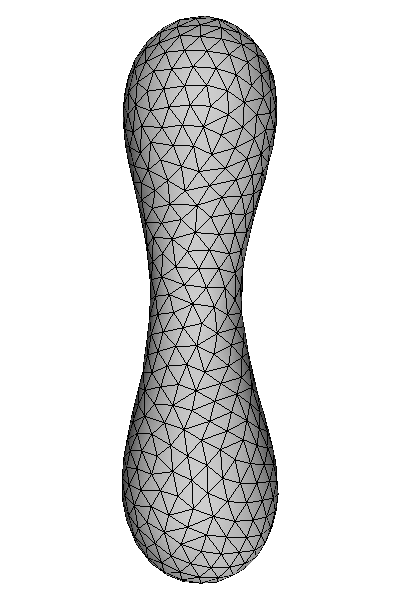}   \\
\raisebox{\cheight\height}{B: $N_t=5120$} &
\raisebox{0.2\height}{\includegraphics[width=\asize\columnwidth]{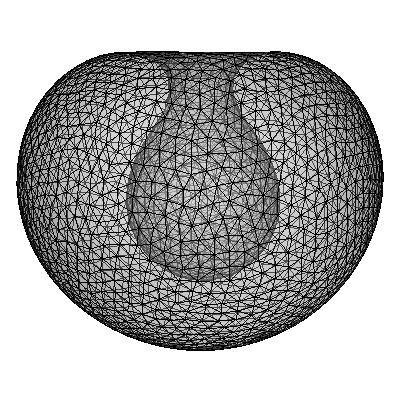} } &
\raisebox{0.2\height}{\includegraphics[width=\asize\columnwidth]{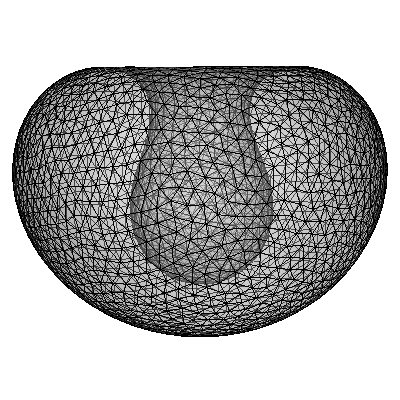} } &
\raisebox{\height}{\includegraphics[width=\asize\columnwidth]{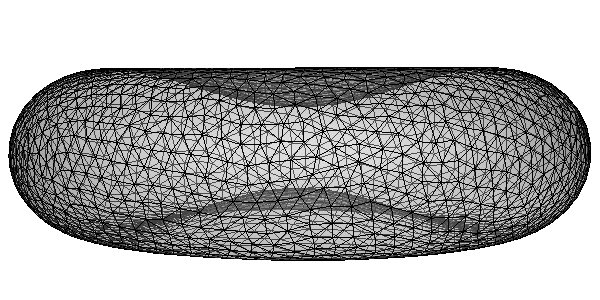} } &
\raisebox{0.2\height}{\includegraphics[width=\asize\columnwidth]{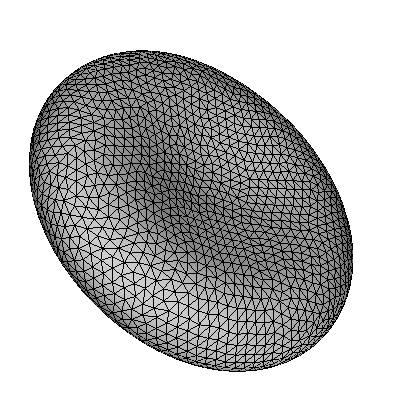} } &
\includegraphics[width=\asize\columnwidth]{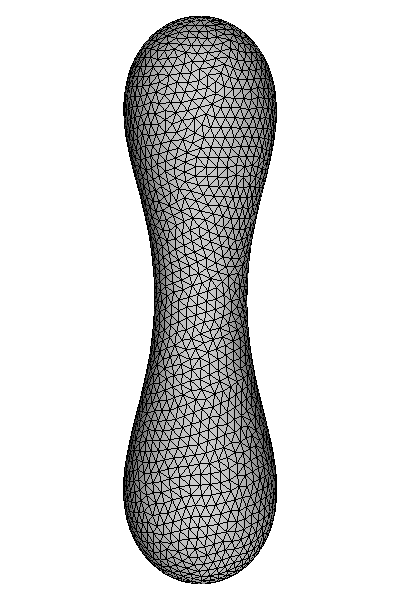}   \\
\raisebox{\cheight\height}{C: $N_t=1280$} &
\raisebox{0.2\height}{\includegraphics[width=\asize\columnwidth]{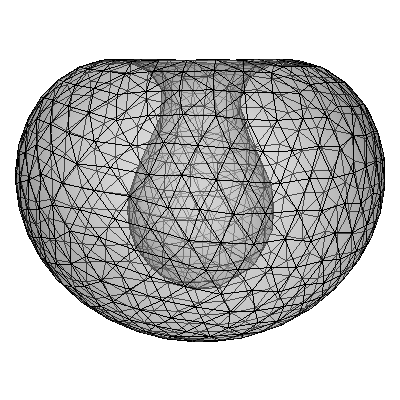} } &
\raisebox{0.2\height}{\includegraphics[width=\asize\columnwidth]{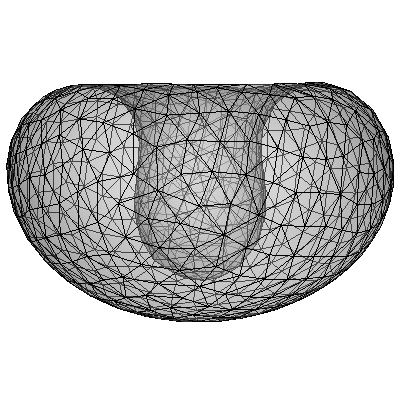} } &
\raisebox{\height}{\includegraphics[width=\asize\columnwidth]{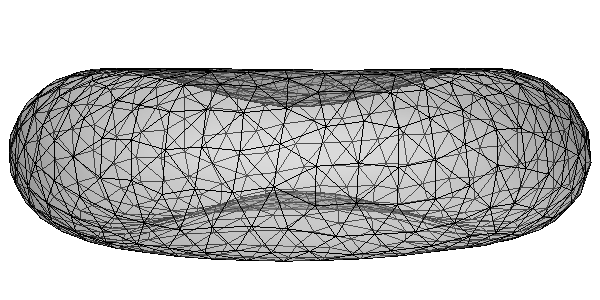} } &
\raisebox{0.2\height}{\includegraphics[width=\asize\columnwidth]{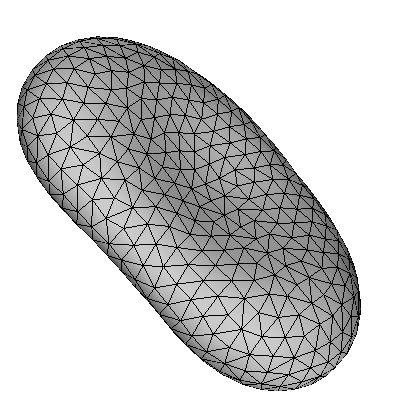} } &
\includegraphics[width=\asize\columnwidth]{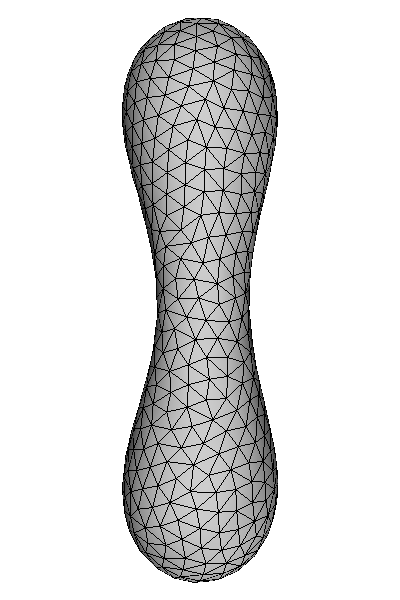}   \\
\raisebox{\cheight\height}{C: $N_t=5120$} &
\raisebox{0.2\height}{\includegraphics[width=\asize\columnwidth]{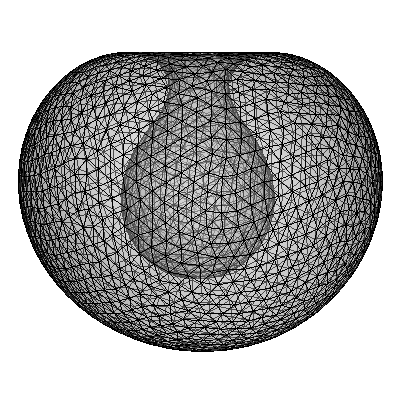} } &
\raisebox{0.2\height}{\includegraphics[width=\asize\columnwidth]{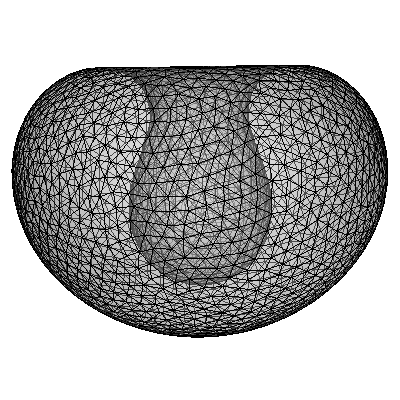} } &
\raisebox{\height}{\includegraphics[width=\asize\columnwidth]{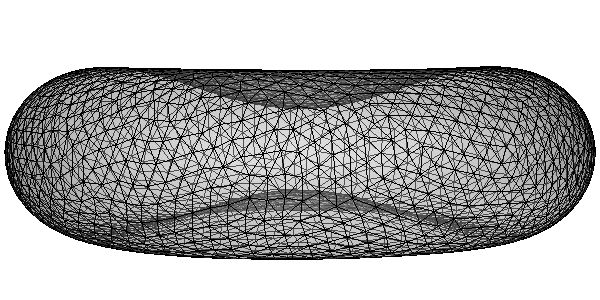} } &
\raisebox{0.2\height}{\includegraphics[width=\asize\columnwidth]{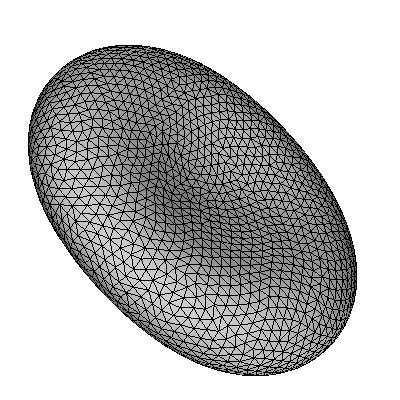} } &
\includegraphics[width=\asize\columnwidth]{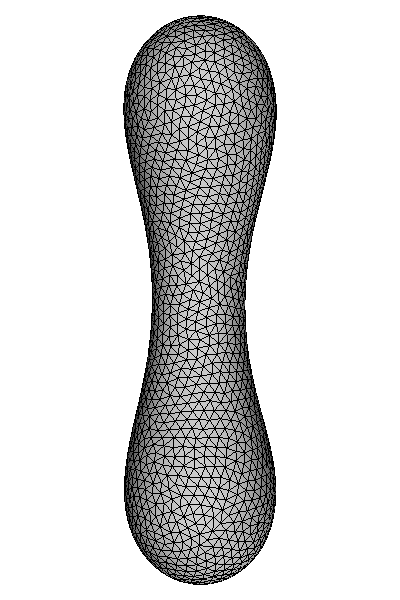}   \\
\raisebox{\cheight\height}{D: $N_t=1280$} &
\raisebox{0.2\height}{\includegraphics[width=\asize\columnwidth]{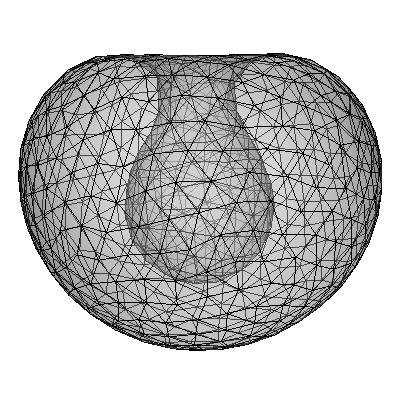}  }&
\raisebox{0.2\height}{\includegraphics[width=\asize\columnwidth]{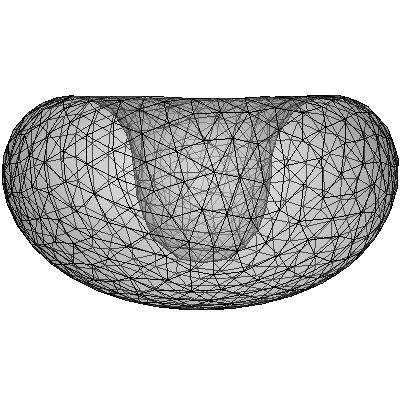} } &
\raisebox{\height}{\includegraphics[width=\asize\columnwidth]{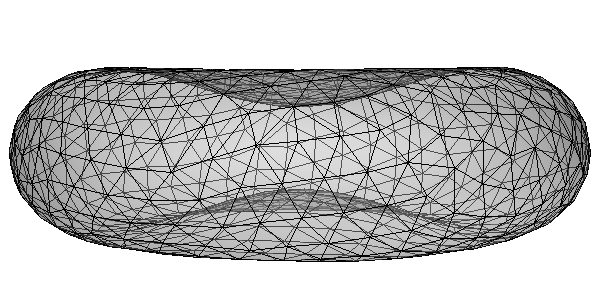} } &
\raisebox{0.2\height}{\includegraphics[width=\asize\columnwidth]{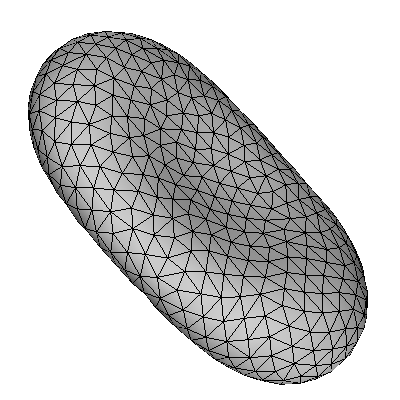} } &
\includegraphics[width=\asize\columnwidth]{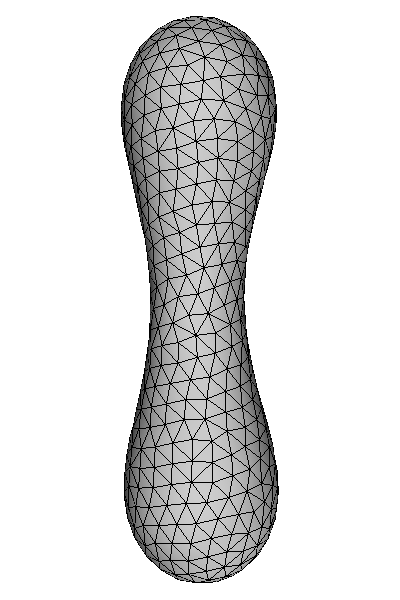}  \\
\raisebox{\cheight\height}{D: $N_t=5120$} &
\raisebox{\cheight\height}{NA} &
\raisebox{0.2\height}{\includegraphics[width=\asize\columnwidth]{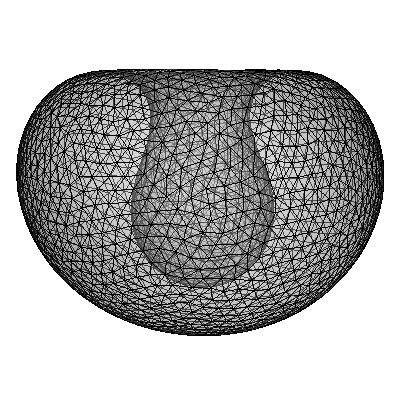} } &
\raisebox{\height}{\includegraphics[width=\asize\columnwidth]{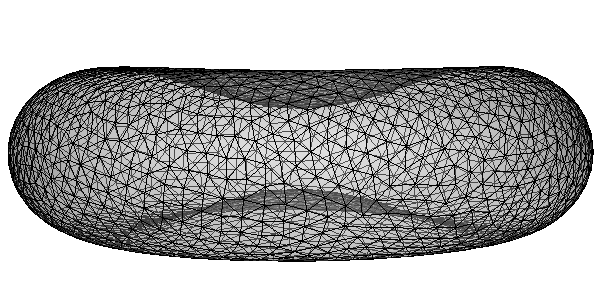} } &
\raisebox{0.2\height}{\includegraphics[width=\asize\columnwidth]{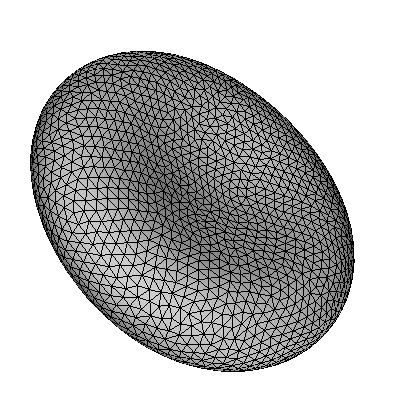} } &
\includegraphics[width=\asize\columnwidth]{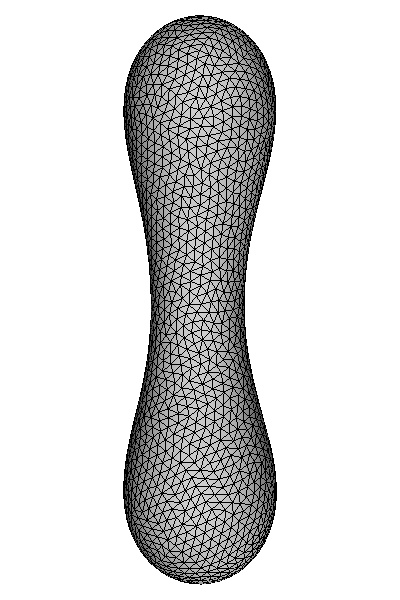}  \\
\end{tabular}
\label{table_ade}
\end{table}

We present equilibrium configurations from scheme B, C and D with
$N_t=1280$ and $5120$ in Table \ref{table_ade}, where values of
both $\Delta a_0$ and $\Delta a_0'$ are considered, as the latter
corresponds to the parameter in Fig 3c of Khairy et al \cite{Khairy2008a}.
With resolutions of $N_t=1280$, all schemes B, C and D resemble closely
the results of the reference.  As $\Delta a_0$  increases, we observe
axi-symmetric stomatocytes with smaller height, as exemplified by the
first two columns of Table~\ref{table_ade}.  This regime corresponds
to the first regime discussed above for Fig. \ref{fig_ade_alpha064},
where $\Delta a$ grows as $\Delta a_0$ increases; As $\Delta a_0$
(and $\Delta a$) further increases, it enters to the second regime,
where $\Delta a$ stays almost plateau.  In this regime, we observe
equilibrium shapes as biconcave-oblate and biconcave-elliptocyte,
which are exemplified by the third and fourth columns of Table
\ref{table_ade}.  In the last regime, $\Delta a$ stays  flat and we
observe prolate-dumbbells as shown for example in the last column of
Table \ref{table_ade}.  These configurations of $N_t=1280$ agree  with
the work of Kairy et al \cite{Khairy2008a}.  However, with $N_t=5120$
the anisotropy of elliptocytes with $0.21 \lesssim 100 \Delta a'_0
\lesssim 0.23$ tends to disappear and the final configurations are
less elongated.  The results on elliptocyte with $N_t=5120$. We note
that the difference between computed energies for $N_t=1280$ and $5120$
is very small as shown in Fig.~\ref{fig_ade_alpha064}.  The  results
from the three different schemes are consistent with each other, which
leads us to suspect that the number of terms in the spherical harmonics
of the reference could be insufficient to resolve this subtle difference.
\subsection{A dynamic simulation of of area-difference-elasticity model}
\label{sec_numerics3}

\begin{figure}
\centering
\begin{subfigure}{0.45\textwidth}
\includegraphics[width=\columnwidth, height=\columnwidth]{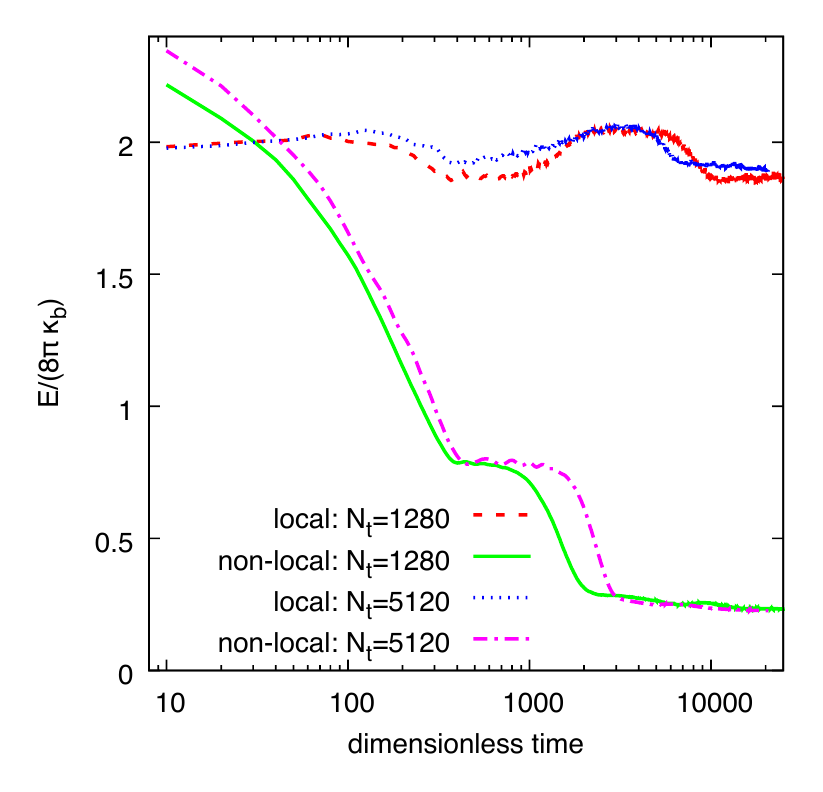}
\caption{}
\label{fig_dynamic_a}
\end{subfigure}
\begin{subfigure}{0.45\textwidth}
\includegraphics[width=\columnwidth, height=\columnwidth]{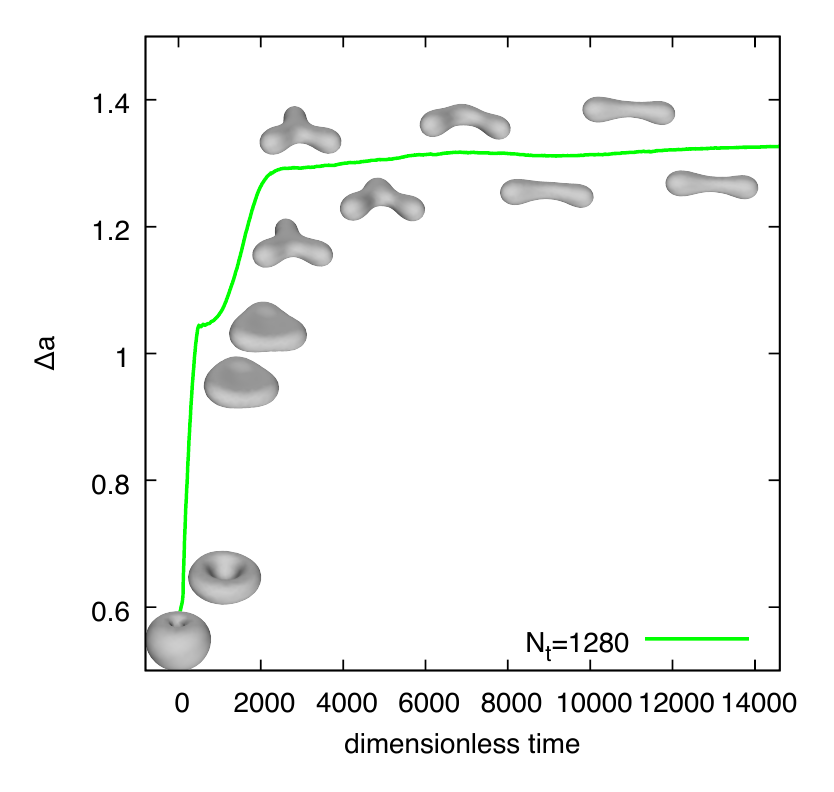}
\caption{}
\label{fig_dynamic_b}
\end{subfigure}
\caption{A dynamic trajectory of the ADE model with stomatocyte as initial
shape and prolate-dumbbell as final shape.  (a) Local
and non-local energy versus time.  (b) Area difference versus time and
representative shapes along the trajectory.  The representative
shapes correspond to $t^*=0, 250, 750, 1000, 2000, 3000,
5000, 7000, 9000, 11000, 13000$ and $\Delta a=0.585, 0.648, 0.96,
1.046, 1.179,1.295, 1.305, 1.307, 1.312, 1.321, 1.325$.}
\label{fig_dynamic}
\end{figure}
Finally, we present  dynamic simulations of lipid bilayer membranes using
the  ADE model discretized by scheme B.  We consider the same parameters
as in section \ref{sec_numerics2_ade}, that correspond to mechanics
activities of a RBC.  We start with a stomatoctye shape, which takes the
final shape of the previous minimization procedure (see results
of $\Delta a_0=1.11$ or $100\Delta a'_0=0.1$ on Table \ref{table_ade}).
We set the reference area-difference as $\Delta a_0=3.34$ or $100\Delta
a'_0=0.3$.

We report the dynamic trajectories of local energy, non-local
energy, area-difference and corresponding shapes on
Fig. \ref{fig_dynamic}.  The in-plane viscous damping force is orthogonal
to the normal direction of the membrane surface and has negligible
effects on the dynamic trajectories.  However, a larger viscous force is
needed to stabilize simulations with a higher resolution and therefore
demands a smaller time step.  The dynamic simulation arrives  at the
prolate-dumbbell shape, as shown on Fig. \ref{fig_dynamic} in agreement
with results on Fig. \ref{fig_ade_alpha064_a} and \ref{fig_ade_alpha064_b}
and on Table \ref{table_ade}, respectively, which are obtained from
independent minimization procedures.   From Fig. \ref{fig_dynamic_a}, we
observe that the local energy does not change significantly over time,
whereas the non-local energy decreases around ten fold.  This confirms
that the ADE is primarily responsible for the transformation of shapes.
Comparing Fig. \ref{fig_dynamic_b} with \ref{fig_ade_alpha064_b}, we
notice that the accessible states of $\Delta a$ are more abundant and
even continuous in the dynamic simulation.  Accordingly, there is a rich
range of shapes including triangle oblates, curved-prolate-like shapes,
and prolate-dumbbells, as represented on Fig. \ref{fig_dynamic_b}.
These intermediate configurations are dynamic (non-equilibrium) and
do not have a  correspondence for the intermediate $\Delta a_0$ values
on Table \ref{table_ade}.  Further comparing with the phase diagram of
the BC model on Fig. \ref{fig_bc_pd}, we also do not find corresponding
equilibrium shapes for the same values of reduced volume and $\Delta a$.
The small discrepancy between the energy trajectories of $N_t=1280$ and
$5120$ are not  surprising, as they start from slightly different values.
\section{Summary} 
\label{sec_summary}
We have presented a comparative study of  four bending models for lipid
bilayer membranes and their respective discretization on triangulated
meshes.  The physical models are the minimal, spontaneous curvature
(SC), bilayer-couple (BC) and area-difference elasticity (ADE) models
and the four schemes are termed as scheme A, B, C, and D respectively.

We find that the total energies computed by scheme A on a prescribed
sphere and biconcave oblate differ significantly from the analytical
values.  The energy computed by scheme A cannot be tuned to match
the Helfrich energy for an arbitrary shape.  For reduced volume $v>0.75$,
scheme A is able to generate prolate-dumbbells at equilibrium, although
the corresponding energies are inaccurate.  However, for $0.59< v <
0.65$, scheme A does not result in biconcave oblates as equilibrium shapes
and it  does not sustain the shape if it is given a biconcave oblate as
the initial shape.  Below $v < 0.59 $, scheme A has numerical
artifacts of budding transitions and vesiculations.  Finally, due to
the lack of direct definition of mean curvature and local area, scheme
A can not be directly extended to discretize the SC, BC, or ADE models.

The energies computed by scheme B, C and D on prescribed sphere
and biconcave oblate match well the analytical values.  However,
the accuracy of the three schemes has a varying dependence on their
resolution.  On  the phase diagram of the minimal model, for $v>0.59$,
all three schemes produce prolate dumbbells and biconcave oblates as the
equilibrium configurations.  They also resolve well the critical energy
values at $v=0.75$ and $0.65$ and $0.59$, with scheme B at $N_t=1280$ but
scheme C and D at $N_t=5120$.  However, for $v<0.59$, Scheme B produces
stomatocytes as equilibrium configurations accurately with $N_t=1280$,
where the inner sphere separates clearly from the outer sphere.  Scheme C
generates acceptable equilibrium configurations, but cannot maintain a
clear physical separation between the inner and outer spheres.  For the
most challenging case of $v=0.25$, scheme C is in trouble to generate a
physically correct configuration.  A higher resolution of $N_t=5120$ does
not help scheme C significantly on the overall equilibrium shape.
Scheme D has stability issues in the regime of stomatocytes that may be
attributed to its lack of explicit energy and momentum conservation.

We have also computed two slices ($h_0=1.2$ and $1.5$) of the phase
diagram for the SC model.  These are challenging benchmarks, as $h_0=1.2$
includes the budding transitions at equilibrium, while $h_0=1.5$ include
both buddings and vesiculations.  We observe that schemes B, C and D
all capture accurately the budding process, but scheme C and D have
difficulties to generate vesiculations.  Furthermore,  with the same
resolution of $N_t=1280$ scheme B computes the energy more accurately
than scheme C and D.  Moreover, the energies computed by scheme C at
budding transitions are completely off the reference values due to the
poor representation of the configuration necks.

As a special case of the ADE model, we considered the BC model, where
a constraint is imposed on the area difference between the two layers
of lipid. We focus on the results of scheme B and  reproduce the phase
diagram for different resolutions.  We find that results of scheme B are
comparable to the reference on the equilibrium shapes and energy values,
which were calculated by Surface Evovler.  We have also considered
the complete energy functional of the ADE model.  We reproduce the
sequence of stomatocytes, biconcave oblates, biconcave elliptocytes
and prolate dumbbells by all three schemes, except for the extreme
case of stomatocyte where scheme D runs into instability again.  As the
last demonstration, we consider a dynamic trajectory of the ADE model from
stomatocyte to prolate-dumbbell.  The values of local energy, non-local
energy, and area-difference along the dynamic trajectory may also serve
as reference for other researchers.

Our results indicate that it is a challenging  task to compute accurately
the bending energy and in particular  forces on a triangulated mesh.
However, due to its generality it is still a very promising path
to pursue.  At the same time we note the level set formulations
introduced in~\cite{Maitre2009} which incorporates some of the energy
models presented in this work. We believe that this is a very promising
direction to account for large membrane deformations and topological
changes. Future work could be directed in comparing level set and
triangulated surface representations.

In summary,  the relatively simple formulation of scheme B, which bypasses
definitions of Laplace-Beltrami operator and unit normal vectors, and
instead defines the integral of mean curvature with respect to area,
exhibits excellent accuracy, robustness, stability.  Moreover as the  the
implementation of scheme B is also almost as simple as that of scheme
A we suggest that it deserves to be more broadly investigated as an
alternative for vesicle and RBC simulations. While scheme B is readily
applicable to the minimal model, in this work we promote and extend its
potentials to further resolve SC, BC and ADE models.  With ADE model
as the most accurate model for lipid bilayer membrane, we envision many
numerical applications on vesicles and RBCs in a dynamic context.

\appendix

\section{Force from variational calculus of energy}
\label{sec_appendix_force_variational_formulation}
In differential geometry, the Cartesian coordinates ${\bf x}$ of a point on a surface may be
expressed in terms of two independent parameters $u_1$ and $u_2$
as ${\bf x}={\bf x}(u_1,u_2)$.
For further derivations, 
we introduce some definitions~\cite{DoCarmo2016, Tu2018}
\begin{align}
{\bf x}_{,i} = \frac{\partial {\bf x}}{\partial u_i}, \quad 
{\bf x}_{,ij}=\frac{\partial^2 {\bf x}}{\partial u_i \partial u_j}, \quad 
g_{ij} = {\bf x}_{,i} \cdot {\bf x}_{,j}, \nonumber \\
g^{ij} = \left(g_{ij} \right)^{-1} \quad or \quad g^{ij} g_{jl} = \delta^i_l, \quad g = \det (g_{ij}),
\end{align}
where free index $i$ or $j$ is either $1$ or $2$,
and summation convention applies to repeated index.
Index $i$ after ``," denotes the partial derivative with respect to $u_i$.
$g_{ij}$ is the covariant metric tensor and $g$ is its determinant.
$g^{ij}$ is the contravariant metric tensor.
An infinitesimal area is then
\begin{align}
dA = g^{1/2} du_1du_2.
\end{align}
The parameterization of $u_1$ and $u_2$ are chosen in such a way
the the normal direction 
\begin{align}
{\bf n} = \frac{{\bf x}_{,1} \times {\bf x}_{,2} }{| {\bf x}_{,1} \times {\bf x}_{,2}|},
\label{eq_appendix_variation_norm}
\end{align}
points inwards of a closed surface.
Therefore, $H=\nabla \cdot {\bf n}/2$ is positive for a sphere.

For a small and continuous perturbation $\psi(u_1,u_2)$ 
along the normal direction ${\bf n}(u_1, u_2)$,
the new position of the point on the parametric surface is given as
\begin{eqnarray}
{\bf x}'(u_1,u_2) = {\bf x}(u_1,u_2) + \psi(u_1,u_2){\bf n}(u_1,u_2).
\end{eqnarray}
Thereafter,  the variations for $dA$ and $H$ are given as
\begin{align}
\delta (dA) &=  2H \psi g^{1/2} du_1 du_2 = 2H \psi dA,  \\
\delta H 
&= -(2H^2-G)\psi - \frac{1}{2}g^{ij}\left(\psi_{,ij}-\Gamma^k_{ij}\psi_{,k}\right) 
= -(2H^2-G)\psi - \frac{1}{2} g^{ij} \nabla_i \psi_{,j},
\end{align}
where $\Gamma^k_{ij}$ is the Christoffel symbol
and $\nabla^2_s$ is the Laplace-Beltrami operator on the surface.
We note a different sign convention used~\cite{Tu2018}. 

To have an explicit expression of $\delta E$, 
we need variational expressions for each moment.
The variation of the zero moment simply reads as
\begin{align}
\delta \mathcal{M}_0 &= \delta A = \dint \delta (dA) =  \dint 2H \psi dA.
\label{eq_appendix_0moment_variation}
\end{align}
However,  to calculate variations of the first and second moments
we need some fundamental equalities.
For any function $f(u_1,u_2)$ we have
\begin{align}
\dint f \psi_{,i} du_1 du_2 = - \dint f_{,i} \psi du_1 du_2 , 
\quad
\dint f \psi_{,ij} du_1 du_2 = \dint f_{,ij} \psi du_1 du_2 ,
\label{eq_appendix_basic_relation1} 
\end{align}
which can be readily proven via integration by parts.
We further notice 
\begin{align}
\left( g^{1/2}  g^{ij} f \right)_{,ij}
&=\left[ \left( g^{1/2} g^{ij} \right)_{,j} f \right]_{,i}
+ \left( g^{1/2} g^{ij}  f_{,j} \right)_{,i} , 
\label{eq_appendix_basic_relation2a} \\
\left[ \left( g^{1/2} g^{ij} \right)_{,j} f \right]_{,i} 
&= - g^{1/2}g^{ij} \left(\Gamma^k_{ij}f \right)_{,k},
\label{eq_appendix_basic_relation2b}  \\
g^{-1/2}\left[ g^{1/2} g^{ij}  f_{,j} \right]_{,i} &= \nabla^2_s f.
\label{eq_appendix_basic_relation2c} 
\end{align}
Therefore,
\begin{align}
& \dint f g^{ij}\left(\psi_{,ij}-\Gamma^k_{ij}\psi_{,k}\right) dA \nonumber \\
=& \dint f g^{ij}\left(\psi_{,ij}-\Gamma^k_{ij}\psi_{,k}\right) g^{1/2} du_1du_2 \nonumber \\
=& \dint  \left[ \left( g^{1/2}  g^{ij} f \right)_{,ij}  
 +\left(  g^{1/2} g^{ij} \Gamma^k_{ij} f  \right)_{,k}  \right] \psi du_1du_2 \nonumber \\
=& \dint  \left\{ \left[ \left( g^{1/2} g^{ij} \right)_{,j} f \right]_{,i}
+ \left( g^{1/2} g^{ij}  f_{,j} \right)_{,i}  
- \left[ \left( g^{1/2} g^{ij} \right)_{,j} f \right]_{,i} \right\} \psi du_1du_2 \nonumber \\
=& \dint \left( g^{1/2} g^{ij}  f_{,j} \right)_{,i}  \psi du_1du_2 \nonumber \\
=& \dint  \nabla^2_s f g^{1/2} du_1du_2 = \dint  \nabla^2_s f dA,
\label{eq_appendix_basic_relation3} 
\end{align}
Eq.~(\ref{eq_appendix_basic_relation1}) is applied 
from the second to the third line.
Eqs.~(\ref{eq_appendix_basic_relation2a})
and (\ref{eq_appendix_basic_relation2b}) are applied 
from the third to the fourth line.
Eq.~(\ref{eq_appendix_basic_relation2c}) is applied 
from the fifth to the sixth line.

The variation of the first moment is readily obtained as
\begin{align}
\delta \mathcal{M}_1 
&= \dint \left[ \delta H dA + H\delta (dA)\right] \nonumber \\
&= \dint \left[ -(2H^2-G)\psi  
-\frac{1}{2}g^{ij}\left(\psi_{,ij}-\Gamma^k_{ij}\psi_{,k}\right)
+2H^2 \psi  \right] dA \nonumber \\
&= \dint \left[ G \psi  
-\frac{1}{2}g^{ij}\left(\psi_{,ij}-\Gamma^k_{ij}\psi_{,k}\right) \right] dA \nonumber \\
&= \dint \left(G - \frac{1}{2} \nabla^2_s 1 \right)  \psi dA  = \dint  G  \psi dA,
\label{eq_appendix_1moment_variation}
\end{align}
where Eq.~(\ref{eq_appendix_basic_relation3})
is applied from the third to the fourth line.
The variation of the second moment is also readily derived as
\begin{align}
\delta \mathcal{M}_2 
&= \dint \left[ 2H\delta H dA + H^2\delta (dA)\right] \nonumber \\
&= \dint \left[-2H (2H^2-G)\psi  
-Hg^{ij}\left(\psi_{,ij}-\Gamma^k_{ij}\psi_{,k}\right)
+2H^3 \psi  \right] dA \nonumber \\
&= \dint \left[-2H (H^2-G)\psi  
-Hg^{ij}\left(\psi_{,ij}-\Gamma^k_{ij}\psi_{,k}\right) \right] dA \nonumber \\
&= \dint \left[-2H \left(H^2-G\right) - \nabla^2_s H \right] \psi dA,
\label{eq_appendix_2moment_variation}
\end{align}
where Eq.~(\ref{eq_appendix_basic_relation3})
is applied from the third to the fourth line.

Since the perturbation $\psi$ is continuous, small and arbitrary, 
the integrand must be the density of virtual work along the normal direction.
Therefore, the magnitudes of the virtual force density 
due to the three moments read as
\begin{align}
\mathcal{F}_0 = 2H, \quad 
\mathcal{F}_1 = G, \quad 
\mathcal{ F}_2 = -2H \left(H^2-G\right) - \nabla^2_s H,
\label{eq_appendix_moments_force}
\end{align}
and they all act along the normal direction ${\bf n}$.
These results also corroborate another two recent derivations 
following different routes~\cite{Barrett2008a, Laadhari2010}.

The variation of energy of Eq.~(\ref{eq_energy_continuum_group1}) read as
\begin{align}
\delta E  &= 2\kappa_b  \delta \mathcal{M}_2    
+\frac{4 \alpha \kappa_b \pi}{A} \mathcal{M}_1 \delta \mathcal{M}_1
- \frac{2 \alpha \kappa_b \pi}{A^2} \mathcal{M}^2_1 \delta A  
 - 4\kappa_{b} H_0  \delta \mathcal{M}_1 \nonumber \\
& -\frac{2 \alpha \kappa_b \pi}{A} \frac{\Delta A_0}{D} \delta \mathcal{M}_1  
+ \frac{2 \alpha \kappa_b \pi}{A^2} \frac{\Delta A_0}{D} \mathcal{M}_1 \delta A 
 + 2 \kappa_b H^2_0 \delta A   
-\frac{\alpha \kappa_b \pi}{2A^2}   \left(\frac{ \Delta A_0}{D}\right)^2 \delta A. 
 \label{eq_appendix_energy_total_group1_variation}
\end{align}
Given the variations of the zero, first and second moments
in Eqs.~(\ref{eq_appendix_0moment_variation}), (\ref{eq_appendix_1moment_variation}),
and (\ref{eq_appendix_2moment_variation}),
we have readily an explicit expression for the variation of the energy
and omit the repetitious details here.
Accordingly,  the magnitude of force density due to 
the energy Eq.~(\ref{eq_energy_continuum_group1}) reads as
\begin{align}
f  &= 
\underbrace{ 2\kappa_b  \mathcal{F}_2 }_{f^H}   
+ \underbrace{ \frac{4 \alpha \kappa_b \pi}{A} \mathcal{M}_1 \mathcal{F}_1 }_{f^{AD}}
- \underbrace{ \frac{2 \alpha \kappa_b \pi}{A^2} \mathcal{M}^2_1 \mathcal{F}_0 }_{f^{AD}}  
 - \underbrace{ 4\kappa_{b} H_0  \mathcal{F}_1 }_{f^H}   \nonumber \\
& - \underbrace{ \frac{2 \alpha \kappa_b \pi}{A} \frac{\Delta A_0}{D} \mathcal{F}_1 }_{f^{AD}}  
+ \underbrace{ \frac{2 \alpha \kappa_b \pi}{A^2} \frac{\Delta A_0}{D} \mathcal{M}_1 \mathcal{F}_0 }_{f^{AD}}
+ \underbrace{  2 \kappa_b H^2_0   \mathcal{F}_0 }_{f^H}   
- \underbrace{ \frac{\alpha \kappa_b \pi}{2A^2}   \left(\frac{ \Delta A_0}{D}\right)^2 \mathcal{F}_0 }_{f^{AD}},
 \label{eq_appendix_force_density_total_group1}
\end{align}
where geometric evaluations of $\mathcal{F}_0$, $\mathcal{F}_1$ and $\mathcal{F}_2$
are given in Eq. (\ref{eq_appendix_moments_force}),
and $f$ acts along the normal direction ${\bf n}$.
We may also obtain the force density according to the historical development 
of each energy term as
\begin{align}
{f}^H &= 2 \kappa_b \left( \mathcal{F}_2 - 2H_0 \mathcal{F}_1 + H^2_0 \mathcal{F}_0 \right)
      = -2\kappa_b \left[ 2(H-H_0)(H^2+H_0H-G)+\nabla^2_sH \right], \nonumber \\
{f}^C &= -2\kappa_b \left[ 2H(H^2-G)+\nabla^2_sH \right], \nonumber \\
{f}^S &= 4\kappa_b H_0\left( H_0H - G\right), \nonumber \\
{f}^{AD} &=  \alpha \kappa_b \pi \left( \frac{4\mathcal{M}_1\mathcal{F}_1}{A} - 
\frac{2 \mathcal{M}^2_1\mathcal{F}_0}{A^2} 
-\frac{2\Delta A_0\mathcal{F}_1}{AD} 
+\frac{2\Delta A_0 \mathcal{M}_1\mathcal{F}_0}{A^2D}
-\frac{1}{2A^2}\left(\frac{\Delta A_0}{D}\right)^2 \mathcal{F}_0
\right) \nonumber \\
&=\alpha \kappa_{b} \pi 
\left[
\left( 2\mathcal{M}_1 - \frac{\Delta A_0}{D} \right) \frac{2G}{A}  
- \left( 2\mathcal{M}_1 - \frac{\Delta A_0}{D} \right)^2 \frac{H}{A^2}
\right],
\label{eq_appendix_force_density_individual}
\end{align}
where the superscripts ``$C$, $H$, $S$, and $AD$" correspond to
the forces of Canham, Helfrich, spontaneous curvature and ADE, respectively.
They all act along the normal ${\bf n}$ direction.
${f}^C$ is just degenerated from ${f}^H$ by setting $H_0=0$.
Finally, given the total energy of Eq.~(\ref{eq_energy_continuum_group1}),
the total force density is simply 
${\bf f}= ({f}^H + {f}^{AD}){\bf n}$
or equivalently $f{\bf n}$.
\section{Discrete force from scheme A}
\label{sec_appendix_force_scheme_a}
The force on vertex $m$ is
\begin{eqnarray}
{\bf F}_{m} 
= -\frac{\partial E}{\partial {\bf x}_m}
=-2\tilde{\kappa}_b \sum^{N_e}_{e: \left<i,j\right>} \sin \left( \theta_{e} - \theta_0 \right)\frac{\partial \theta_e}{\partial {\bf x}_m}.
\label{eq_appendix_force_a}
\end{eqnarray}
Similarly, for the linearized version the force is
\begin{eqnarray}
{\bf F}^{1st}_{m} 
= -\frac{\partial E^{1st}}{\partial {\bf x}_m}
=-2\tilde{\kappa}_b \sum^{N_e}_{e: \left<i,j\right>} \left( \theta_{e} - \theta_0 \right)\frac{\partial \theta_e}{\partial {\bf x}_m}.
\label{eq_appendix_force_a1st}
\end{eqnarray}
Therefore, the primitive element for calculation of force is
the partial derivative of the dihedral angle's supplementary angle with respect to position,
that is, ${\partial \theta_e}/{\partial {\bf x}_m}$~\cite{Bridson2005}.
\section{Discrete force for scheme B}
\label{sec_appendix_force_scheme_b}
Force on vertex ${\bf x}_m$ is the negative derivative of the discrete energy
in Eq.~(\ref{eq_energy_discrete_group2})
\begin{align}
{\bf F}_m=-\frac{\partial E }{\partial {\bf x}_m }  
&= \underbrace{ -2\kappa_b  \frac{\partial \mathcal{M}_2}{\partial {\bf x}_m}  
- \frac{4 \alpha \kappa_b \pi}{A} \mathcal{M}_1 \frac{\partial \mathcal{M}_1}{\partial  {\bf x}_m} }_{{\bf F}^{AD}_m} 
+ \underbrace{ \frac{2 \alpha \kappa_b \pi}{A^2} \mathcal{M}^2_1 \frac{\partial A}{\partial {\bf x}_m} }_{{\bf F}^{AD}_m}
+ \underbrace{ 4\kappa_{b} H_0  \frac{\partial  \mathcal{M}_1 }{\partial {\bf x}_m} }_{{\bf F}^H_m} \nonumber \\
&+ \underbrace{ \frac{2 \alpha \kappa_b \pi}{A} \frac{\Delta A_0}{D} \frac{\partial  \mathcal{M}_1 }{\partial {\bf x}_m}
-  \frac{2 \alpha \kappa_b \pi}{A^2} \frac{\Delta A_0}{D} \mathcal{M}_1 \frac{\partial A}{\partial {\bf x}_m} }_{{\bf F}^{AD}_m} 
 - \underbrace{ 2 \kappa_b H^2_0 \frac{\partial A}{\partial {\bf x}_m}   }_{{\bf F}^H_m}
+ \underbrace{ \frac{\alpha \kappa_b \pi}{2A^2}   \left(\frac{ \Delta A_0}{D}\right)^2 \frac{\partial A}{\partial {\bf x}_m} }_{{\bf F}^{AD}_m} . 
 \label{eq_appendix_force_total_group_discrete}
\end{align}
Equivalently, ${\bf F}_m={\bf F}^H_m + {\bf F}^{AD}_m$,
due to force of Helfrich and ADE energy.
We need to calculate the derivatives of the three moments as
\begin{align}
\frac{\partial \mathcal{M}_0}{\partial {\bf x}_m}
& =\frac{\partial A}{\partial {\bf x}_m}  = \sum^{N_v}_i \frac{\partial A_i}{\partial {\bf x}_m},  \nonumber \\
\frac{\partial \mathcal{M}_1}{\partial {\bf x}_m}
&=\sum^{N_v}_i \left( \frac{\partial H_i}{\partial {\bf x}_m} A_i 
+H_i \frac{\partial A_i}{\partial {\bf x}_m} \right), \nonumber \\
\frac{\partial \mathcal{M}_2}{\partial {\bf x}_m}  
&= \sum^{N_v}_i \left( 2H_i \frac{\partial H_i}{\partial {\bf x}_m} A_i 
+H^2_i \frac{\partial A_i}{\partial {\bf x}_m} \right).
\end{align}
The constituting elements of derivatives are
$\partial H_i / \partial {\bf x}_m$,
and $\partial A_i/\partial {\bf x}_m$,
which are given explicitly as follows.

Given $H_i$ in Eq.~(\ref{eq_cm_b}),
its derivative with respective to ${\bf x}_m$ reads
\begin{eqnarray}
\frac{\partial H_i}{\partial {\bf x}_m}
=\frac{1}{4A_i} \sum^{N^{i}_e}_{e: \left<i,j\right>} 
\left( \frac{\partial l_e}{\partial {\bf x}_m} \theta_e 
+ l_e\frac{\partial \theta_e}{\partial {\bf x}_m} \right)
-\frac{H_i}{A_i} \frac{\partial A_i}{\partial {\bf x}_m},
\label{eq_appendix_mean_curvature_derivative_j}
\end{eqnarray}
where summation runs over $N^i_e$ neighboring edges around vertex $i$.
Given $A_i$ in Eq.~(\ref{eq_area_b}),
its derivative with respective to ${\bf x}_m$ reads
\begin{eqnarray}
\frac{\partial A_i}{\partial {\bf x}_m}
=\frac{1}{3}\sum^{N^{i}_t}_{t:\left<i,j,k\right>} \frac{\partial A^{t}}{\partial {\bf x}_m},
\label{eq_appendix_area_local_derivative_j}
\end{eqnarray}
where summation runs over $N^i_t$ neighboring triangles around vertex $i$.
It remains a few primitive derivatives 
\begin{eqnarray}
\frac{\partial l_e}{\partial {\bf x}_m}, \quad  
\frac{\partial \theta_e}{\partial {\bf x}_m}, \quad
\frac{\partial A^t}{\partial {\bf x}_m}
\end{eqnarray}
to be evaluated.
We consider them as basic elements,
which pose no difficulty and therefore, omit the technical details here.
\section{Discrete force for scheme C}
\label{sec_appendix_force_scheme_c}
{\def\sv#1{\sum_i^{N_v} #1}
\def\sa#1{\sum^{N_t^i}_{t:<i,j,k>} #1}
\def\tb{c_j}
\def\tc{c_k}
\def\eb{\mathbf{e}_j}
\def\ec{\mathbf{e}_k}
\def\sb{s_j}
\def\sc{s_k}
\def\lp{\mathbf{l}}
\def\dot#1#2{#1 \cdot #2}
\def\n{\mathbf{n}}
\def\u{\mathbf{u}}
\def\dn{\frac{d|\n|}{d\n}}
\def\ldn{\mathbf{L}}
\def\ang{\phi_{i}}
\def\S{E_A}
\def\Q{E_h}
\def\deltad#1{ \boldsymbol{\delta} #1}
\def\deltat#1{ \boldsymbol{\delta}_t #1}
It it convenient to write an expression for the discrete total energy in the following form
\begin{align}
  E = \sv{E_i} = \sv{E(A_i, h_i)},
\end{align}
where $h_i = A_i H_i$. 
To compute local area $A_i$, 
we  introduce several scalar and vector quantities for every triangle (omitting index $i$) as
\begin{align}
  \eb^t = \mathbf{x}^t_i - \mathbf{x}^t_j, \quad
  \ec^t = \mathbf{x}^t_i - \mathbf{x}^t_k,   \quad
  \sb^t = |\eb^t|, \quad   \sc^t = |\ec^t|, \quad
  \tb^t = \cot \phi_{j}^t, \quad  \tc^t = \cot \phi_{k}^t,
\end{align}
so that the local area is simply
\begin{align}
  A &=   \frac{1}{8} \sa{ \tb^t \sc^t + \tc^t \sb^t },
\end{align}
where summation runs over all neighboring triangles.
To compute $h_i$, triangle's normal $\u^t$ is further required (omitting index $i$)
\begin{align}
  h = \dot{\lp}{\n}, \quad
  \lp = \frac{1}{2} \sa{ \tb^t \ec^t + \tc^t \eb^t }, \quad
  \n =  \frac{\mathbf{m}}{|\mathbf{m}|}, \quad \mathbf{m} = \sa{ \ang^t \u^t }.
\end{align}

Since discrete energy is a function of all vertex coordinates in neighboring triangles,
a complete explicit expression for the force is extremely tedious. 
We give an implicit expression by considering the variation of $E(A, h)$ 
for the perturbation of every triangle $t$ as
\begin{align}
  \deltat{E} = \deltat{E(A, h)} = \S \deltat{A} + \Q \deltat{h},
\end{align}
where $\S$ and $\Q$ are the corresponding partial derivatives.
After some transformations and omitting index $t$, we obtain
\begin{align}
  \deltad{E} &=
     \dot{\frac{\Q \ang \ldn}{2}} \deltad{\u}  
  + \frac{\Q \dot{\ldn}{\n}}{2} \deltad{\ang}
  + \frac{ \S \sc}{8} \deltad{\tb} 
  -  \dot{\frac{\Q\dot{\ec}{\n}}{4}}{\deltad{\tb}} 
  + \frac{ \S \sb}{8} \deltad{\tc} \nonumber \\
  &- \dot{\frac{\Q\dot{\eb}{\n}}{4}}{\deltad{\tc}} 
  + \frac{ \S \tc }{8} \deltad{\sb}
  + \frac{ \S \tb }{8} \deltad{\sc}
  + \dot{\frac{ \Q \tc \n}{4}} \deltad{\eb}
  + \dot{\frac{ \Q \tb \n}{4}} \deltad{\ec},
\end{align}
where $\ldn = \dot{\lp}{\dn}$. Primitive variational can be expressed
as linear functions of $\deltad{x_i}$, $\deltad{x_j}$, and $\deltad{x_k}$, 
which lead to an explicit expression for the forces 
on all vertices in the triangle due to energy at vertex $i$.
It is required to specify expressions for $E$ and its derivatives. 
In our case, constitute moments are the essential terms
\begin{align}
  A     = E(A, h) = A, \quad
  A H   = E(A, h) = h, \quad
  A H^2 = E(A, h) = h^2/A.
\end{align}
}
\section{Analytical solutions for a prescribed shape}
\label{sec_appendix_maxima}
For verification of the energy and force on a triangulated mesh,
we calculate these quantities analytically for a prescribed biconcave-oblate shape,
which is descibed by an emipircal function~\cite{Evans1972}. 
We employ the algebra software "Maxima" and
parameterize the membrane surface  as
\begin{align}
 x_1 = \sin(u_1) \sin(u_2), \quad
 x_2 = \sin(u_1) \cos(u_2), \quad
x_3  = \cos(u_1),
\label{eq_appendix_parameterization}
\end{align}
where $(x_1, x_2, x_3)$ are the Cartesian coordinates
and $(u_1, u_2)$ are the parametric coordinates.
Hence, normal unit vector ${\bf n}$ of the surface
defined in Eq.~(\ref{eq_appendix_variation_norm})
of \ref{sec_appendix_force_variational_formulation}
points inwards of a closed surface.
Given the definitions of moments in Eq. (\ref{eq_moments_continuum})
and force density in Eq.  (\ref{eq_moments_force}),
we apply Eq. (\ref{eq_appendix_parameterization})
to compute analytically the energies and forces 
in the minimal, SC and ADE models.
\section{Energy and force due to area constraint}
\label{sec_appendix_area}
Due to the weak compressibility,
we consider an extra energy to penalize its global area variation 
\begin{align}
E^A_g = \kappa_{ag} \frac{\left(A - A_0\right)^2}{A_0}, 
\end{align}
where $A= \sum^{N_t}_{t:<i,j,k>} A^t$ and $A_0$ are the current and original total area, respectively.
To regularize each triangle area,  we introduce another energy
on penalizing the local area as 
\begin{align}
E^A_l =\kappa_{al} \sum^{N_t}_{t:<i,j,k>} \frac{\left(A^t - A^t_0\right)^2}{A^t_0},
\end{align}
where $A^t$ and $A^t_0$ are the current and target area of the triangle with index $t$.
The strength of the penalization is controlled 
by $\kappa_{ag}$ and $\kappa_{al}$, respectively. 
The force is 
\begin{align}
{\bf F}^A = -\frac{\partial E^A}{\partial {\bf x}_m} 
= -\kappa_{ag} \frac{2\left(A - A_0\right)}{A_0} \sum^{N_t}_{t:<i,j,k>} \frac{\partial A^t}{\partial {\bf x}_m} 
- \kappa_{al} \sum^{N_t}_{t:<i,j,k>} \frac{2\left(A^t - A^t_0\right)}{A^t_0}\frac{\partial A^t}{\partial {\bf x}_m} ,
\end{align}
where index $m$ for position ${\bf x}_m$
is any one of the three vertices $i$, $j,$ and $k$.

\section{Energy and force due to volume constraint}
\label{sec_appendix_volume}
Due to osmotic pressure, a close membrane regulates its volume accordingly. 
Therefore, we consider an extra energy on penalizing the total volume variation as
\begin{align}
E^V = \kappa_v \frac{\left(V-V_0\right)^2}{V_0},
\end{align}
where $V$ and $V_0$ are the current and target volume enclosed by the membrane, respectively.
Total volume $V=\sum^{N_t}_{t:<i,j,k>} V^t$, where $V^t$ is the volume of the tetrahedron 
formed by triangle $t$ and the origin, that is, 
$V^t=({\bf a} \times {\bf b}) \cdot {\bf c}/6$,
where ${\bf a}$, ${\bf b}$ and ${\bf c}$ are vectors of the three edges of triangle $t$.
A coefficient $\kappa_v$ controls the strength of the penalization
to achieve the target volume or its conjugate osmotic pressure.
The force is 
\begin{align}
{\bf F}^V=-\frac{\partial E^V}{\partial {\bf x}_m} 
= -\kappa_{v} \frac{2\left(V - V_0\right)}{V_0} \sum^{N_t}_{t:<i,j,k>} \frac{\partial V^t}{\partial {\bf x}_m},
\end{align}
where index $m$ for position ${\bf x}_m$
is any one of the three vertices $i$, $j,$ and $k$.
\section{Numerical results on force for a prescribed shape}
\label{sec_appendix_numerics_force}
We present the force in this section with a caveat; the computed energy
is robust to a moderate change of the triangulated mesh, whereas the
corresponding force is highly sensitive.  We shall bear this in mind
when interpreting the numerical results.

\begin{figure}
\centering
\begin{subfigure}{0.45\textwidth}
\includegraphics[width=\columnwidth]{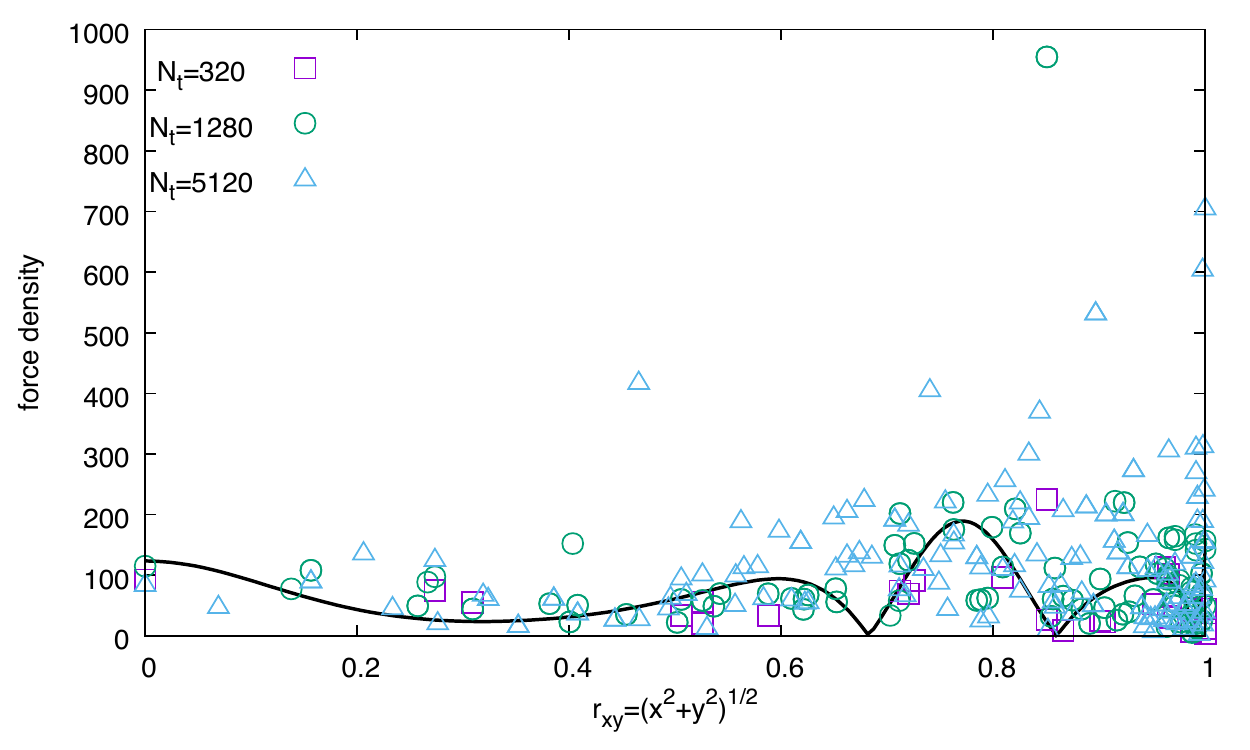}
\caption{Scheme A}
\end{subfigure}
\begin{subfigure}{0.45\textwidth}
\includegraphics[width=\columnwidth]{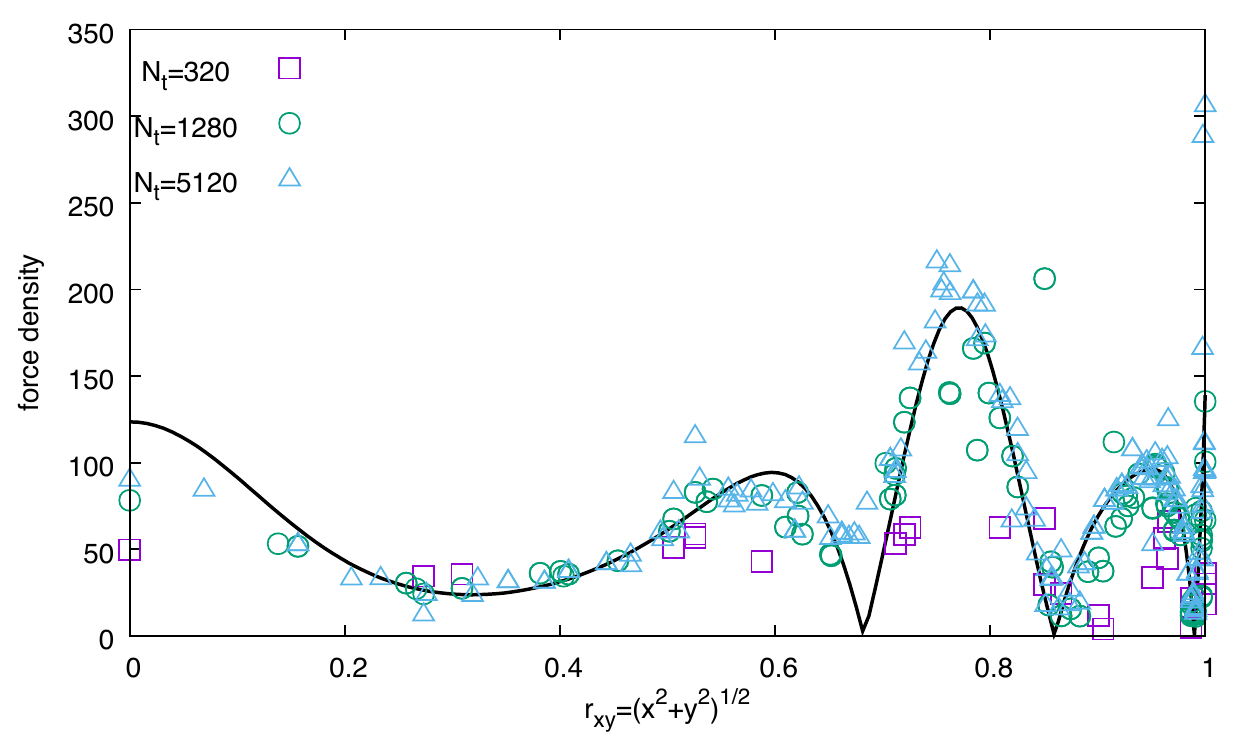}
\caption{Scheme B}
\end{subfigure}
\begin{subfigure}{0.45\textwidth}
\includegraphics[width=\columnwidth]{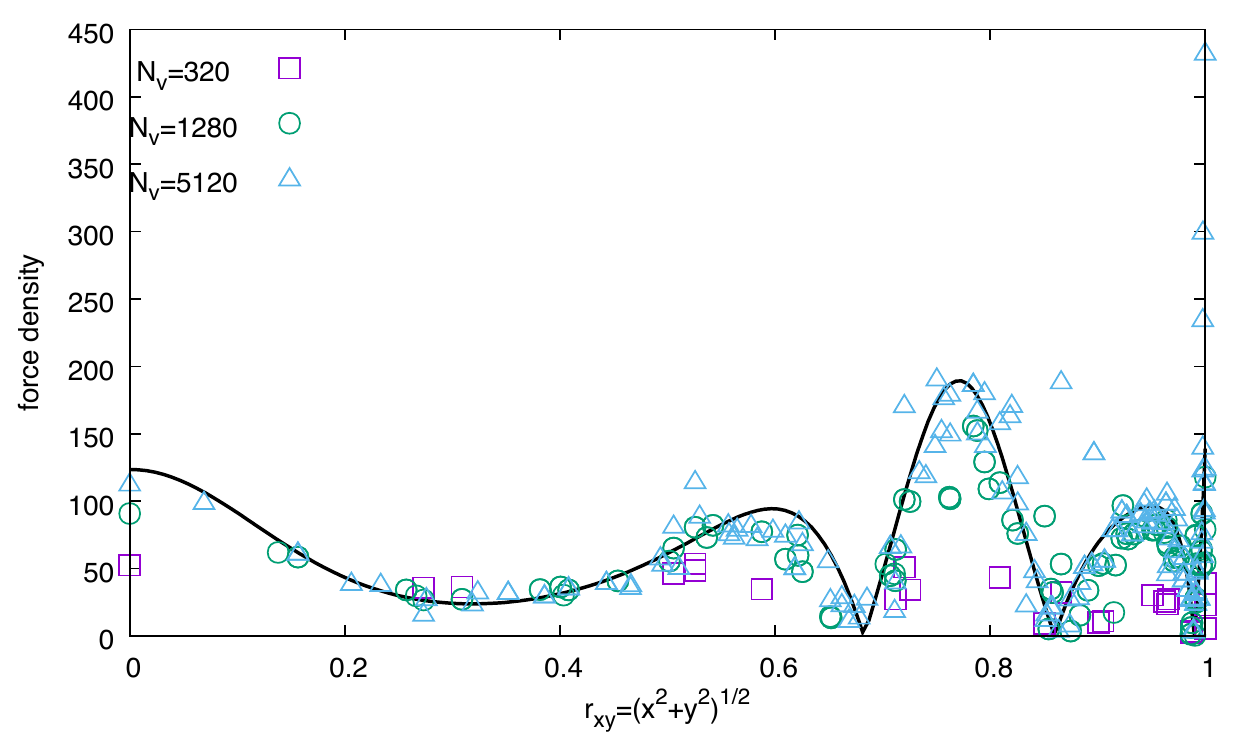}
\caption{Scheme C}
\end{subfigure}
\begin{subfigure}{0.45\textwidth}
\includegraphics[width=\columnwidth]{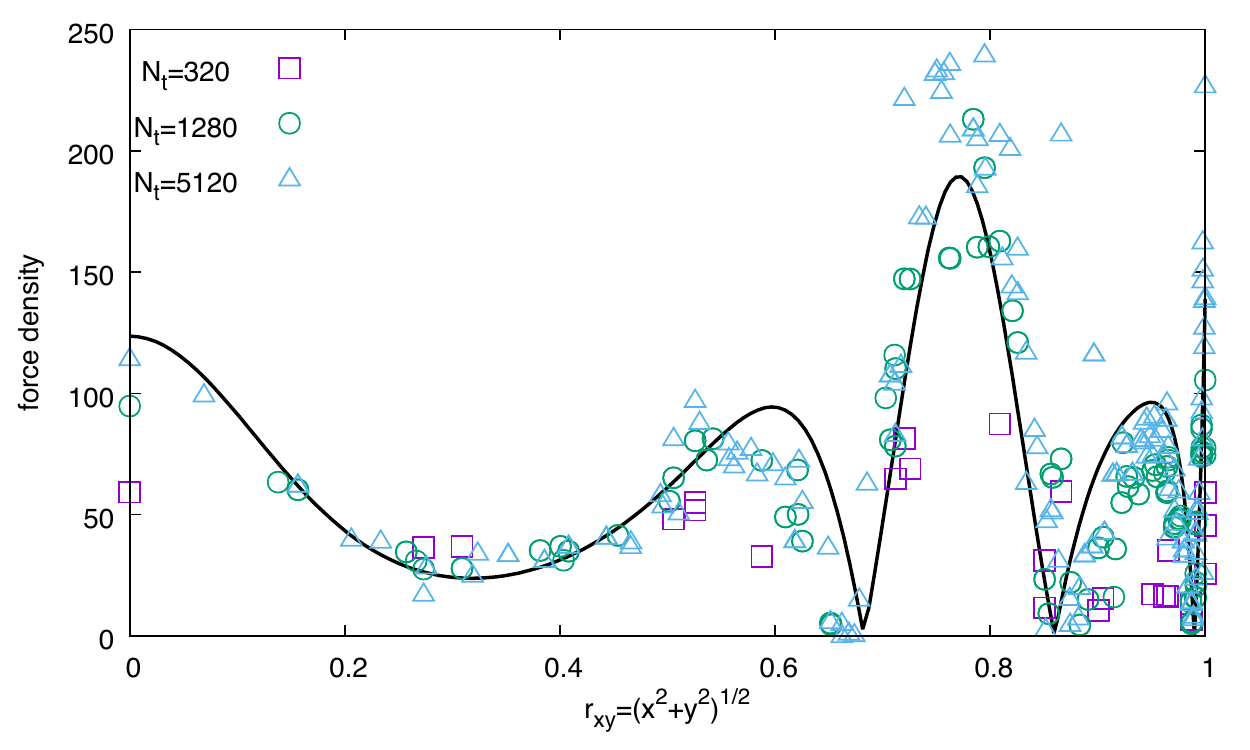}
\caption{Scheme D}
\end{subfigure}
\caption{Force density (absolute value) versus radial distance of vertices from
axis for a biconcave oblate with  $\kappa_b=1$, $h_0=0$ and $\alpha=0$.
For scheme A $\tilde{\kappa}_b = \sqrt{3}\kappa_b$ and $\theta_0=0$.}
\label{fig_minimal_force}
\end{figure}

\begin{figure}
\centering
\begin{subfigure}{0.45\textwidth}
\includegraphics[width=\columnwidth]{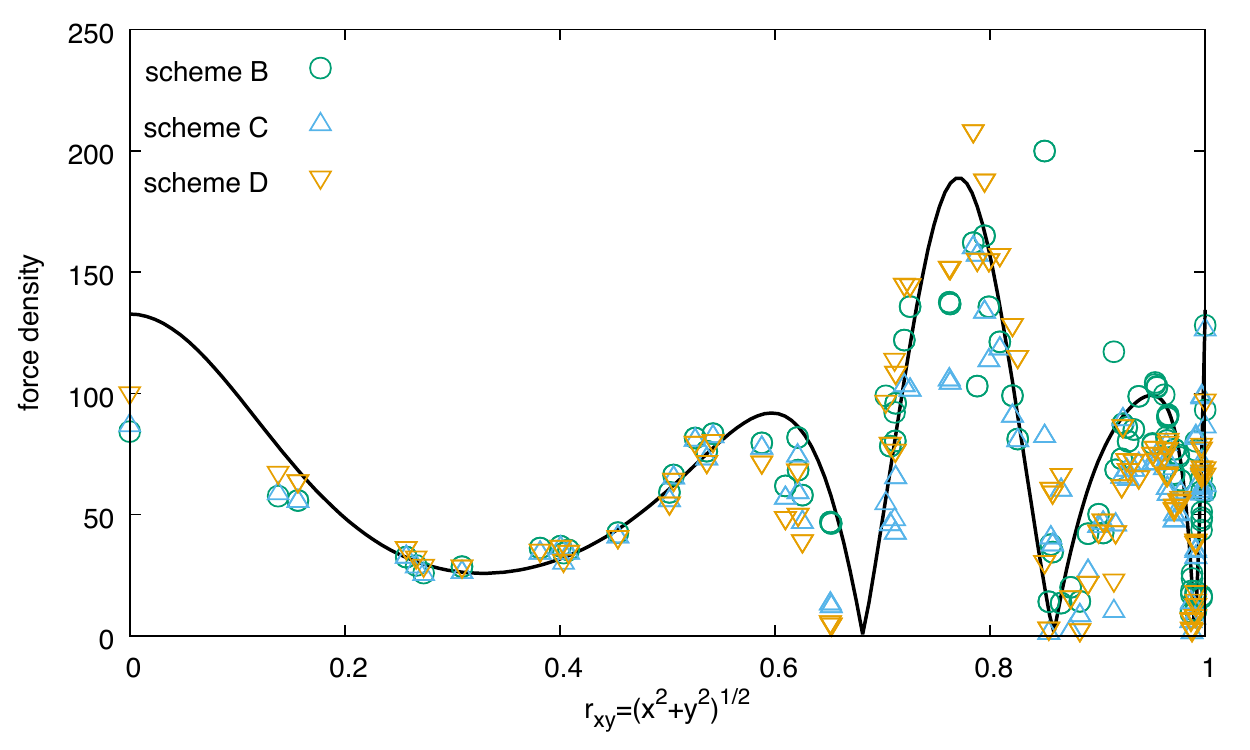}
\caption{}
\end{subfigure}
\begin{subfigure}{0.45\textwidth}
\includegraphics[width=\columnwidth]{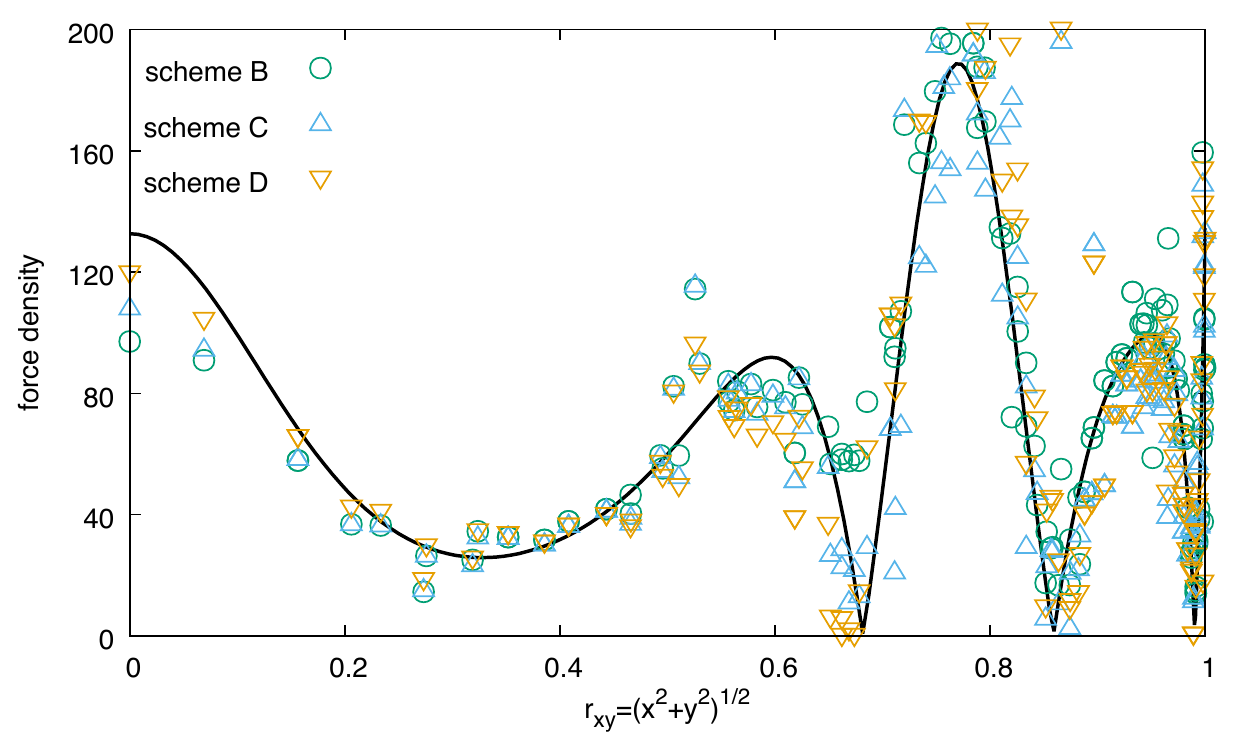}
\caption{}
\end{subfigure}
\caption{Force density (absolute value) versus radial distance of vertices from
axis of symmetry for a biconcave oblate with $\kappa_b=1$, $h_0=-0.5$
and $\alpha=0$.
(a) $N_t=1280$ plotted for clarity with every $4$ vertices in arbitrary order; 
(b) $N_t=5120$ and plotted for clarity with every $16$ vertices in arbitrary
order.}
\label{fig_helfrich_force}
\end{figure}

\begin{figure}
\begin{subfigure}{0.45\textwidth}
\includegraphics[width=\columnwidth]{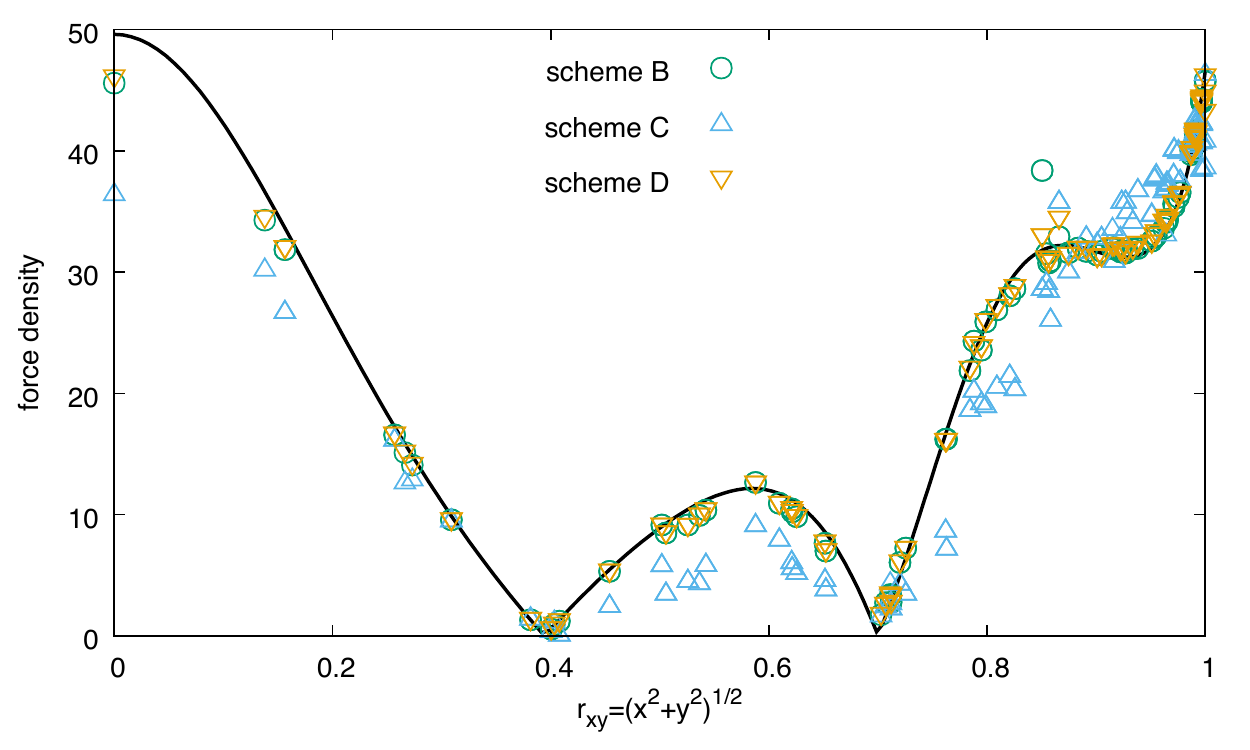}
\caption{}
\end{subfigure}
\begin{subfigure}{0.45\textwidth}
\includegraphics[width=\columnwidth]{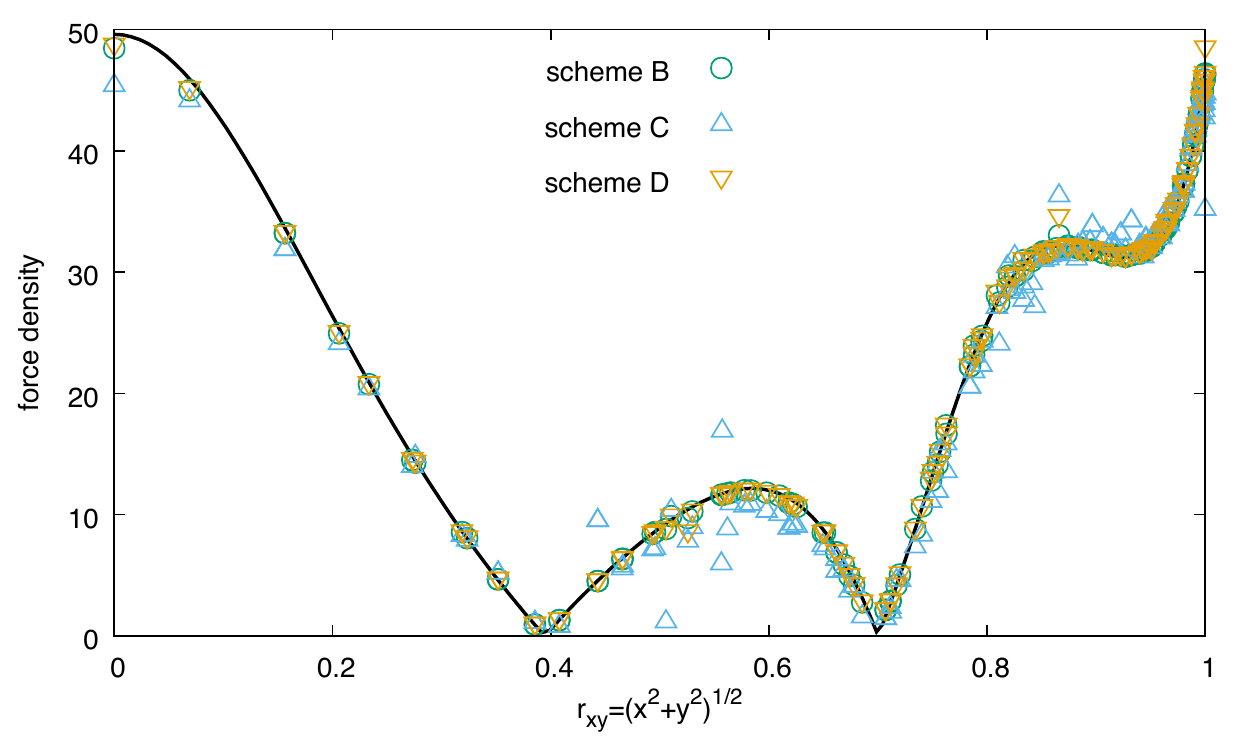}
\caption{}
\end{subfigure}
\caption{Force density (absolute value) versus radial distance of vertices from
axis of symmetry for a biconcave oblate with  $\alpha\kappa_{b}=1$. Force
due to non-local ADE energy. (a) $N_t=1280$ plotted for clarity with
every $4$ vertices in arbitrary
 order; (b) $N_t=5120$ and plotted for clarity with every $16$ vertices
 in arbitrary order.}
\label{fig_ade_force}
\end{figure}
For each scheme,  we present the force density with three resolutions in
Fig.~\ref{fig_minimal_force}, where analytical lines are computed with
software Maxima.  There is no convergence of force in the strict sense as
reported previously.  However, results from scheme B, C, D with $N_t \ge
1280$ follow the reference line reasonably close, with a few exceptions
at turning points of the curvature.  The results of scheme A are off
the reference.  We may draw similar conclusions from the results of
Fig.~\ref{fig_helfrich_force} that all three extended schemes are noisy,
but follow the reference lines closely.  For the force due to ADE energy,
we observe that scheme C is less accurate in comparison with the other
two schemes, as show on Fig.~\ref{fig_ade_force}.

\section*{Acknowledgement}
This work is supported by ERCl Advanced Investigator Award 341117.
X.B. benefits from discussions with Prof. J\"ulicher on scheme B.  
There is no numerical detail in Lim et al.~\cite{Lim2002}.
However, after we have finished this work,
X. B. learned via a private communication with Prof. Wortis
that they also applied scheme B in the context of Monte Carlo simulations,
and details of which are presented in a book~\cite{Lim2008}.


\end{document}